\ifdefined\MainOnly
  \def\OPTIONAppendix{0}
\else
  \def\OPTIONAppendix{1}
\fi

\def\OPTIONArxiv{0}%
\def\OPTIONConf{2}

\ifnum\OPTIONConf=1%
  \ifnum\OPTIONArxiv=0
\documentclass{llncs}

\setcopyright{none}             

\author{Matthew A. Hammer}
\affiliation{
  \institution{University of Colorado Boulder}
  \department{Department of Computer Science}}
\author{Joshua Dunfield}
\affiliation{
  \institution{University of British Columbia}
  \department{Department of Computer Science}
}
\author{Dimitrios J. Economou}
\affiliation{
  \institution{University of Colorado Boulder}
  \department{Department of Computer Science}
}
\author{Monal Narasimhamurthy}
\affiliation{
  \institution{University of Colorado Boulder}
  \department{Department of Computer Science}
}

  \else
         \documentclass[preprint,nonatbib]{sigplanconf}
         \usepackage[breaklinks]{hyperref}
         \usepackage{breakurl}
  \fi%
  \usepackage{joshuadunfield}
  \usepackage{euler}
\fi

\ifnum\OPTIONConf=2
   \documentclass[a4paper]{llncs}
   \usepackage[breaklinks]{hyperref}
   \usepackage{breakurl}
   \usepackage{joshuadunfield}

   \usepackage{amsthm}
   \usepackage{amsmath,amssymb,amsthm}
   \usepackage{thmtools,thm-restate}

   \usepackage{euler}
   \author{Matthew A.\ Hammer\INST{1}
     \and Joshua Dunfield\INST{2} \and
     \\  Dimitrios J.\ Economou\INST{1}
     \and Monal Narasimhamurthy\INST{1}
   }
   \INSTitute{%
     University of Colorado Boulder
     \and
     Queen's University (Canada)
   }
\fi%

\ifnum\OPTIONConf=0%
  \documentclass{purple}
  \usepackage{joshuadunfield}
  \usepackage{euler}
\fi

\usepackage{srcltx}

\usepackage{pst-node,graphicx,eepic}

\ifnum\OPTIONConf=0%
    \usepackage{fullpage}
\fi

\usepackage[overload]{textcase}

\usepackage{listings}

\ifnum\OPTIONConf=1%
  \setcitestyle{nosort}    
\else
  \usepackage[authoryear]{natbib}
  \bibpunct{(}{)}{;}{a}{}{,}
\fi

\usepackage{pgf,tikz}
\usetikzlibrary{shapes,arrows,matrix,fit,backgrounds,calc,decorations,decorations.pathmorphing}

\ifnum\OPTIONConf=0%
    \usepackage{charter}
\fi
\ifnum\OPTIONConf=1%
\fi
\ifnum\OPTIONConf=2%
\fi

\ifnum\OPTIONConf=2%
  \usepackage{latexsym}
\else
  \usepackage{amsmath,amssymb,amsthm}
  \usepackage{thmtools,thm-restate}
\fi

\usepackage{multirow}
\usepackage{enumerate}
\usepackage{comment}
\usepackage{mathpartir} \usepackage{proof}
\usepackage{framed}
\usepackage{rotating}
\usepackage{relsize}
\usepackage{wrapfig}
\usepackage{llproof}
\usepackage{rulelinks}

\usepackage{latexsym}
\usepackage{amsmath,amsthm,amssymb}
\usepackage{thmtools,thm-restate}

\usepackage{fancybox}

\def\MathparLineskip{\lineskiplimit=0.9em\lineskip=0.8em plus 0.2em}

\ifnum\OPTIONConf=0%
    \declaretheoremstyle[
      bodyfont=\sl
    ]{mytheoremstyle}
\fi

\usepackage{semantic}   
\usepackage{braket}     
\usepackage{mathpartir} 
\usepackage{rotating} 
\usepackage{proof}
\usepackage{pdflscape}
\usepackage{fancyvrb} 
\usepackage{amsmath,amsthm,amssymb}
\usepackage{thmtools,thm-restate}
\usepackage{mathtools}
\usepackage{color}

\let\MathRightArrow\Rightarrow 
\usepackage{marvosym} 
\def\Rightarrow{\MathRightArrow}

\ifnum\OPTIONConf=2
 \makeatletter

 \let\c@lemma\undefined

 \let\c@property\undefined

 \let\c@example\undefined

 \let\c@proposition\undefined

 \let\c@remark\undefined

 \let\c@definition\undefined

 \makeatother
\fi
\declaretheoremstyle[
  bodyfont=\sl
]{mytheoremstyle}

\ifnum\OPTIONConf=1

\else
    \declaretheorem[name=Theorem, style=mytheoremstyle]{theorem}
    \declaretheorem[name=Lemma, style=mytheoremstyle]{lemma}
    \declaretheorem[name=Corollary, style=mytheoremstyle, sibling=theorem]{corollary}
    \declaretheorem[name=Conjecture, style=mytheoremstyle, sibling=lemma]{conjecture}
    \declaretheorem[name=Example]{example}
    \makeatletter
    \declaretheorem[name=Proposition,style=mytheoremstyle,
        sibling=lemma,
      ]{proposition}
    \makeatother
    \declaretheorem[name=Definition,style=mytheoremstyle]{definition}
\fi
\declaretheorem[name=Property, style=mytheoremstyle]{property}

\declaretheorem[name=Remark]{remark}

\newcommand{\Label}[1]{\label{#1}} 
\newcommand{\FLabel}[1]{\label{#1}} 
\newcommand{\D}{\mathcal{D}}
\newcommand{\Ss}{\mathcal{S}}

\newcommand{\derives}{::}

\ifnum\OPTIONConf=0
\newtheorem*{thm*}{Theorem}

\fi

\newcommand{\such}{\mathrel{|}}

\newcommand{\Type}{\textsf{type}} 

\newcommand{\type}{\Type}

\newcommand{\St}{S} 
\newcommand{\StoreType}[1]{{#1}~\normalfont\textsf{store-type}} 

\newcommand{\sort}{\gamma} 
\newcommand{\namesort}{\boldsf{Nm}}
\newcommand{\namesetsort}{\boldsf{NmSet}}
\newcommand{\unitsort}{\boldsf{1}}
\newcommand{\Nm}{\namesort}
\newcommand{\NmSet}{\boldsf{NmSet}}

\newcommand{\effsubsym}{{\preceq}}
\newcommand{\effsub}{\mathrel{\effsubsym}}

\newcommand{\xwfeff}{\text{wf-effects}}
\newcommand{\wfeff}{\;\xwfeff}

\newcommand{\xctype}{\text{ctype}}
\newcommand{\ctype}{~\xctype}
\newcommand{\xtefftype}{\text{efftype}}
\newcommand{\tefftype}{~\xtefftype}

\newcommand{\xwfprop}{\text{prop}}
\newcommand{\wfprop}{~\xwfprop}

\newcommand{\appscope}[1]{\star(#1)}

\newcommand{\Unit}{\tyname{unit}}
\newcommand{\unitty}{\Unit}
\newcommand{\unit}{\textvtt{()}}
\newcommand{\unitexp}{\unit}

\newcommand{\unitindex}{\unit}

\newcommand{\IdxNmFnApp}[2]{#1[[#2]]}

\newcommand{\runonboldsf}{\sffamily\bfseries\selectfont}
\newcommand{\boldsf}[1]{\text{\normalfont\runonboldsf #1}}

\newcommand{\xF}{\boldsf{F}}
\newcommand{\F}{\xF\,}
\newcommand{\xU}{\boldsf{U}}

\newcommand{\disj}{\mathrel{\bot}}

\newcommand{\Value}{{~\normalfont\textsf{val}}}
\newcommand{\mapset}[3]{#1 [[ #2 ]] \leadsto #3 }

\newcommand{\trueprop}{\boldsf{tt}}
\newcommand{\andpropsym}{\boldsf{and}}
\newcommand{\andprop}{\mathrel{\andpropsym}}
\newcommand{\impty}{\supset}

\newcommand{\convsym}{=_{\beta}}
\newcommand{\conv}{\mathrel{\convsym}}
\newcommand{\convPf}[3]{\Pf{#1}{\conv}{#2}{#3}}
\newcommand{\continueconvPf}[2]{\Pf{}{\conv}{#1}{#2}}

\newcommand{\Allsym}{\forall}
\newcommand{\All}[1]{\Allsym{#1}.\,}

\newcommand{\DAll}[2]{\Allsym{{#1} \such {#2}}.\,}

\newcommand{\idxapp}[2]{{#1}({#2})}

\newcommand{\Split}[4]{\keyword{split}\Lparen{#1}, {#2}.{#3}.{#4}\Rparen}
 \newcommand{\Proj}[2]{\keyword{prj}_{#1}{#2}}
\newcommand{\Case}[5]{\keyword{case}\Lparen{#1}, {#2}.{#3}, {#4}.{#5}\Rparen}

\newcommand{\Scope}[2]{\keyword{scope}\Lparen{#1}, {#2}\Rparen}

\mathlig{::=}{\bnfas}
\mathlig{||-}{\Vdash}
\mathlig{|}{\;|\;}
\mathlig{[[}{\texttt{\upshape[}}
      \mathlig{]]}{\texttt{\upshape]}}
\mathlig{@@}{{\cdot}}
\mathlig{**}{\times}
\mathlig{|>}{\rhd}
\mathlig{->}{\arr}
\mathlig{-->}{\impty}
\mathlig{=>}{\Rightarrow}
\mathlig{<=}{\Leftarrow}

\mathlig{@>}{\stackrel{{}_{{}_{{}_\namesort}}}{\arr}}
\mathlig{@@>}{\stackrel{{}_{{}_{{}_\textsf{idx}}}}{\arr}}
\mathlig{*>}{\curvearrowright}

\newcommand{\effseqsym}{\mathsf{then}}
\newcommand{\effseq}{\,\effseqsym\,}

\newcommand{\effcoalsym}{\mathsf{after}}
\newcommand{\effcoal}{\,\effcoalsym\,}

\newcommand{\Lparen}{\texttt{(}}
\newcommand{\Rparen}{\texttt{)}}
\newcommand{\NmBin}[2]{\textvtt{$\left<\!\left<\hspace{-1pt}\right.\right.\!$}#1\textvtt{,}\;#2\textvtt{$\!\left.\left.\hspace{-1pt}\right>\!\right>$}}

\newcommand{\RefNr}[1]{\textsf{Ref}(#1)}
\newcommand{\Line}[1]{\textsf{Line}[[#1]]}

\newcommand{\SetNat}[1]{\textsf{Set}[[#1]]}
\newcommand{\SeqNat}[1]{\textsf{Seq}[[#1]]}
\newcommand{\VecNat}[0]{\textsf{Vec}}
\newcommand{\Level}[0]{\textsf{Lev}}

\newcommand{\SeqBinCons}[0]{\texttt{SeqBin}}
\newcommand{\SeqLeafCons}[0]{\texttt{SeqLf}}

\newcommand{\SetBinCons}[0]{\texttt{SetBin}}
\newcommand{\SetLeafCons}[0]{\texttt{SetLf}}


\newcommand{\chkcolor}{dBlue}
\newcommand{\syncolor}{dRed}

\newcommand{\chk}{\mathrel{\mathcolor{\chkcolor}{\Leftarrow}}}
\newcommand{\uncoloredsyn}{{\Rightarrow}}
\newcommand{\syn}{\mathrel{\mathcolor{\syncolor}{\uncoloredsyn}}}

\mathlig{<<}{\langle}
\mathlig{>>}{\rangle}
\mathlig{==>}{\syn}
\mathlig{<==}{\chk}
\mathlig{||}{\Downarrow}
\mathlig{!!}{\Downarrow}

\newcommand{\e}{\epsilon}

\newcommand{\ambns}{M}

\newcommand{\disjoint}{\mathrel{\bot}}

\newcommand{\tbrack}[1]{\texttt{\upshape[}{#1}\texttt{\upshape]}}

\newcommand{\leafname}{\textvtt{leaf}}

\newcommand{\Mv}{V}
\newcommand{\ntevalsym}{\Downarrow_{\textit{M}}}
\newcommand{\nteval}{\mathrel{\ntevalsym}}

\newcommand{\xrefv}{\keyword{ref}}
\newcommand{\xthunk}{\keyword{thunk}}

\newcommand{\xname}{\keyword{name}}
\newcommand{\xName}{\tyname{Nm}}

\newcommand{\xRef}{\tyname{Ref}}
\newcommand{\xThunk}{\tyname{Thk}}

\newcommand{\refv}[1]{\xrefv\;{#1}}
\newcommand{\thunk}[1]{\xthunk\;{#1}}

\newcommand{\name}[1]{\xname\;{#1}}
\newcommand{\Name}[1]{\xName\tbrack{#1}}

\newcommand{\xnamefn}{\keyword{nmfn}}

\newcommand{\namefn}[1]{\xnamefn\;{#1}}
\newcommand{\nametm}[1]{\namefn{#1}}

\newcommand{\Ref}[1]{\xRef\tbrack{#1}\,}
\newcommand{\Thk}[1]{\xThunk\tbrack{#1}\,}

\newcommand{\Thunk}[2]{\keyword{thunk}\Lparen{#1},{#2}\Rparen}

\newcommand{\Refe}[2]{\keyword{ref}\Lparen{#1},{#2}\Rparen} 

\newcommand{\List}[3]{\tyname{List}\tbrack{#1;#2}\,{#3}}
\newcommand{\ListDiff}[3]{\tyname{List}\diffbox{\tbrack{#1;#2}}\,{#3}}

\newcommand{\Int}[0]{\tyname{Int}}
\newcommand{\Nat}[0]{\tyname{Nat}}
\newcommand{\Bool}[0]{\tyname{Bool}}
\newcommand{\Listi}[2]{\tyname{List}\tbrack{#1}\,{#2}}

\let\Force\undefined
\newcommand{\Force}[1]{\keyword{force}\Lparen{#1}\Rparen}
\newcommand{\Get}[1]{\keyword{get}\Lparen{#1}\Rparen}

\newcommand{\Ret}[1]{\keyword{ret}\Lparen{#1}\Rparen}
\newcommand{\Let}[3]{\keyword{let}\Lparen{#1},{#2}.{#3}\Rparen}

\newcommand{\PreSt}[3]{{#1}\vdash^{#2}_{#3}}

\newcommand{\emptystore}{\cdot}

\newcommand{\DHasEffects}[3]{#1~\textsf{\upshape reads}~#2~\textsf{\upshape writes}~#3}
\newcommand{\DByHasEffects}[4]{#1~\textsf{by}~#2~\textsf{\upshape reads}~#3~\textsf{\upshape writes}~#4}
\newcommand{\mergeRds}{\mathrel{\cup}}

\newcommand{\te}{t}

\newcommand{\NotInScope}[1]{}
\newcommand{\REVISEME}[1]{{\color{blue}{(Text to revise goes here; See latex source)}}}

\definecolor{light-gray}{gray}{0.95}
\definecolor{white}{gray}{1}

\newcommand{\diffbox}[1]{\colorbox{dLighterBlue}{\!\ensuremath{#1}\!}}
\newcommand{\samebox}[1]{\colorbox{white}{\!\ensuremath{#1}\!}}

\newcommand{\ByStoreTypingRuleApp}{By rule (\Figref{fig:store-typing})}

\newcommand{\ByRWSetRule}{By \Defnref{def:rw}}

\let\inst\undefined
\newcommand{\inst}[2]{{#1}\tbrack{#2}}

\newcommand{\extractsym}{\textit{extract}}
\newcommand{\extract}[1]{\extractsym({#1})}

\newcommand{\extractassnssym}{\textit{extract-assns}}
\newcommand{\extractassns}[1]{\extractassnssym({#1})}
\newcommand{\extractctxsym}{\textit{extract-ctx}}
\newcommand{\extractctx}[1]{\extractctxsym({#1})}

\newcommand{\RuleHead}[1]{\text{\raisebox{1em}[0pt]{\ensuremath{\mathsz{\ifnum\OPTIONConf=1 14pt\else 18pt \fi}{#1}}}}~~~~~}

\newcommand{\inj}[1]{\keyword{inj}_{#1}\,}
\newcommand{\Inj}[1]{\inj{#1}}

\newcommand{\rulename}[1]{\text{\normalfont\textsf{#1}}}

\newrulecommand{TAbort}{\targettyperulename{Abort}}

\definecolor{darkgreen}{rgb}{0,0.5,0}
\definecolor{darkpurple}{rgb}{0.5,0,0.5}

\newcommand{\code}[1]{\lstinline|#1|}

\begin{document}

\ifnum\OPTIONConf=1
    \title{Refinement~types~for~precisely named cache locations}

\else
    \title{Refinement~types~for~precisely named cache locations}
\fi

%

\iffalse
\newcommand{\MattSays}[1]{\relax}
\newcommand{\MattSaysTODO}[1]{\relax}
\else
\newcommand{\MattSays}[1]{{\color{blue}{\textbf{Matt:}~#1}} \color{black}}
\newcommand{\MattSaysTODO}[1]{{\color{blue}{\textbf{TODO (Matt:)}~#1}} \color{black}}
\fi

\maketitle
\begin{abstract}
Many programming language techniques for incremental computation
employ programmer-specified \emph{names} for cached information.
At runtime, each name identifies a ``cache location'' for a dynamic
data value or a sub-computation; in sum, these cache location choices
guide change propagation and incremental (re)execution.

We call a cache location name \emph{precise} when it identifies
\emph{at most one} value or subcomputation; we call all other names
\emph{imprecise}, or \emph{ambiguous}.
At a minimum, cache location names must be precise to ensure that
change propagation works correctly; yet, reasoning statically about
names in incremental programs remains an open problem.

As a first step, this paper defines and solves the \emph{precise name
  problem}, where we verify that incremental programs with explicit
names use them precisely.
To do so, we give a refinement type and effect system, and prove it
sound (every well-typed program uses names precisely).
We also demonstrate that this type system is expressive by verifying
example programs that compute over efficient representations of
incremental sequences and sets.
Beyond verifying these programs, our type system also \emph{describes}
their dynamic \emph{naming strategies}, e.g., for library
documentation purposes.


\end{abstract}


\section{Introduction}
\label{sec:intro}

Language-based incremental computation techniques such as
\emph{self-adjusting computation} and Nominal Adapton~\citep{Hammer15}
strive to improve the asymptotic time complexity of programs using
\emph{general-purpose} incremental computing techniques that combine
function call caching, dynamic dependency graphs, and change
propagation~\citep{AcarThesis,Acar06,Acar06ML,Acar09,AcarLeyWild09,
  Hammer08,Hammer09,Ley-Wild08,Ley-Wild09,Chen12}.

In practice, each such system's \emph{cache implementation} exposes a
programming model where the incremental algorithm employs explicit
\emph{names}, where each name identifies a cache location for dynamic
data (a dynamic pointer allocation), or a cache location for
dynamic sub-computations (e.g., the arguments, dependencies, and
results of a recursive function call).
For an evaluation derivation~$\mathcal{D}$, we say that an
allocated pointer name is \emph{precise} when it has at most one
definition in~$\mathcal{D}$, and otherwise we say that a name is
\emph{ambiguous} (or \emph{imprecise}) when it identifies two or more
data structures and/or sub-computations.

This paper defines a verification problem, \emph{the precise name
  problem}, for programs that compute with explicit pointer names: prove that for
all possible inputs, in every execution, each name allocated by the
program is \emph{precise}, and not \emph{ambiguous}.
That is to say, the problem is to verify that every evaluation
derivation~$\mathcal{D}$ of a well-typed incremental program defines
each name uniquely.

Of course, across successive incremental (re-)evaluations of an
incremental program~(derivations~$\mathcal{D}_1, \mathcal{D}_2,
\ldots$), each name~$n$ may be associated with different values, when
and if this content changes incrementally (e.g., first value~$v_1$, then
later, $v_2$ such that $v_1 \ne v_2$).
In these situations, name~$n$ is still precise exactly when, \emph{in
  each execution}~$\mathcal{D}_i$, the incremental program only sees a
single, unique value~$v_i$ at the location named~$n$.

Hence, the goal of this paper is to \emph{statically} reason about one
execution at a time (not incremental re-executions) and show that
\emph{dynamically}, in each such execution, each name is used
precisely (uniquely).
Before outlining our approach and contributions, we give background on
programming with names.

\subsection{Background: Incremental computation with names}


In practice, incremental programs employ explicit cache location names
(1) to choose cached allocation names \emph{deterministically}, and (2) to witness
\emph{dynamic independence} between subproblems, thereby improving incremental reuse.

\paragraph{Deterministic allocation via precise names.}
The first role of explicit names for incremental computing concerns \emph{deterministic store allocation},
which permits us to give a meaningful definition to \emph{cached}
allocation.
To understand this role, consider these two evaluation rules (each of
the judgement form~$\sigma; e !! \sigma'; v$) for reference cell
allocation:
\begin{mathpar}
\Infer{alloc$_1$}
{ \ell \notin \textsf{dom}(\sigma) }
{ \sigma; \texttt{ref}_1(v) \Downarrow \sigma\{\ell \mapsto v\}; \textsf{ref}~\ell }
\and
\Infer{alloc$_2$}
{ n \notin \textsf{dom}(\sigma) }
{ \sigma; \texttt{ref}_2 (n, v) \Downarrow \sigma\{n \mapsto v\}; \textsf{ref}~n }
\end{mathpar}
The left rule is conventional: $\textsf{ref}_1$ allocates a value~$v$ at a store
location~$\ell$; because the program does not determine~$\ell$, the
implementor of this rule has the freedom to choose~$\ell$ any way that
they wish.
Consequently, this program is \emph{not} deterministic, and hence, \emph{not} a function.
Because of this fact, it is not immediately obvious what it means to
\emph{cache} this kind of dynamic allocation using a technique like
\emph{function caching}~\citep{Pugh89}, or techniques based on it.
To address this question, the programmer can determine a name~$n$
for the value~$v$, as in the right rule.
The point of this version is to expose the naming choice directly to
the programmer, and their incremental algorithm.

In some systems, the programmer chooses this name~$n$ as the hash value of value~$v$.
This naming style is often called ``hash-consing''.
We refer to it as \emph{structural naming}, since by using
it, the name of each ref cell reflects the \emph{entire structure} of
that cell's content.\footnote{The structural hashing approach is closely
  related to Merkle trees, the basis for revision-control systems like \code{git}.}
By contrast, in a \emph{nominal} incremental system, the programmer
generally chooses $n$ to be related to the \emph{evaluation context of
  using~$v$}, and often, to be \emph{independent} of the {value~$v$}
itself, providing \emph{dynamic independence} for subcomputations may
otherwise be treated as dependent.
Notably, since structural names lack \emph{any} independence with the
content that they name, structural names cannot provide dynamic independence.
We give one example of such a naming strategy below, and many more in
\Secref{overview}.

\paragraph{Dynamic independence via names.}
Programmers of incremental computations augment ordinary algorithms
with names, permitting incremental computing techniques like
memoization and change propagation to exploit dynamic
  independence in the cached computation.

As a simple illustrative example, consider the left version of
\code{rev} below, the recursive program that reverses a list.
After being reversed, suppose the input list to \code{rev} undergoes
incremental insertions and removals of list elements.  To respond to
these input changes, we wish to update the output (the reversed list)
by using the algorithm for \code{rev} below, along with
general-purpose incremental computing techniques, including
memoization of recursive calls to \code{rev}, marked by the
\code{memo}~keyword.

\begin{tabular}{p{0.45\textwidth}p{2mm}p{0.45\textwidth}}
\begin{minipage}{0.5\textwidth}
\begin{lstlisting}[basicstyle=\fontsize{8}{9}\ttfamily]
rev : List -> List -> List
rev l r = match l with
 | Nil       => r
 | Cons(h,t) => 

    memo(rev !t (Cons(h,r)))
\end{lstlisting}
\end{minipage}
&
&
\begin{minipage}{0.5\textwidth}
\begin{lstlisting}[basicstyle=\fontsize{8}{9}\ttfamily]
rev : List -> List -> List
rev l r = match l with
 | Nil       => r
 | Cons(n,h,t) => 
    let rr = ref(n, r) in 
    memo(rev !t (Cons(h,rr)))
\end{lstlisting}
\end{minipage}
\\
Accumulator~\code{r} consists of (depends on) the \emph{entire list prefix}.
&&

Ref cell~\code{rr} affords dynamic independence:
\code{rr} is never accessed by \code{rev}.
\end{tabular}
\\[2mm]

\noindent
The function~\code{rev} reverses a list of \code{Cons} cells, using an
accumulator~\code{r} that holds the reversed prefix of the list.
In the \code{Cons} case, the function~\code{rev} calls itself on the
(dereferenced) tail~\code{!t}, pusing the list element~\code{h} onto
the accumulator~\code{r}.
To incrementalize the code for~\code{rev}, we memoize the recursive
call to \code{rev}, which involves matching its arguments with those
of cached calls.

During incremental (re)executions, we wish to exploit the fact that
each recursive step of~\code{rev} is, in a sense, independent of the
other recursive steps.
However, the left version's use of~\code{memo} fails to achieve this
effect on its own, requiring an additional name~\code{n} and
reference cell~\code{rr}, as shown on the right.

To understand why, consider how change propagation re-evaluates and
how memoization reuses the cached calls to function~\code{rev}.
Specifically, after an input change, change propagation re-evaluates a
step of~\code{rev} where the input changed.  Next, the
accumulator~\code{Cons(h,r)} contains any incremental changes to head
value~\code{h}, as well as any changes in the prefix of the input list
(now reversed in the previous accumulator value~\code{r}).
This fact poses a problem for memoization since such accumulator
changes will repeatedly prevent us from \code{memo}-matching the
recursive call to~\code{rev}, which requires that all arguments match
that of a prior execution.
Consequently, change propagation will re-evaluate all the calls that follow the
position of an insertion or removal: This change is recorded in the
accumulator, which has changed structurally.

This memoization issue with~\code{rev} is an example of a general
pattern, also found in the output accumulators of quicksort and
quickhull, and more generally, in many cases when a recursive
algorithm allocates and consumes dynamic data structures.
To address this memoization issue, past researchers suggest that we
introduce nominal \code{ref}~cells into the accumulator, each
allocated with a name associated with their corresponding
input~\code{Cons} cell~\citep{Hammer15}.
In place of ``names'', earlier work variously used the terms
\emph{(allocation) keys}~\citep{AcarThesis,Hammer08,AcarLeyWild09} and
\emph{indices}~\citep{Acar06,Acar06ML}, but the core idea is the same.

In this updated (right hand) version, each name~\code{n} localizes any
change to the accumulator argument~\code{r}, since the dynamic
dependency graph will record the fact that~\code{rev} allocates, but
never accesses, these reference cells.
More precisely, these reference cells witness that, in fact, the
recursive steps of running~\code{rev} on a list are independent.
In \Secref{qh}, we show that quickhull (for computing convex hull) has
a similar accumulator structure; again, we use names to separate the
subproblems, yet still efficiently accumulate their results.

\subsection{Static reasoning for dynamic cache location names}

Statically, we wish to reason about incremental programs with explicit
names.
At a minimum, cache location names must be precise to ensure that
caching and change propagation work correctly together.
Beyond precision, other name-based considerations, such as a cache
eviction policy, are relevant too.
In larger systems, as we build algorithms that compose, we seek a way
of statically organizing and documenting different cache location name
choices made throughout (e.g., see the quick hull examples in
\Secref{sec:qh}).

Meanwhile, existing literature lacks static reasoning tools for cache
location names~(see \Secref{sec:related} for details).
As a first step in this direction, we give a refinement type and
effects system, toured in the next two sections, and defined formally
in \Secref{sec:typesystem}.
As our examples demonstrate, these (dependent) refinement types
provide a powerful descriptive tool for statically organizing and
documenting choices about cache location names.

%

\paragraph{Contributions:}
\begin{itemize}
\item
We define the \emph{precise name} verification problem for programs
that use names to identify (and \emph{cache}) their dynamic data and
computations.

\item 
We present a refinement type and effect system to describe and verify
programs that use explicit allocation names.


\item 
  To refine the types, we employ separate name and index term languages, for
  statically modeling names and name sets, respectively~(\Secref{sec:typesystem}).
  The name-set indices were inspired by set-indexed types in the style
  of DML \citep{Xi99popl}; we extend this notion with polymorphism over
  kinds and index functions; the latter was inspired by abstract
  refinement types \citep{Vazou13}.

\item
 In \Secref{examples}, we tour a collections library for sequences and sets.
 These programs demonstrate the descriptive power system.

\item 
  To validate our approach theoretically, we give a definition of \emph{precise
    effects} for an evaluation derivation~$\mathcal{D}$.  
  We prove
  that for all well-typed programs, every evaluation
  derivation~$\mathcal{D}$ consists of precise
  effects~(\Secref{sec:metatheory}).

\item 
  Drawing closer to an implementation, we give a bidrectional
  version of the type system that we show is sound and complete
  with respect to our main type assignment system~(\Secref{apx:bidir}).
  %
\end{itemize}

%

\section{Overview of our approach}
\label{sec:overview}
\label{overview}

Our type system statically approximates the \emph{name set} of each
first-class name, and the \emph{write set} of each sub-computation.
By reasoning about these sets statically, we rule out programs
that contain dynamic type errors, enforce precise names
where desired, and provide descriptive affordances to the
programmer, so that she can specify her program's nominal effects to
others (e.g., in the types of a module signature).

Without a suitable type system, programmer-chosen allocation names can
quickly give rise to imperative state, and unfortunately, to the
potential for unintended errors that undermine the type safety of a
program.

\begin{tabular}{p{0.45\textwidth}p{2mm}p{0.45\textwidth}}
\\[-1mm]
Re-allocation at \emph{different} types:
&&
Re-allocation at the \emph{same} type:
\\[-4mm]
\begin{lstlisting}[basicstyle=\fontsize{8}{9}\ttfamily]
let r1 = ref(n, (3,9))
let r2 = ref(n, 27)
\end{lstlisting}
&&
\begin{lstlisting}[basicstyle=\fontsize{8}{9}\ttfamily]
let r1 = ref(n, 0)
let r2 = ref(n, 27)
\end{lstlisting}
\\[-2mm]
\end{tabular}

\noindent 
What if a common name~\code{n} is used simultaneously for two
reference cells of two different types?
Should a program that re-allocates a name at different types be
meaningful? (e.g., see the first program to left, above).
We say, ``no'', such programs are not meaningful: They use a
name~\code{n} imprecisely at two \emph{different} types, undermining
the basic principle of type safety.

On the other hand, suppose a name~\code{n} is re-allocated at the \emph{same} type
(e.g., if \code{r1} and \code{r2} had the same type, as in the second
program), we may consider this behavior meaningful, though \emph{imperative}.
In this case, we want to know that the program is imperative, and that
we should not cache its behavior in contexts where we expect precise
names.

\paragraph{Complementary roles: Editor and Archivist.}
To distinguish imperative name allocation from name-precise
computation, our type system introduces two \emph{roles}, which we
term \emph{editor} and \emph{archivist}, respectively; specifically,
we define the syntax for roles as~$r ::= \textsf{ed}~|~\textsf{ar}$.
The archivist role~(\textsf{ar}) corresponds to computation whose dependencies we
cache, and the editor role~(\textsf{ed}) corresponds to computation that feeds the
archivist with input changes, and demands any changed output that is
relevant; in short, the editor represents the world outside the cached
computation.

\noindent
Consider the following typing rules, which approximate our full type system:
%
%
{ \small
\begin{mathpar}
\Infer{}
{ 
\Gamma |- v_n : \Name{X} 
\\
\Gamma |- v : A 
}
{ \Gamma |- \textsf{ref}( v_n, v ) : \RefNr{A} |> r(X) }
~\hfill~
\Infer{}
{ 
\Gamma |- e_1 : A |> \textsf{ar}(X)
\\\\
\Gamma, x : A |- e_2 : B |> \textsf{ar}(Y)
\\\\
\Gamma |- (X \disj Y) \equiv Z : \namesetsort
}
{\Gamma |- \Let{e_1}{x}{e_2} : B |> \textsf{ar}(Z)}
~\hfill~
\Infer{}
{ 
\Gamma |- e_1 : A |> \textsf{ed}(X)
\\\\
\Gamma, x : A |- e_2 : B |> \textsf{ed}(Y)
\\\\
\Gamma |- (X \cup Y) \equiv Z : \namesetsort
}
{\Gamma |- \Let{e_1}{x}{e_2} : B |> \textsf{ed}(Z)}
\end{mathpar} 
}
The rules conclude with the judgement form~$\Gamma |- e : A |> r(X)$,
mentioning the written set with the notation~$|> r(X)$, where the
set~$X$ approximates the set of written names, and $r$ is the role.

The first rule types a reference cell allocation named by
programmer-chosen value~$v_n$, whose type~$\Name{X}$ is \emph{indexed
  by} an approximate name set~$X$ from which name value $v_n$ is
drawn; in the rule's conclusion, this name set~$X$ serves as the allocation's
\emph{write set}.

The second and third rules give \texttt{let}~sequencing:
The second rule enforces the archivist role, where names are precise.
The third rule permits the editor role, where names allocated
later may \emph{overwrite} names allocated earlier.
The two rules each include a premise that judges the equivalence of
name set~$Z$~(at sort~$\namesetsort$) with sets $X$ and $Y$, which
index the write sets of $e_1$ and $e_2$, respectively.

For names to be precise, as in the archivist (middle) rule, the
written sets~$X$ and $Y$ must be disjoint, which we notate as~$X{\disj}Y$.
Otherwise, $X$ and $Y$ may overlap, permitting $e_2$ to overwrite one
or more names also written by~$e_1$, as in the third rule.
Though not shown here, we permit the archivist to \emph{nest}
within the editor, as a subroutine, but not vice versa.

In the remainder of this paper, we focus on verifying the
\emph{archivist's computations}, and we temporarily ignore the editor
role, for space considerations.

\paragraph{Name functions, higher-order apartness, write scope.}

Names are \emph{not} linear resources in most incremental programs.
In fact, it is common for precise programs to consume a name more than
once (e.g., see listings for quickhull in \Secref{qh}).
To disambiguate multiple writes of the same name, implementations
commonly employ \emph{distinct memo tables}, one for each unique function.
In this paper, we employ a more general notion: \emph{name functions}.

In particular, we define a higher-order account of \emph{separation}, or
\emph{apartness} (the term we use throughout) to compose precise
programs.
Just as apart names~$n \disj m$ do not overwrite each other (when
written, the two singleton write sets are disjoint), \emph{apart} name
functions~$f \disj g$ give rise to \emph{name spaces} that do not
overwrite each other, i.e., $f \disj g \iff \forall x. f\,x \disj
g\,x$. That is, $f\disj g$ when the images of $f$ and $g$ are always
disjoint name sets, for any preimage.

To streamline programs for common composition patterns, our type
system employs a special, ambient name space, the \emph{write scope}.
As illustrated in~\Secref{examples}, name functions and write
scopes naturally permit name-precise sub-computations to compose into
larger (name-precise) computations:
Just as functions provide the unit of composition for functional
programming, (apart) name functions provide the unit of composition for
(precise) naming strategies.

\section{Examples}
\label{examples}
\label{sec:examples}

In this section, we demonstrate larger example programs and their
types, focusing on a collections library of sequences and sets.

\subsection{Examples of datatype definitions: Sequences and sets}

We present nominal datatype definitions for sequences and finite maps,
as first presented in~\citet{Hammer15}, but without types to enforce
precision.

We represent sequences as \emph{level trees}. Unlike lists, level trees
permit efficient random editing~\citep{Hammer15,Headley16}, and their balanced
binary structure is useful for organizing efficient incremental
algorithms~\citep{Pugh89}.
We represent sets and finite maps as binary hash tries, which we build
from these level trees.

\paragraph{Level tree background:} 
A level tree is a binary tree whose internal binary nodes are each
labeled with a level, and whose leaves consist of sequence elements.
A level tree \emph{well-formed} when every path from the root to a
leaf passes through a sequence of levels that decrease monotonically.
Fortunately, it is simple to pseudo-randomly generate random levels
that lead to probabilistically balanced level trees~\citep{Pugh89}.
Because they are balanced, level trees permit efficient persistent
edits (insertions, removals) in logarithmic time, and zipper-based
edits can be even be more efficient~\citep{Headley16}.
Further, level trees assign sequences of elements (with interposed
levels) a single canonical tree that is \emph{history independent},
making cache matches more likely.

For nominal level trees, we use names to identify allocations,
permitting us to express incremental insertions or removals in
sequences as $O(1)$ re-allocations (store mutations).
By contrast, without names, each insertion or removal from the input
sequence generally requires~$O(\log(N))$ fresh pointer allocations,
which generally cascade during change propagation, often giving rise
to larger changes \citep{Hammer15}.

\paragraph{Sequences as nominal level trees.}
%
%
Below, type $\SeqNat{X}$ refines the ``ordinary type'' for
sequences~$\textsf{Seq}$; it classifies sequences whose names are
drawn from the set~$X$, and has two constructors for binary nodes and
leaves:
{%
\small%
\[
\begin{array}{@{}r@{\;}c@{\;}l@{}l}
\SeqBinCons &: & \All{X{\disj}Y{\disj}Z:\namesetsort} &
\Name{X} \hspace{0.7pt}{->}\hspace{0.7pt} \Level \hspace{0.7pt}{->}\hspace{0.7pt} \RefNr{\SeqNat{Y}} \hspace{0.7pt}{->}\hspace{0.7pt} \RefNr{\SeqNat{Z}} \hspace{0.7pt}{->}\hspace{0.7pt}
\SeqNat{X{\disj}Y{\disj}Z}
\\
\SeqLeafCons &{:}& \All{X:\namesetsort} & \VecNat -> \SeqNat{X}
\end{array}
\]
}%
For simplicity here, we focus on monomorphic sequences at base type (a
vector type~$\VecNat$ for vectors of natural numbers); in the
appendix, we show refinement types for polymorphic sequences whose
type parameters may be higher-kinded (parameterized by names, name
sets, and name functions).

In the type of constructor $\SeqBinCons$,
the notation $X {\disj} Y {\disj} Z$ denotes a set of names, and it
asserts \emph{apartness} (pair-wise disjointness) among the
three sets.
Binary nodes store a natural number \emph{level}, classified by type
$\Level$, a name, and two incremental pointers to left and right
sub-trees, of types $\RefNr{\SeqNat{Y}}$ and $\RefNr{\SeqNat{Z}}$, respectively.
The type $\Name{X}$ classifies first-class names drawn from the
set~$X$.
In this case, the apartness~$X\disj Y \disj Z$ means that the name
from~$X$ at the $\SeqBinCons$ node is distinct from those in its
subtrees, whose names are distinct from one another (name sets $Y$ vs
$Z$).
Generally, computations that consume a data structure assume that
names are non-repeating within a sequence (as enforced by this type);
however, computations conventionally reuse input names to name
corresponding constructors in the output, as in the \code{filter},
\code{trie} and \code{quickhull} examples below.

In the $\SeqLeafCons$ case, we store short vectors of elements to
capitalize on low-level operations that exploit cache coherency.  (In
empirical experiments, we often choose vectors with one thousand
elements).
Notably, this leaf constructor permits any name set~$X$; we find it
useful when types over-approximate their names, e.g., as used in the
type for \code{filter} below, which generally filters away some input names, but
does not reflect this fact in its output type.

\paragraph{Sets as nominal hash tries.}
In addition to sequences, we use balanced trees to represent sets and
finite maps.
A \emph{nominal hash trie} uses the hash of an element to determine a
path in a binary tree, whose pointers are named.
The type for~$\SetNat{X}$ is nearly the same as that of $\SeqNat{X}$, above;
we just omit the level at the binary nodes:
{%
\small%
\[
\begin{array}{rcll}
\SetBinCons &: & \All{X{\disj}Y{\disj}Z:\namesetsort} &
\Name{X} -> \RefNr{\SetNat{Y}} -> \RefNr{\SetNat{Z}} ->
\SetNat{X{\disj}Y{\disj}Z}
\\
\SetLeafCons &{:}& \All{X:\namesetsort} & \VecNat -> \SetNat{X}
\end{array}
\]
}%
The natural number vectors at the leaves at meant to represent
``small'' sets of natural numbers.
(In this section, we focused on giving types for sequences and sets of
natural numbers; \Secref{sec:polymorphism} of the supplement describes
polymorphic type definitions).

\paragraph{A collections library.}
\Secref{sec:structuralrec} gives simple structurally recursive
algorithms over level trees that we use as subroutines for later
implementing quickhull.
In \Secref{sec:trie}, we give an incrementally efficient conversion
algorithm~\code{trie} for building binary hash tries from level trees;
it uses nested structural recursion to convert a binary tree of type
$\SeqNat{X}$ into one of type $\SetNat{X}$.
\Secref{qh} gives two variants of the divide-and-conquer quickhull
algorithm, which only differ in their naming strategy.
In the context of these examples, we illustrate the key patterns for
typing the collections of \citet{Hammer15}, and relate these patterns
to the relevant typing rules of our system.

\subsection{Examples of structural recursion: Reducing and filtering sequences}
\label{sec:structuralrec}

\Figref{fig:max-listing} gives the type and code for \code{max} and
\code{filter}, which compute the maximum element of the sequence, and
filter the sequence by a predicate, respectively.
Both algorithms use structural recursion over the $\SeqNat{X}$ datatype,
defined above.

\begin{figure}[t]
\begin{tabular}{cc}

\begin{minipage}{0.43\textwidth}
\begin{lstlisting}[basicstyle=\fontsize{7}{8}\ttfamily]
max : $\All{X:\namesetsort}$
      $\SeqNat{X} -> \Nat$
    $|>~ \big( \lam{x:\namesort} \{ x@@1 \} \disj \{ x@@2 \} \big)  [[ X ]]$

max seq = match seq with
| SeqLf(vec) => vec_max vec
| SeqBin(n,_,l,r) =>
 let (_,ml) = memo[n@1](max !l)
 let (_,mr) = memo[n@2](max !r)
 if ml > mr then ml else mr

vec_max   : $\VecNat -> \Nat$
vec_filter: $\VecNat -> (\Nat -> \Bool) -> \VecNat$
\end{lstlisting}
\end{minipage}

&

\begin{minipage}{0.57\textwidth}
\begin{lstlisting}[basicstyle=\fontsize{7}{8}\ttfamily]
filter : $\All{X:\namesetsort}$
         $\SeqNat{X} -> (\Nat -> \Bool) -> \SeqNat{X}$
       $|>~ \big( \lam{x:\namesort} \{ x@@1 \} \disj \{ x@@2 \} \big)  [[ X ]]$

filter seq pred = match seq with
| SeqLf(vec) => SeqLf(vec_filter vec pred)
| SeqBin(n, lev, l, r) =>
 let (rl,sl) = memo[n@1](filter !l pred)
 let (rr,sr) = memo[n@2](filter !r pred)
 match (is_empty sl, is_empty sr)
 | (false,false) => SeqBin(n, lev, rl, rr)
 | (_,true) => sl 
 | (true,_) => sr
\end{lstlisting}
\end{minipage}
\end{tabular}

\caption{Recursive functions over sequences: Finding the max element, and filtering elements by a predicate.}
\label{fig:max}
\label{fig:max-listing}
\label{fig:filter-listing}
\end{figure}

The computation of \code{max} is somewhat simpler.  In the leaf case
(\SeqLeafCons), \code{max} uses the auxiliary
function~\code{vec_max}. In the binary case (\SeqBinCons), \code{max}
performs two memoized recursive calls, for its left and right
sub-trees.
We access these reference cells using~\code{!}, as in SML/OCaml notation.

In our core calculus, named thunks are the unit of function caching.
To use them, we introduce syntactic sugar for memoization, where
\code{memo[n@1]($\cdots$)} expands to code that allocates a named
thunk (here, named~\code{n@1}) and demands its output value:
%
\code{let x = thunk(n@1,$\cdots$) in forceref(x)}.
%
The primitive~\code{forceref} forces a thunk, and returns a pair
consisting of a reference cell representation of the forced thunk, and
its cached value; the reference cell representation of the thunk is
useful in \code{filter}, and later examples, but is ignored
in~\code{max}.
The code listing for \code{max} uses two instances of \code{memo},
with two \emph{distinct} names, each based on the name of the binary
node~\code{n}.

\paragraph{A formal structure for names.}
Our core calculus defines names as binary trees, $n ::=
\leafname~|~\NmBin{n}{n}$.
In practice, we often want to build names from primitive data, such as
string constants, natural numbers and bit strings; without loss of
generality, we assume embeddings of these types as binary trees.
For instance, we let $1$ be shorthand for
$\NmBin{\leafname}{\leafname}$, the name with one binary node;
similarly, $2$ stands for $\NmBin{\leafname}{\NmBin{\leafname}{\leafname}}$
and $0$ for $\leafname$.
In the code listing, we use \code{n@1} as shorthand for $\NmBin{n}{1}$.
More generally, we use \code{n@m} as a right-associative shorthand for
name lists of binary nodes, where \code{1@2@3} represents the
name~$\NmBin{1}{\NmBin{2}{3}}$.

The following deductive reasoning rules help formalize notions of name
equivalence and apartness for name terms:
\begin{mathpar}
\runonfontsz{9pt}
\Infer{}
{\Gamma |- M_1 \equiv N_1 : \namesort
  \\\\
\Gamma |- M_2 \equiv N_2 : \namesort
}
{\Gamma |-
  \!\!\arrayenvl{
  ~~~\NmBin{M_1}{M_2}
  \\ \equiv \NmBin{N_1}{N_2} : \namesort}}
~
\Infer{}
{\Gamma |- M_1 \disj N_1 : \namesort
}
{\Gamma |-
  \NmBin{M_1}{M_2}
  \disj \NmBin{N_1}{N_2} : \namesort
}
~
\Infer{}
{
  \Gamma |- M_1 \equiv N : \namesort
}
{\Gamma |- \NmBin{M_1}{M_2} \disj N : \namesort}
\end{mathpar}
We use the context~$\Gamma$ to store assumptions about equivalence and
apartness (e.g., from type signatures) and judge pairs of name terms
at a common sort~$\Nm$.
The first rule says that two binary names are equivalent if their left
and right sub-names are equivalent.
The second rule says that two binary names are apart (distinct) if
their left names are apart.
The third rule says that a binary name is always distinct from a name
equivalent to its left sub-name.
In the supplement (\Secref{sec:nametermlang} and
\Secref{sec:indextermlang}), we give more rules for deductive
apartness and equivalence of name terms and index terms; here and
below, we give a few selected examples.

We turn to the type of \code{max} given in the listing.
For all sets of names~$X$, the function finds the natural number in a
sequence (index by~$X$).
The associated effect (after the $|>$) consists of the following write set:
\[
   \big( \lam{x:\namesort} \{ x@@1 \} \disj \{ x@@2 \} \big)  [[ X ]] 
   ~~~
   \equiv
   ~~~
   \big( \lam{x:\namesort} \{ x@@1 \} [[ X ]] \big) \disj \big( \lam{x:\namesort} \{ x@@2 \} [[ X ]] \big)
\]
Specifically, this is an index term that consists of mapping a function
over a name set~$X$, and taking the (disjoint) union of these results.
Intuitively (but informally), this term denotes the following set
comprehension:
$\bigcup_{x \in X} \{ x@@1, x@@2 \}$.
That is, the set of all names~$x$ in~$X$, appended with either one of
name constants~$1$ and~$2$.
More generally, the index form~$f[[ X ]]$ employs a name set mapping
function~$f$ of sort $\Nm @@> \NmSet$, where index~$X$ is a name
set and has sort~$\NmSet$.
Here, $f :\equiv \Lam{x}\{ x@@1, x@@2 \}$.

The following table of rules give equivalence and apartness reasoning
(left vs.\ right columns) for index function abstraction and application
(top vs.\ bottom rows); by convention, we use $i$ and $j$ for general indices that may not be name sets,
and use $X$, $Y$ and $Z$ for name sets.
\begin{mathpar}
\small
\Infer{}
{\Gamma, \big(a \equiv b : \sort_1\big) |- i \equiv j                   : \sort_2 }
{\Gamma                                 |- \Lam{a}{i} \equiv \Lam{b}{j} : \sort_1 @@> \sort_2 }
~~~~
\Infer{}
{\Gamma, \big(a \equiv b : \sort_1\big) |- i \disjoint j                   : \sort_2 }
{\Gamma                                 |- \Lam{a}{i} \disjoint \Lam{b}{j} : \sort_1 @@> \sort_2 }
\\
\Infer{}
{
\arrayenvcl{
 \Gamma |- i \equiv j : \namesort @@> \namesetsort
  \\
 \Gamma |- X \equiv Y : \namesetsort
}
}
{\Gamma |- i [[ X ]] \equiv j [[ Y ]] : \namesetsort}
~~~~
\Infer{}
{\arrayenvcl{
  \Gamma |- i \disjoint j : \namesort @@> \namesetsort
  \\
 \Gamma |- X \equiv    Y : \namesetsort
 }
}
{\Gamma |- i [[ X ]] \disjoint j [[ Y ]] : \namesetsort}
\end{mathpar}
To be equivalent, function bodies must be equivalent under the
assumption that their arguments are equivalent; to be apart, their
bodies must be apart.
To be equivalent, function applications must have equivalent functions
and arguments; to be apart, the functions must be apart, with the same
argument.
These application rules are specialized to set comprehensions, using a function from names to name sets, and a starting name set.
An analogous pair of application rules reason about ordinary function application, with arbitrary input and output sorts (not shown here).

\paragraph{Allocating structurally recursive output.}
The type and listing for~\code{filter} is similar to that of~\code{max}.
The key difference is that \code{filter} builds an output data
structure with named reference cells; it returns a filtered sequence,
also of type~$\SeqNat{X}$.
Meanwhile, its write set is the same as that of~\code{max}.

Rather than allocate redundant cells (which we could, but do not),
each reference cell of filtered output arises from the \code{memo}
shorthand introduced above.
In particular, when the left and right filtered sub-trees are
non-empty, \code{filter} introduces a \code{SeqBin} code, with two
named pointers, \code{rl} and \code{rr}, the left and right references
introduced by memoizing function calls' results.
This ``trick'' works because the output is built via structural
recursion, via memoized calls that are themselves structurally
recursive over the input.
Below, we nest structural recursions.

\subsection{Example of nested structural recursion: Sequences into maps}
\label{sec:trie}

\begin{figure}[t]
\begin{tabular}{cc}

\begin{minipage}{0.43\textwidth}
\runonfontsz{7pt}
\begin{lstlisting}
trie :
  $\All{X:\namesetsort}$
    $\SeqNat{X} -> \SetNat{X}$
 $|>~ \big( \lam{x{:}\namesort} ( \lam{y{:}\namesort} \{ x@@y \} ) [[ \widehat{\textsf{join}}(X)]] \big) [[ X ]]$

trie seq = match seq with
| SeqLf(vec) => trie_lf vec
| SeqBin(n,_,l,r) =>
 let (tl,_) = memo[n@1](trie !l)
 let (tr,_) = memo[n@2](trie !r)
 let trie =[n] join n tl tr
 trie

$\textrm{where:}$
 $\widehat{\textsf{join}}(X) :\equiv \big( \lam{x{:}\namesort}\{x@@1\} \disj \{x@@2\} \big)^\ast [[ X ]]$
\end{lstlisting}
\end{minipage}

&

\begin{minipage}{0.57\textwidth}
\runonfontsz{7pt}
\begin{lstlisting}
join : $\All{X{\disj}Y:\namesetsort}$
       $\SetNat{X} -> \SetNat{Y} -> \SetNat{\widehat{\textsf{join}}(X{\disj}Y)}$
    $|>~\widehat{\textsf{join}}(X{\disj}Y)$

join n l r = match (l,r) with
| SetLf(l), SetLf(r) => join_vecs n l r
| SetBin(_,_,_), SetLf(r) =>
             join n@1 l (split_vec n@2 r)
| SetLf(l), SetBin(_,_,_) =>
             join n@1 (split_vec n@2 l) r
| SetBin(ln,l0,l1), SetBin(rn,r0,r1) => 
 let (_,j0) = memo[ln@1](join ln@2 l0 r0)
 let (_,j1) = memo[rn@1](join rn@2 l1 r1)
 SetBin(n, j0, j1)
\end{lstlisting}
\end{minipage}

\end{tabular}
\caption{Nested structural recursion: Converting sequences into maps}
\label{fig:trie}
\end{figure}

\Figref{fig:trie} defines functions~\code{trie} and~\code{join}.
The function~\code{trie} uses structural recursion, following a
pattern similar to~\code{max} and \code{filter}, above.
However, compared to \code{max} and \code{filter}, the body of the
\code{SeqBin} case for~\code{trie} is more involved: rather than
compare numbers, or create a single datatype constructor, it
invokes~\code{join} on the recursive results of calling~\code{trie} on
the left and right sub-sequences.
When a \code{trie} represents a set, the function~\code{join} computes
the set union, following an algorithm inspired by~\citet{Pugh89}, but
augmented with names.
Given two tries with apart name sets~$X$ and $Y$, \code{trie}
constructs a trie with names~$\widehat{\textsf{join}}(X{\disj}Y)$,
where $\widehat{\textsf{join}}$ is a function over name sets, whose
definition we discuss below.

\paragraph{The ambient write scope: a distinguished name space.}
To disambiguate the names allocated by the calls to~\code{join}, the
code for~\code{trie} uses a name space~\code{n}, with the
notation~\code{let trie =[n] $e$}.
In particular, this notation controls the ambient write scope, the
name space for allocations that write the store.
Informally, it says to perform the writes of the sub-computation~$e$
by first pre-pending the name~\code{n} to each allocation name.

To type the ambient write scope, we use a typing judgement
form~$\Gamma |-^N e : A |> X$ that tracks the ambient name space~$N$,
a name term of sort~$\Nm @> \Nm$.
(This arrow sort over names is more restricted than $\Nm @@> \NmSet$:
it only involves name construction, and lacks set structure.)
Consider the following typing rules for write scopes (left rule) and
name function application (right rule):
\begin{mathpar}
\small
    \Infer{}
        {
          \Gamma |- v_1 : \Name{X}
          \\
          \Gamma |- v_2 : A
        }
        {
          \Gamma |-^\ambns
          \Refe{v_1}{v_2}
          :
          \Ref{\IdxNmFnApp{M}{X}}{A}
          |>
          M[[X]]
        }
    ~~~~
    \Infer{}
        {
          \arrayenvbl{
          \Gamma |- v : (\namesort @@> \namesort)[[N']]
          \\
          \Gamma |-^{N \circ N'} e : A |> W
          }
       }
       {
         \Gamma |-^N \Scope{v}{e} : A |> W
       }
\end{mathpar}

The left rule uses the name term~$M$ to determine the final
allocation name for a reference cell.
As indicated by the conclusion's type and write effect, it computes a name in
set~$M[[X]]$, where index term~$X$ gives the name set of the argument.
The thunk rule (not shown here) is similar: it uses the ambient write
scope to determine the allocated name.

The right rule extends the ambient write scope~$N$ by name function
composition with~$N'$, for the duration of evaluating a given
subterm~$e$.
The notation in the code listing~\code{let x =[n]($e$)} is
shorthand for 
$\Let{x}{\Scope{\Lam{a}{n@@a}}{e}}{\cdots}$. 
That is, we prepend name~$n$ to each written name from evaluating
subterm~$e$.
Our collections library commonly uses this shorthand to disambiguate
names that are re-used across sequenced calls, as in the listings
for~\code{trie}, and in quickhull, below.

\paragraph{Types for recursive name sets.}
Sometimes we need to specify the names of a set recursively, as in the
code listing for~\code{join}, which is parameterized by an argument
name~\code{n}.
In some recursive calls to~\code{join}, this name argument ``grows''
into \code{n@1} or \code{n@2}.
In general, each name may grow a variable number of times.
This generality is required: The algorithm for \code{trie} allocates a
final binary hash trie whose structure is generally \emph{not}
isomorphic to the initial level tree; it is generally \emph{larger}.

To capture these recursive name-growth patterns, we introduce another
special variant of index terms, for describing sets that we ``grow''
inductively by applying a function zero or more times to an initial
set.  We use this index form to define $\widehat{\textsf{join}}$ from
\Figref{fig:trie}:
\begin{mathpar}
\runonfontsz{9pt}
\Infer{}
{
\arrayenvcl{
 \Gamma |- i \equiv j : \namesort @@> \namesetsort
  \\
 \Gamma |- X \equiv Y : \namesetsort
}
}
{
  \arrayenvl{
  \Gamma |- i^\ast [[ X ]] \equiv j^\ast [[ Y ]] : \namesetsort
  \\
  ~}
}
~
\Infer{}
{
\arrayenvcl{
  \Gamma |- i \equiv j : \namesort @@> \namesetsort
  \\
  \Gamma |- X \equiv Y : \namesetsort
 }
}
{\Gamma |-\!\!\!
  \arrayenvl{
    ~~~i^\ast [[ X ]]
    \\
    \equiv
    i^\ast [[ j [[ Y ]] ]] \union Y : \namesetsort}
}
~
\Infer{}
{
\arrayenvcl{
 \Gamma |- i \disjoint j : \namesort @@> \namesetsort
 \\
 \Gamma |- X \disjoint Y : \namesetsort 
}
}
{
  \arrayenvl{
  \Gamma |- i^\ast [[ X ]] \disjoint j^\ast [[ Y ]] : \namesetsort
  \\
  ~}
}
\end{mathpar}
The first rule matches that of the non-recursive form.  
The second rule is unique to recursive applications; it witnesses one
unfolding of the recursion.
The final rule says that two uses of recursive application are apart
when the starting sets and functions are apart.

\subsection{Examples of naming divide-and-conquer: Quickhull}
\label{sec:qh}
\label{qh}
\label{quickhull}
\label{sec:quickhull}

\Figref{fig:qh} defines two quickhull functions~\code{qh1}
and~\code{qh2}.
They consist of the same sequence-processing steps, but use differing
naming strategies, and thus exhibit different caching and change
propagation behavior.
Each version computes the furthest point~\code{p} from the current line~\code{ln} by using
\code{max_pt}, which is similar to \code{max} in \Figref{fig:max}.
This point~\code{p} defines two additional lines, \code{(ln.0,p)} and
\code{(p,ln.1)}, which quickhull uses to filter the input points
before recurring on these new lines.

We use the typing ingredients introduced above to give two naming
strategies, each with distinct cache behavior.
To summarize the naming strategies of~\code{max_pt} and \code{filter}
(each uses structural recursion over level trees), we introduce the
common shorthand~$\widehat{\textsf{bin}}[[ X ]] :\equiv
\big(\lam{x}\{x@@1,x@@2\}\big)[[ X ]]$.

The left naming strategy of~\code{qh1} uses the identity of the partition lines
to introduce three name spaces (for the calls to \code{max_pt} and
\code{filter}), and three allocations, for the two calls to
\code{qh1}, and the output hull point~\code{p} that is computed by
\code{max_pt}.
In this case, the library for computational geometry provides the lines'
names, where we exploit that quickhull does not process the
same line twice.
We elide some details in this reasoning, for brevity.

The right version~(\code{qh2}) ignores the names of lines.
Instead, it names everything based on a binary path in the
quickhull call graph:
The two recursive calls to~\code{qh2} each extend
write scope, first with name~\code{1}, then with~\code{2}.

The two versions are identical, \emph{except} for having different
naming strategies, and by extension, different cache behavior.
We elide the details, but in brief, the first strategy is better at
changes that disrupt and then restore the hull (inserting and then
removing a new hull point) but it also lacks a simple cache
\emph{eviction} strategy (when to ``forget'' about work for a line?).
By overwriting earlier computations based on call graph position, the
second naming strategy is cruder, but gives a natural cache eviction
strategy.

\begin{figure}[t]
\begin{minipage}{0.48\textwidth}
\begin{lstlisting}[basicstyle=\fontsize{7}{8}\ttfamily]
qh1 :
  $\All{X{\disj}Y{\disj}Z:\namesetsort}$
    $\Line{X} -> \SeqNat{Y} -> \SeqNat{Z} -> \SeqNat{Y{\disj}Z}$
  $|>~ \big( \lam{x{:}\namesort} ( \lam{y{:}\namesort} \{ x@@y \} ) [[ \widehat{\textsf{qh}_1}(X)]] \big) [[ X ]]$

qh1 ln pts h = 
 let p =[ln@1] max_pt ln pts
 let l =[ln@2] filter (ln.0,p) pts
 let r =[ln@3] filter (p,ln.1) pts
 let h = memo[ln@1](qh1 (p,ln.1) r h)
 let h = push[ln@2](p,h)
 let h = memo[ln@3](qh1 (ln.0,p) l h)
 h

$\widehat{\textsf{qh}_1}(X) :\equiv
\arrayenvl{
~~\big( \lam{a} \{1@@a, 2@@a, 3@@a\} \big)
[[ \widehat{\textsf{bin}}[[ X ]] ]]
\\ \disj  \{1,2,3\}
}
$
\end{lstlisting}
\end{minipage}
\!
\begin{minipage}{0.49\textwidth}
\begin{lstlisting}[basicstyle=\fontsize{7}{8}\ttfamily]
qh2 :
  $\All{X{\disj}Y{\disj}Z:\namesetsort}$
    $ \Line{X} -> \SeqNat{Y} -> \SeqNat{Z} -> \SeqNat{Y{\disj}Z}$
  $|>~ \big( \lam{y{:}\namesort} \{ 1@@y \} \disj \{ 2@@y \} \big)^\ast [[ \widehat{\textsf{qh}_2}(Y{\disj}Z) ]]$

qh2 ln pts h = 
 let p =[3@1] max_pt ln pts
 let l =[3@2] filter (ln.0,p) pts
 let r =[3@3] filter (p,ln.1) pts
 let h = memo[1]([1](qh2 (p,ln.1) r h))
 let h = push[2](p,h)
 let h = memo[3]([2](qh2 (ln.0,p) l h))
 h

$\widehat{\textsf{qh}_2}(X) :\equiv
\arrayenvl{
~~\big( \lam{a} \{3@@1@@a, 3@@2@@a, 3@@3@@a\} \big)
[[ \widehat{\textsf{bin}} [[ X ]] ]]
\\ \disj \{1,2,3\}
}
$
\end{lstlisting}
\end{minipage}

\caption{Two naming strategies for quickhull, same algorithmic structure.}
\label{fig:qh}
\end{figure}

%

\section{Program Syntax}
\Label{sec:progsyntax}

The examples from the prior section use an informally defined variant
of ML, enriched with a (slightly simplified) variant of our proposed
type system.
In this section and the next, we focus on a core calculus for
programs and types, and on making these definitions precise.

\begin{figure}[t]
  \centering
 
\begin{grammar}
  Values
  & $v$
      &$\bnfas$&
      $x
      \bnfalt \unitexp
      \bnfalt \Pair{v_1}{v_2}
      \bnfalt \inj{i}{v}
      \bnfalt \name{n}
      \bnfalt \namefn{M}
      \bnfalt \refv{n}
      \bnfalt \thunk{n}
      $
  \\[1ex]
  Terminal exprs.
  & $t$
      &$\bnfas$&
              $\Ret{v}
      \bnfalt \lam{x} e
      $
  \\[1ex]
  Expressions
  & $e$
      &$\bnfas$&
           $t
      \bnfalt \Split{v}{x_1}{x_2}{e}
      \bnfalt \Case{v}{x_1}{e_1}{x_2}{e_2}
      \bnfaltBRK e\;v
      \bnfalt \Let{e_1}{x}{e_2}
      \bnfalt \Thunk{v} e
      \bnfalt \Force{v}
      \bnfalt \Refe{v} v
      \bnfalt \Get{v}
      \bnfaltBRK \Scope{v}{e}
      \bnfalt v_M\;v
      $
\vspace*{-1.0ex}
\end{grammar}

  \caption{Syntax of expressions}
  \label{fig:expr}
\end{figure}


\subsection{Values and Expressions}

\Figref{fig:expr} gives the grammar of
values $v$
and expressions $e$.
We use call-by-push-value (CBPV) conventions
 in this syntax, and in
the type system that follows.
There are several reasons for this.
First,
CBPV can be interpreted as a ``neutral'' evaluation order that
includes both call-by-value or call-by-name, but prefers neither in
its design.
Second, since we make the unit of memoization a thunk,
and CBPV
makes explicit the creation of thunks and closures,
it exposes exactly the structure that we extend to a general-purpose
abstraction for incremental computation.
In particular, thunks are the means by which we cache results and track
dynamic dependencies.
%

Values $v$ consist of variables, the unit value, pairs, sums, and several
special forms (described below).

We separate values from expressions, rather than considering values to be a subset of 
expressions.  Instead, \emph{terminal expressions} $t$ are a subset of expressions.
A terminal expression $t$ is either $\Ret{v}$---the expression
that returns the value $v$---or a $\lambda$.
Expressions $e$ include terminal expressions,
elimination forms for pairs, sums, and functions
(\keyword{split}, \keyword{case} and $e\,v$, respectively);
let-binding (which evaluates $e_1$ to $\Ret{v}$
and substitutes $v$ for $x$ in $e_2$);
introduction (\keyword{thunk}) and elimination (\keyword{force})
forms for thunks; and introduction (\keyword{ref})
and elimination (\keyword{get}) forms for pointers (reference
cells that hold values).

The special forms of values are
names $\name{n}$,
name-level functions $\namefn{M}$,
references (pointers), and thunks.
References and thunks include a name $n$,
which is the name of the reference or thunk,
\emph{not} the contents of the reference or thunk.

The syntax described above follows that of prior work on
Adapton, including \citet{Hammer15}.
We add the notion of a \emph{name function},
which captures the idea of a namespace and other simple
transformations on names. 
%
The construct $\Scope{v}{e}$ construct controls monadic state for the current
name function, composing it with a name function $v$
within the dynamic extent of its subexpression $e$.
Name function application $M\;v$ permits programs to
compute with names and name functions that reside within the type indices.
Since these name functions always terminate, they do not affect a
program's termination behavior.


We do not distinguish syntactically between value pointers (for
reference cells) and thunk pointers (for suspended expressions); the
store maps pointers to either of these.

%

\subsection{Names}
\label{sec:names}

\begin{figure}[t]
\begin{grammar}
  Names
  & $m, n$
     &$\bnfas$&
           $\leafname$  & leaf name
   \\ 
   ~~(binary trees)
&&& $\bnfaltbrk \NmBin{n_1}{n_2}$ & binary name composition
 \\[1ex]
  Name terms
 &\!\!\!\!\!\!\!$M, N$
  &$\bnfas$& $n \bnfalt \NmBin{M_1}{M_2}$ & literal names, binary name composition
 \\ 
 \multicolumn{2}{l}{(STLC+names)\!\!\!}
 && $\bnfaltbrk a \bnfalt \lam{a} M \bnfalt \idxapp{M}{N}$
       & variable, abstraction, application
  \\[1ex]
      Name term values\!\!\!\!\!\!\!
     &$\Mv$
          &$\bnfas$&
          \multicolumn{2}{@{}l@{}}{%
          $n
          \bnfalt \lam{a} M
          $
          }
\\[1ex]
  Name term sorts
  & $\sort$
      &$\bnfas$&
      $\namesort$  &\!\!\!\!\! name; inhabitants $n$
   \\ &&& $\bnfaltbrk \sort @> \sort$  &\!\!\!\!\! name term function; inhabitants $\lam{a} M$
\\[1.0ex]
  Typing contexts
  & $\Gamma$
     &$\bnfas$&
           $
           \cdot
           \bnfalt
           \Gamma, a : \sort
           \bnfalt
           \cdots
           $  
           &\!\!\!\!\! full definition in \Figureref{fig:syntax-types}
\end{grammar}

  \vspace*{-1.2ex}

  \caption{Syntax of name terms: a $\lambda$-calculus over names, as binary trees}
  \label{fig:syntax-name-terms}
\end{figure}


\begin{figure}[t]
  \centering

  \judgbox{\Gamma |- M : \sort}
          {Under $\Gamma$,
            name term $M$ has sort $\sort$
          }
  \vspace{-2.0ex}
  \begin{mathpar}
    \Infer{\!M-const}
        {\strut}
        {\Gamma |- n : \namesort}
    \and
    \Infer{\!M-var}
        {(a : \sort) \in \Gamma}
        {\Gamma |- a : \sort}
    \and
    \Infer{\!M-bin}
        {
          \Gamma |- M_1 : \namesort
          \\\\
          \Gamma |- M_2 : \namesort
        }
        {\Gamma |- \NmBin{M_1}{M_2} : \namesort}
   \\
    \Infer{\!M-abs}
        {\Gamma, a : \sort' |- M : \sort}
        {
          \arrayenvbl{
          \Gamma |- (\lam{a} M) : (\sort' @> \sort)
          }
        }
  \and
  \Infer{\!M-app}
        {
          \Gamma |- M : (\sort' @> \sort)
          \\
          \Gamma |- N : \sort'
        }
        {\Gamma |- M(N) : \sort}
  \vspace*{-2.5ex}
  \end{mathpar}


  \vspace*{-1.0ex}

  \caption{Sorting rules for name terms $M$
  }
  \label{fig:sorting-name-terms}
\end{figure}


\begin{figure}[t]
  \centering

  \judgbox{M \nteval \Mv}
          {Name term $M$ evaluates to name term value $\Mv$
          }
  \vspace*{-1.5ex}
  \begin{mathpar}
    \Infer{teval-value}  
       {}
       {\Mv \nteval \Mv}
    ~~
    \Infer{teval-bin}
        {
          M_1 \nteval n_1
          \\
          M_2 \nteval n_2
        }
        {
          \NmBin{M_1}{M_2} \nteval \NmBin{n_1}{n_2}
        }
    ~~
    \Infer{teval-app}
          {
         \arrayenv{
          M \nteval \lam{a} M'
          \\
          N \nteval \Mv
          \\{}
          [V / a]M' \nteval \Mv'
         }
        }
       {M(N) \nteval \Mv'}
  \vspace*{-3ex}
  \end{mathpar}
  
  \caption{Evaluation rules for name terms}
  \label{fig:eval-name-terms}
\end{figure}


\Figureref{fig:syntax-name-terms}
shows the syntax of literal names, name terms, name term values, and name term sorts.
Literal names $m$, $n$ are simply binary trees: either an empty leaf $\leafname$
or a branch node $\NmBin{n_1}{n_2}$.
Name terms $M$, $N$ consist of
literal names $n$ and branch nodes $\NmBin{M_1}{M_2}$,
abstraction $\lam{a} M$ and application $\idxapp{M}{N}$.

Name terms are classified by sorts $\sort$: sort $\namesort$
for names $n$,
and $\sort @> \sort$ for (name term) functions.

The rules for name sorting $\Gamma |- M : \sort$
are straightforward  (\Figureref{fig:sorting-name-terms}),
as are the rules for name term evaluation $M \nteval \Mv$
(\Figureref{fig:eval-name-terms}). 
We write $M \conv M'$ when name terms $M$ and $M'$
are convertible, that is, applying any series of $\beta$-reductions
and/or $\beta$-expansions changes one term into the other.


\section{Type System}
\label{sec:typesystem}

The structure of our type system is inspired by Dependent ML \citep{Xi99popl,Xi07}.
Unlike full dependent typing, DML is separated into
a \emph{program level} and a less-powerful \emph{index level}.
The classic DML index domain is integers with linear inequalities,
making type-checking decidable.
Our index domain includes names, sets of names, and functions over names.
Such functions constitute a tiny domain-specific language that is
powerful enough to express useful transformations of names,
but preserves decidability of type-checking.

Indices in DML have no direct computational content.
For example, when applying a function on vectors that is indexed by
vector length, the length index is not directly manipulated at run time.
However, indices can indirectly reflect properties of run-time values.
The simplest case is that of an indexed \emph{singleton type}, such as $\Int[[k]]$.
Here, the ordinary type $\Int$ and the index domain of integers are in one-to-one
correspondence; the type $\Int[[3]]$ has one value, the integer 3.

While indexed singletons work well for the classic index domain of integers,
they are less suited to names---at least for our purposes.
Unlike integer constraints, where integer literals are common in types---%
for example, the length of the empty list is $0$---%
literal names are rare in types.
Many of the name constraints we need to express look like
``given a value of type $A$ whose name in the set $X$,
this function produces a value of type $B$ whose name is in the set $f(X)$''.
A DML-style system can express such constraints, but the types become verbose:
\begin{mathdispl}
   \arrayenvl{
     \forall a : \namesort.\; \forall X : \namesetsort.\;
     (a \in X) \impty
     \big(
          A[[a]] -> B[[f(a)]] 
     \big)
    }
\end{mathdispl}
The notation is taken from one of DML's descendants, Stardust \citep{DunfieldThesis}.
The type is read ``for all names $a$ and name sets $X$,
such that $a \in X$, given some $A[[a]]$ the function returns
$B[[f(a)]]$''.

We avoid such locutions by indexing single values by name sets, rather than names.
For types of the shape given above, this removes half the quantifiers
and obviates the $\in$-constraint attached via $\impty$:
%
$
   \arrayenvl{
     \forall X : \namesetsort.\;
         A[[X]] -> B[[f(X)]]
    }
$.
%
This type says the same thing as the earlier one, but
now the approximations
are expressed within the indexing of $A$ and $B$.
Note that $f$, a function on names, is interpreted pointwise:
$f(X) = \{ f(N) \such N \in X \}$.

(Standard singletons will come in handy for index functions on names,
where one usually needs to know the specific function.)

For aggregate data structures such as lists, indexing by a name set denotes
an \emph{overapproximation} of the names present.
That is, the proper DML type
\begin{mathdispl}
   \arrayenvl{
     \forall Y : \namesetsort.\; \forall X : \namesetsort.\;
     (Y \subseteq X) \impty
     \big(
          A[[Y]] -> B[[f(Y)]] 
     \big)
    }
\end{mathdispl}
can be expressed by
$
   \arrayenvl{
     \forall X : \namesetsort.\;
          A[[X]] -> B[[f(X)]] 
    }
$.


%
Following call-by-push-value \citep{Levy99,LevyThesis},
we distinguish \emph{value types} from \emph{computation types}.
Our computation types will also model effects, such as the allocation of a thunk with
a particular name.


\begin{figure}[t]
\begin{grammar}
  Index exprs.\ 
  & $i, j,$
      &$\bnfas$& $a$ & index variable
 \\ 
& $X, Y, Z,$
&& $\bnfaltbrk \{ N \}$ & singleton name set
 \\ 
& $R, W$
&& $\bnfaltbrk \emptyset ~|~ X \disj Y$ & empty set, separating union
\\ &&& $\bnfaltbrk X \union Y$ & union (not necessarily disjoint)
 \\ &&& $\bnfaltbrk \unitindex \bnfalt \Pair{i}{i} \bnfalt \Proj{1}{i} \bnfalt \Proj{2}{i}$
 & unit, pairing, and projection
 \\ &&& $\bnfaltbrk \lam{a} i \bnfalt i(j)$ & function abstraction and application
 \\ &&& $\bnfaltbrk M [[ i ]] \bnfalt i [[ j ]] \bnfalt i^\ast [[ j ]] $ & name set mapping and set building
 \\[1ex]
  Index sorts
  & $\sort$ 
     &$\bnfas$& 
  $\cdots~\bnfaltbrk \namesetsort$  & name set sort
  \\
  &&& $\bnfaltbrk \unitsort$      &unit index sort; inhabitant $\unitexp$
  \\
  &&& $\bnfaltbrk \sort * \sort$  &product index sort; inhabitants $\Pair{i}{j}$
  \\
& & & $\bnfaltbrk \sort_1 @@> \sort_2$ & index functions over name sets 
\end{grammar}

\vspace*{-3.0ex}

  \caption{Syntax of indices, name set sort}
  \label{fig:syntax-indices}
\end{figure}


\begin{figure}[t]
  \centering
 
\begin{grammar}
  Kinds
  & $K$
      &$\bnfas$&
      $\type$ & kind of value types
      \\ &&& $\bnfaltbrk \type => K$ & type argument (binder space)
      \\ &&& $\bnfaltbrk \sort => K$ & index argument (binder space)
  \\[1ex]
  Propositions
  & $P$
     &$\bnfas$&
     $\trueprop \bnfalt P \andprop P$ & truth and conjunction
     \\ &&& $\bnfaltbrk i \disj j : \gamma$ & index apartness
     \\ &&& $\bnfaltbrk i \equiv j : \gamma$ & index equivalence
  \\[1ex]
  Effects
  & $\e$
      &$\bnfas$&
               $<<W; R>>$
 \\[1ex]
  Value types
  & $A, B$
      &$\bnfas$&
            $\alpha \bnfalt d \bnfalt \Unit$ & type variables, type constructors, unit
      \\ &&& $\bnfaltbrk A + B \bnfalt A ** B$ & sum, product
      \\ &&& $\bnfaltbrk \Ref{i}{A}$  & named reference cell
      \\ &&& $\bnfaltbrk \Thk{i}{E}$  & named thunk (with effects) 
      \\ &&& $\bnfaltbrk  A[[i]]$  & application of type to index
      \\ &&& $\bnfaltbrk  A\;B$  & application of type constructor to type
      \\ &&& $\bnfaltbrk  \Name{i}$  & name type (name in name set $i$)
      \\ &&& $\bnfaltbrk  (\namesort @> \namesort)[[M]]$  & name function type (singleton)
\vspace*{-3ex}
\end{grammar}

\begin{grammar}
  Computation types
  & $C, D$
      &$\bnfas$&
      $\F A \bnfalt A -> E$  & li$\xF$t, functions
  \\[1ex]
  \dots with effects
  & $E$
      &$\bnfas$&
      $C |> \e$  & effects
 \\ &&& $\bnfaltbrk \All{\alpha : K} E  $   & type polymorphism
 \\ &&& $\bnfaltbrk (\DAll{a : \sort}{P} E)  $   & index polymorphism
 \\[1ex]
  Typing contexts
  & $\Gamma$
    &$\bnfas$&
    $\cdot$
  \\ &&& $\bnfaltbrk \Gamma, a : \sort$ & index variable sorting
  \\ &&& $\bnfaltbrk \Gamma, \alpha : K$ & type variable kinding
  \\ &&& $\bnfaltbrk \Gamma, d : K$ & type constructor kinding
  \\ &&& $\bnfaltbrk \Gamma, N : A$ & ref pointer
  \\ &&& $\bnfaltbrk \Gamma, N : E$ & thunk pointer
  \\ &&& $\bnfaltbrk \Gamma, x : A$ & value variable
  \\ &&& $\bnfaltbrk \Gamma, P$ & proposition $P$ holds
\end{grammar}
\vspace*{-2.5ex}

  \caption{Syntax of kinds, effects, and types}
  \label{fig:syntax-types}
\end{figure}


\begin{figure}
  \judgbox{\Gamma |- i : \sort}
          {Under $\Gamma$,
            index $i$ has sort $\sort$
          }
  \begin{mathpar}
    \Infer{sort-var}
        {
          (a : \sort) \in \Gamma
        }
        {
          \Gamma |- a : \sort
        }
    ~~~~
    \Infer{sort-unit}
        {}
        {
          \Gamma |- \unitindex : \unitsort
        }
    ~~~~
    \Infer{sort-pair}
        {
          \Gamma |- i_1 : \sort_1
          \\
          \Gamma |- i_2 : \sort_2
        }
        {
          \Gamma |- \Pair{i_1}{i_2} : (\sort_1 * \sort_2)
        }
    \and
    \Infer{sort-proj}
        {
          \Gamma |- i : \sort_1 * \sort_2
        }
        {
          \Gamma |- \Proj{b}{i} : \sort_b
        }
    ~~~~
    \Infer{sort-empty}
        { }
        {
          \Gamma |- \emptyset : \namesetsort
        }
    ~~~~
    \Infer{sort-singleton}
        {
          \Gamma |- N : \namesort
        }
        {
          \Gamma |- \{ N \} : \namesetsort
        }
    \and
    \Infer{\!sort-union}
        {
          \arrayenvbl{
            \Gamma |- X : \namesetsort
            \\
            \Gamma |- Y : \namesetsort
          }
        }
        {
          \Gamma |- (X \union Y) : \namesetsort
        }
    ~~
    \Infer{\!sort-sep-union}
        {
          \arrayenvbl{
            \Gamma |- X : \namesetsort
            \\
            \Gamma |- Y : \namesetsort
          }
          ~~~
          \extract{\Gamma} ||- X \disj Y : \namesetsort
        }
        {
          \Gamma |- (X \disj Y) : \namesetsort
        }
   \and
    \Infer{sort-abs}
    {
      \Gamma, a : \sort_1 |- i : \sort_2
    }
    {
      \Gamma |- (\Lam{a}{i}) : (\sort_1 @@> \sort_2)
    }
   ~~~~
     \Infer{sort-apply}
        {
          \Gamma |- i : \sort_1 @@> \sort_2
          \\
          \Gamma |- j : \sort_1
        }
        {
          \Gamma |- i(j) : \sort_2
       }        
     \\
     \Infer{\!sort-map}
          {
          \arrayenvbl{
          \Gamma |- M : \namesort @> \namesort
          \\
          \Gamma |- j : \namesetsort
          }
        }
        {
          \Gamma |- M[[j]] : \namesetsort
        }
    ~~
    \Infer{\!sort-build}
          {
          \arrayenvbl{
          \Gamma |- i : \namesort @@> \namesetsort
          \\
          \Gamma |- j : \namesetsort
          }\!\!\!\!\!\!
        }
        {
          \Gamma |- i[[j]] : \namesetsort
        }
    ~~
    \Infer{\!sort-star}
          {
          \arrayenvbl{
          \Gamma |- i : \namesort @@> \namesetsort
          \\
          \Gamma |- j : \namesetsort
          }\!\!\!\!
        }
        {
          \Gamma |- i^\ast[[j]] : \namesetsort
        }
        \vspace*{-2.5ex}
  \end{mathpar}

  \caption{Sorts statically classify name terms $M$, and the name indices $i$ that index types}
  \label{fig:sorting}
\end{figure}



\subsection{Index Level}

\Figureref{fig:syntax-indices} gives the syntax of
index expressions
and
index sorts (which classify indices).
We use several meta-variables for index expressions;
by convention, we use $X$, $Y$, $Z$, $R$ and $W$ only for sets of names---index
expressions of sort $\NmSet$.

%
%
%
%
%

\paragraph{Name sets.}
If we give a name to each element of a list,
then the entire list should carry the set of those names.
We write $\{N\}$ for the singleton name set,
$\emptyset$ for the empty name set,
and $X \disjoint Y$ for a union of two sets $X$ and $Y$
that requires $X$ and $Y$ to be disjoint;
this is inspired by the separating conjunction
of separation logic \citep{Reynolds02}.
While disjoint union dominates the types that we believe programmers need,
our effects discipline requires non-disjoint union,
so we include it ($X \union Y$) as well.

\paragraph{Variables, pairing, functions.}
An index $i$ (also written $X$, $Y$, \dots when the index is a set of names)
is either an index-level variable $a$,
a name set (described above: $\{N\}$, $X \disjoint Y$ or $X \union Y$),
the unit index $\unitindex$,
a pair of indices $\Pair{i_1}{i_2}$,
pair projection $\Proj{b}{i}$ for $b \in \{1,2\}$,
an abstraction $\lam{a} i$, application $i(j)$,
or name term application $M[[i]]$.

Name terms $M$ are \emph{not} a syntactic subset of indices $i$,
though name terms can appear inside indices (for example, singleton name sets $\{M\}$).
%
Because name terms are not a syntactic subset of indices (and name sets
are not name terms), the application form $i(j)$ does not allow us to
apply a name term function to a name set.  Thus, we also need name term
application $M[[i]]$, which applies the name function $M$ to each
element of the name set $i$.
The index-level map form $i[[j]]$ collects the
output sets of function~$i$ on the elements of the input set~$j$.
The Kleene star variation $i^\ast[[j]]$ applies the function~$i$ zero or more
times to each input element in set~$j$.

%

\paragraph{Sorts.}

We use the meta-variable $\sort$ to classify indices as well as name terms.
We inherit the function space ${@>}$ from the name term sorts (\Figureref{fig:syntax-name-terms}).
The sort $\namesetsort$ (\Figureref{fig:syntax-indices}) classifies indices that are name sets.
The function space ${@@>}$ classifies functions over \emph{indices} (e.g., tuples of name sets), not merely name terms.
The unit sort and product sort classify tuples of index expressions.


Most of the sorting rules in \Figureref{fig:sorting} are straightforward,
but rule `sort-sep-union' includes a premise
$\extract{\Gamma} ||- X \disj Y : \namesetsort$,
which says that $X$ and $Y$ are \emph{apart} (disjoint).

\paragraph{Propositions and extraction.}

Propositions $P$ are conjunctions of atomic propositions
$i \equiv j : \sort$
and
$i \disj j : \sort$, which express equivalence and apartness
of indices $i$ and $j$.  For example, $\{n_1\} \disj \{n_2\} : \NmSet$
implies that $n_1 \neq n_2$.
Propositions are introduced into $\Gamma$ via index polymorphism
$\DAll{a:\sort}{P} E$, discussed below.

The function $\extract{\Gamma}$
(\Figureref{fig:extract} in the appendix)
looks for propositions $P$, which become equivalence and apartness assumptions.
It also translates $\Gamma$ into the relational context used in the definition
of apartness.
We give semantic definitions of equivalence and apartness in the appendix
(Definitions \ref{def:super-semantic-equivalence-i}
and \ref{def:super-semantic-apartness-i}).

%

%

\subsection{Kinds}

We use a simple system of \emph{kinds} $K$ (\Figureref{fig:kinding} in the appendix).
Kind $\type$ classifies value types, such as $\unitty$ and $(\Thk{i}{E})$.

Kind $\type => K$ classifies type expressions that are parametrized by a type.
Such types are called \emph{type constructors} in some languages.
  
Kind $\sort => K$ classifies type expressions parametrized by an index.
For example, the $\tyname{Seq}$ type constructor from \Sectionref{examples}
takes a name set, \eg $\tyname{Seq}[[X]]$.
Therefore, this (simplified variant of) $\tyname{Seq}$ has kind $\namesetsort => \type$.
A more general $\tyname{Seq}$ type would also track its pointers (not just its names), and permit any element type, and would thus have kind:
$\namesetsort
=>
\big(
  \namesetsort => (\type => \type)
\big)$


\subsection{Effects}

Effects are described by $<<W; R>>$,
meaning that the associated code may write names in $W$,
and read names in $R$.

Effect sequencing (\Figureref{fig:comp-typing})
is a (meta-level) partial function over
a pair of effects:
the judgment
$\Gamma |- \e_1 \effseq \e_2 = \e$,
means that $\e$ describes the combination
of having effects $\e_1$ followed by effects $\e_2$.
Sequencing is a partial function because the effects are only
valid when
(1) the writes of $\e_1$ are disjoint from the writes of $\e_2$,
and
(2) the reads of $\e_1$ are disjoint from the writes of $\e_2$.
Condition (1) holds when each cell or thunk is not written more than once
(and therefore has a unique value).
Condition (2) holds when each cell or thunk is written before it is read.

Effect coalescing,
written ``$E \effcoal \e$'',
combines ``clusters'' of effects:
\[
\big(C |> <<\{n_2\}; \emptyset>>\big)
\effcoal
<<\{n_1\}; \emptyset>>
~=~
C |> (<<\{n_1\}; \emptyset>> \effseq <<\{n_2\}; \emptyset>>)
~=~
C |> <<\{n_1, n_2\}; \emptyset>>
\]
%
%
Effect subsumption $\e_1 \effsub \e_2$ holds when
the write and read sets of $\e_1$ are subsets of the respective sets of $\e_2$.

\subsection{Types}

\begin{figure}[t]
  \centering

  \judgbox{\Gamma |- v : A}{Under assumptions $\Gamma$, value $v$ has type $A$}
  \vspace*{-1.2ex}
  \begin{mathpar}
    \Infer{unit}
        {}
        {\Gamma |- \unitexp : \Unit}
    ~~~~~
    \Infer{var}
        {(x : A) \in \Gamma}
        {\Gamma |- x : A}
    ~~~~~
    \Infer{pair}
        {
          \Gamma |- v_1 : A_1
          \\
          \Gamma |- v_2 : A_2
        }
        {\Gamma |- \Pair{v_1}{v_2} : (A_1 ** A_2)}
    \\
    \Infer{name}
       {
         \Gamma |- n \in X
       }
       {
         \Gamma |- (\name{n}) : \Name{X}
       }
    ~~~~~
    \Infer{\!namefn}
       {
         \Gamma |- M_v : \namesort @> \namesort
         \\
         M_v \conv M
       }
       {
         \Gamma |- (\namefn{M_v}) : (\namesort @> \namesort)[[M]]
       }
   \\
    \Infer{\!ref}
       {
         \Gamma |- n \in X
         \\
         \Gamma(n) = A
       }
       {
         \Gamma |- (\refv{n}) : (\Ref{X}{A})
       }
    ~~~~~
    \Infer{\!thunk}
       {
         \Gamma |- n \in X
         \\
         \Gamma(n) = E
       }
       {
         \Gamma
         |-
         (\thunk{n})
         :
        (\Thk{X}{E})
       }
  \vspace*{-2.5ex}
  \end{mathpar}
  
  \caption{Value typing}
  \label{fig:value-typing}
\end{figure}

\begin{figure}[htbp]
  \centering

  $~$\hspace*{-2.0ex}\begin{minipage}[t]{0.64\linewidth}
  \judgbox{\Gamma |- (\e_1 \effseq \e_2) = \e}{Effect sequencing}
  \begin{mathpar}
  \hspace*{-5.0ex}
    \Infer{}
       {
       \arrayenvbl{
        \extract \Gamma |- W_1 \disj W_2
         \\
         \extract \Gamma |- R_1 \disj W_2
       }
      ~~
       \arrayenvbl{
         \extract \Gamma |- W_1 \cup W_2 \equiv W_3\!\!\!\!\!
         \\
        \extract \Gamma |- R_1 \cup R_2 \equiv R_3
       }
       }
       {
         \Gamma |-
         <<W_1; R_1>> \effseq <<W_2; R_2>>
         =
         <<
         W_3;
         R_3
         >>
      }    
  \end{mathpar}
  \end{minipage}
  \hspace*{-1ex}
\begin{minipage}[t]{0.34\linewidth}
  \hspace*{-6.0ex}\judgbox{\Gamma |- \e_1 \effsub \e_2}{Effect subsumption}
  \begin{mathpar}
   ~~\Infer{}
       {
         \Gamma |- (X_1 \disj Z_1) \equiv Y_1 : \NmSet
         \\
         \Gamma |- (X_2 \disj Z_2) \equiv Y_2 : \NmSet
       }
       {
         \Gamma
         |-
         <<X_1; X_2>> \effsub <<Y_1; Y_2>>
      }    
  \end{mathpar}
  \end{minipage}

  \medskip

  \judgbox{\Gamma |- (E \effcoal \e) = E'}{\small Effect coalescing}
  \vspace*{-3.5ex}
  \begin{mathpar} \hfill
    \Infer{}
       {
         \Gamma |- (\e_1 \effseq \e_2) = \e
       }
       {
         \Gamma |-
         \big(
           (C |> \e_2) \effcoal \e_1
         \big)
         =
         (C |> \e)
       }    
    ~~
      \Infer{}
         {
           \Gamma |- (E \effcoal \e) = E'
         }
         {
           \Gamma |- 
           (\All{\alpha : K} E) \effcoal \e
           \,=\,
           (\All{\alpha : K} E')
           \\\\
           \Gamma |- 
           (\DAll{a : \sort}{P} E) \effcoal \e
           \,=\,
           (\DAll{a : \sort}{P} E')
         }
  \vspace*{-0.8ex}
  \end{mathpar}

  \judgbox{\Gamma |-^\ambns e : E}
          {Under $\Gamma$, within namespace~$\ambns$,
            computation $e$ has type-with-effects $E$
          }
  \vspace*{-1.0ex}
  \begin{mathpar}
    \Infer{eff-subsume}
        {
          \Gamma |-^\ambns e : (C |> \e_1)
          \\
          \Gamma |- \e_1 \effsub \e_2
        }
        {
          \Gamma |-^\ambns e : (C |> \e_2)
        }
    \vspace*{-2.0ex}
    \\
    \Infer{split}
        {
          \arrayenvbl{
          \Gamma |-
          v : (A_1 ** A_2)
          \\
          \Gamma, x_1:A_1, x_2:A_2 |-^\ambns e : E
          }
        }
        {
          \Gamma |-^\ambns \Split{v}{x_1}{x_2}{e} : E
        }
    \and
    \Infer{case}
        {
          \Gamma |-
          v : (A_1 + A_2)
          \\
          \arrayenvbl{
            \Gamma, x_1:A_1 |-^\ambns e_1 : E
            \\
            \Gamma, x_2:A_2 |-^\ambns e_2 : E
          }
        }
       {
          \Gamma |-^\ambns \Case{v}{x_1}{e_1}{x_2}{e_2} : E
       }
    \and
    \Infer{ret}
        {
          \Gamma |- v : A
        }
        {
          \Gamma |-^\ambns
          \Ret{v}
          :
              (\F A)  |>  <<\emptyset; \emptyset>>
        }
    ~~~
    \Infer{let}
        {
          \arrayenvbl{
            \Gamma |-^M e_1 : (\F A) |> \e_1
            \\
            \Gamma, x:A |-^M e_2 : (C |> \e_2)
          }
          \\
          \Gamma |- (\e_1 \effseq \e_2) = \e
        }
        {
          \Gamma |-^M \Let{e_1}{x}{e_2}
          :
          (
             C |> \e
          )
        }
    \and
    \Infer{lam}
        {
          \Gamma,x:A |-^\ambns e : E
        }
        {
          \Gamma |-^\ambns
          (\lam{x} e)
          :
          \big(
              (A -> E) |> <<\emptyset; \emptyset>>
          \big)
       }
    ~~~~
    \Infer{app}
        {
          \arrayenvbl{
          \Gamma |- (E \effcoal \e_1) = E_1
          \\
          \Gamma |-^\ambns e
              : \big(
                 (A -> E) |> \e_1
                \big)
          }
          \\
          \Gamma |- v : A
        }
        {
          \Gamma |-^\ambns (e\;v) : E_1
        }
    \and
    \Infer{thunk}
        {
          \Gamma |- v : \Name{X}
          \\
          \Gamma |-^\ambns e : E
        }
        {
          \Gamma |-^\ambns
          \Thunk{v}{e}
          :
          \big(
             \F (\Thk{M[[X]]}{E})
          \big)
          |>
          <<M[[X]]; \emptyset>>
        }
    \and
    \Infer{force}
        {
          \Gamma |- v : \Thk{X}{(C |> \e)}
          \\
          \Gamma |- (<<\emptyset; X>> \effseq \e) = \e'
        }
        {
          \Gamma |-^\ambns
          \Force{v}
          :
          (
            C |> \e'
          )
        }
    \and
    \Infer{ref}
        {
          \Gamma |- v_1 : \Name{X}
          \\
          \Gamma |- v_2 : A
        }
        {
          \Gamma |-^\ambns
          \Refe{v_1}{v_2}
          :
             \F (\Ref{M[[X]]}{A})
          |>
          <<M[[X]]; \emptyset>>
        }
    ~~~
    \Infer{get}
        {
          \Gamma |- v : \Ref{X}{A}
        }
        {
          \Gamma |-^\ambns
          \Get{v}
          :
          (
             \F A
          )
          |>
          <<\emptyset; X>>
        }
    \and
    \Infer{name-app}
        {
          \arrayenvbl{
            \Gamma |- v_M : (\namesort @> \namesort)[[M]]
            \\
            \Gamma |- v : \Name{i}
          }
        }
        {
          \Gamma |-^N (v_M\;v) : \F (\Name{M[[i]]}) |> <<\emptyset; \emptyset>>
        }
    ~~~
    \Infer{scope}
        {
          \arrayenvbl{
          \Gamma |- v : (\namesort @> \namesort)[[N']]
          \\
          \Gamma |-^{N \circ N'} e : C |> <<W; R>>
          }
       }
       {
         \Gamma |-^N \Scope{v}{e} : C |> <<W; R>>
       }
    \\
    \Infer{\!AllIndexIntro}
       {
         \Gamma, a:\sort, P |-^\ambns t : E
       }
       {
         \Gamma |-^\ambns t : (\DAll{a : \sort}{P} E)
       }
    ~~~
    \Infer{\!AllIndexElim}
       {
         \arrayenvbl{
         \extract{\Gamma} ||- [i/a]P
         \\  
         \Gamma |-^\ambns e : (\DAll{a : \sort}{P} E)
         }
         \\
         \Gamma |- i : \sort
       }
       {
         \Gamma |-^\ambns
         e 
         : [i/a]E
       }
    \\
    \Infer{AllIntro}
       {
         \Gamma, \alpha : K |-^\ambns t : E
       }
       {
         \Gamma |-^\ambns
         t
         : (\All{\alpha : K} E)
       }
    \and
    \Infer{AllElim}
       {
         \Gamma |-^\ambns e : (\All{\alpha : K} E)
         \\
         \Gamma |- A : K
       }
       {
         \Gamma |-^\ambns
         e 
         : [A/\alpha]E
       }
  \vspace*{-2.5ex}
  \end{mathpar}
  
  \caption{Computation typing}
  \label{fig:comp-typing}
\end{figure}


The value types (\Figureref{fig:syntax-types}), written $A$, $B$, include
standard sums $+$ and products $**$;
a unit type;
the type $\Ref{i}{A}$ of references named $i$ containing a value of type $A$;
the type $\Thk{i}{E}$ of thunks named $i$ whose contents have type $E$ (see below);
the application $A[[i]]$ of a type to an index;
the application $A\;B$ of a type $A$ (\eg a type constructor $d$) to a type $B$;
the type $\Name{i}$;
and a singleton type $(\namesort @> \namesort)[[M]]$ where $M$ is a function on names.

As usual in call-by-push-value, computation types $C$ and $D$ include
a connective $\xF$, which ``lifts'' value types to computation types:
$\F A$ is the type of computations that, when run,
return a value of type $A$.
(Call-by-push-value usually has a connective dual to $\xF$, written $\xU$,
that ``th\textbf{U}nks'' a computation type into a value type;
in our system, $\mathsf{Thk}$ plays the role of $\xU$.)

Computation types also include functions, written $A -> E$.
In standard CBPV, this would be $A -> C$, not $A -> E$.
We separate computation types alone, written $C$, from computation types with
effects, written $E$; this decision is explained in \Appendixref{apx:remarks}.

Computation types-with-effects $E$ consist of $C |> \e$, which is the bare computation type $C$
with effects $\e$, as well as universal quantifiers (polymorphism) over types
($\All{\alpha : K} E$)
and indices
($\DAll{a : \sort}{P} E$).
In the latter quantifier, the proposition $P$ lets us express quantification over
disjoint sets of names.

\paragraph{Value typing rules.}
The typing rules for values (\Figureref{fig:value-typing})
for unit, variables and pairs are standard.
Rule `name' uses index-level entailment to check that the name $n$ is in the name set $X$.
Rule `namefn' checks that $M_v$ is well-sorted, and that $M_v$ is convertible to $M$.
Rule `ref' checks that $n$ is in $X$, and that $\Gamma(n) = A$, that is,
the typing $n : A$ appears somewhere in $\Gamma$.
Rule `thunk' is similar to `ref'.
 
\paragraph{Computation typing rules.}
Many of the rules that assign computation types (\Figureref{fig:comp-typing})
are standard---for call-by-push-value---with the addition of effects and the namespace $M$.
The rules `split' and `case' have nothing to do with namespaces or effects, so they
pass $M$ up to their premises, and leave the type $E$ unchanged.
Empty effects are added by rules `ret' and `lam', since both \textkw{ret} and
$\lambda$ do not read or write anything.
The rule `let' uses effect sequencing to combine the effects of $e_1$ and the let-body $e_2$.
The rule `force' also uses effect sequencing, to combine the effect of forcing the thunk
with the read effect $<<\emptyset; X>>$.

The only rule that modifies the namespace is `scope', which composes the given
namespace $N$ (in the conclusion) with the user's $v = \namefn{N'}$ in the second premise (typing $e$).


\subsection{Bidirectional Version}

The typing rules in Figures \ref{fig:value-typing}
and \ref{fig:comp-typing}
are declarative:
they define what typings are valid, but not how to derive those typings.
The rules' use of names and effects annotations
means that standard unification-based techniques,
like Damas--Milner inference, are not easily applicable.
For example, it is not obvious when to apply chk-AllIntro,
or how to solve unification constraints over names and name sets.

We therefore formulate bidirectional typing rules that directly
give rise to an algorithm.  For space reasons, this system is presented
in the supplementary material (\Appendixref{apx:bidir}).
We prove (in \Appendixref{apx:bidirectional-proofs}) that
our bidirectional rules are sound and complete with respect
to the type assignment rules in this section:

Soundness (Thms.\ \ref{thmrefsoundbival}, \ref{thmrefsoundbicomp}):
Given a bidirectional derivation for an annotated expression $e$,
there exists a type assignment derivation for $e$ without annotations.

Completeness (Thms.\ \ref{thmrefcomplbival}, \ref{thmrefcomplbicomp}):
Given a type assignment derivation for $e$ without annotations,
  there exist two annotated versions of $e$: one that synthesizes,
  and one that checks.  (This result is sometimes called \emph{annotatability}.)



\section{Dynamic Semantics}

\def\DYNSEMFIGMODE{1}
\ifnum\DYNSEMFIGMODE=1
\begin{figure}[t]
\else
\begin{figure}[h]
\fi

\begin{grammar}
Pointers & $p,q$ & $\bnfas$ & $n$ & name constants
\\[0.2ex]
Stores
& $\St$
    & $\bnfas$ &
    $\emptystore$ & empty store
\\ &&& $\bnfaltbrk$ $\St, p {:} v$ & $p$ points to value $v$
\\ &&& $\bnfaltbrk$ $\St, p {:} e\,{@}\,M$
         & $p$ points to thunk $e$, run in scope~$M$
\end{grammar}
\vspace{-1.3ex}

  \textbf{Notation}:
  \begin{tabular}[t]{l@{~~}l@{~~}l@{~~}l@{~~}l@{}}
  $\St\{p{\mapsto}v\}$
  &
  and
  &
  $\St\{p{\mapsto}e@M\}$
  &
  \emph{extend} $\St$ at $p$
  &
  when $p \notin \textsf{dom}(\St)$
  \\
   $\St\{p{\mapsto}v\}$ 
  &
  and
  &
  $\St\{p{\mapsto}e@M\}$
  &
  \emph{overwrite} $\St$ at $p$
  &
  when $p \in \textsf{dom}(\St)$
   \end{tabular}
\vspace{1ex}

\judgbox{\PreSt{\St_1}{M}{m} e !! \St_2 ; t}{
  Under store $\St$ in namespace $M$ at current node $m$, \\
  expression
  $e$ produces new store $\St_2$ and result $t$}
\begin{mathpar}
  \small
  \noindent
\ifnum\DYNSEMFIGMODE=2
    \Infer{\!split}
        {\PreSt{\St_1}{M}{m} [v_2/x_2][v_1/x_1]e             !! \St_2 ; e' }
        {\PreSt{\St_1}{M}{m} \Split{(v_1, v_2)}{x_1}{x_2}{e} !! \St_2 ; e' }
    ~~
    \Infer{\!case}
        {\PreSt{\St_1}{M}{m} [v_i/x_i]e_i                          !! \St_2 ; e' }
        {\PreSt{\St_1}{M}{m} \Case{\Inj{i}{v}}{x_1}{e_1}{x_2}{e_2} !! \St_2 ; e' }
    \\
    \Infer{let}
      {
          \PreSt{\St_1} {M}{m} e_1              !! \St_1' ; \Ret{v} 
          \\\\
          \PreSt{\St_1'} {M}{m} [v/x]e_2         !! \St_2' ; e_2'
      }
      {\PreSt{\St_1'}{M}{m} \Let{e_1}{x}{e_2} !! \St_2' ; e_2' }
    ~~~
    \Infer{app}
        {
            \PreSt{\St_1} {M}{m} e_1      !! \St_1' ; \lam{x} e_2
            \\\\
            \PreSt{\St_1} {M}{m} [v/x]e_2 !! \St_2' ; e_2'
        }
        {\PreSt{\St_1'}{M}{m} e_1\;v    !! \St_2' ; e_2' }
   \\
\fi
    \Infer{\!scope}
        {
          \PreSt{\St_1} {M_1 \circ M_2}{m} e !! \St_2 ; e'
        }
        { \PreSt{\St_1} {M_1}{m} \Scope{M_2}{e} !! \St_2 ; e' }
    ~~~
    \Infer{\!name-app}
        {
            M_1 \nteval \lam{a} M_2
            \\
            [n/a]M_2 \nteval p
        }
        {
          \PreSt{\St}{M}{m}
              M_1\;(\name{n})
          !! \St ; \Ret{\name{p}}
        }
\vspace*{-3.3ex}
\\
    \Infer{\!thunk}
      {
        (M\,n) \nteval p
        \\
        \St_1\{p{\mapsto}e\,{@}\,M\} = \St_2
      }
      {
        \arrayenvl{
        \PreSt{\St_1} {M}{m} \Thunk{\name{n}} e !! \St_2; \Ret{\thunk{p}}
        \\~
        }
      }
    ~~~~
    \Infer{\!ref}
        {
          \arrayenvbl{ ~\\
          (M\,n) \nteval p
          }
          ~~~~
          \St_1\{p{\mapsto}v\} = \St_2 
        }
        {
          \PreSt{\St_1} {M}{m} 
          \arrayenvl{
          ~~~\Refe{\name{n}}{v}
          \\
          !! \St_2; \Ret{\refv{p}}
          }
        }
   \vspace*{-2.9ex}
    \\
    \Infer{\!force}
      {
        S(p) = e~@~M_0
        \\\\
        \PreSt{\St_1}{M_0}{p} e !! \St_2; t
      }
      {
        \PreSt{\St_1} {M}{m} \Force{\thunk{p}} !! \St_2; t
      }
    ~~~
    \Infer{\!get}
      { \St(p) = v }
      { \PreSt{\St} {M}{m} \Get{\refv{p}} !! \St ; \Ret{v} }
    ~~~
  \Infer{\!term}
        {}
        {\PreSt{\St}{M}{m} t !! \St ; t }
 \vspace*{-2.7ex}
\end{mathpar}
\ifnum\DYNSEMFIGMODE=1
  \caption{Excerpt from the dynamic semantics (see also \Figureref{fig:dynamics})}
  \label{fig:dynamics-excerpt}
\else
  \caption{Dynamic semantics, complete}
  \label{fig:dynamics}
\fi

\end{figure}



\paragraph{Name terms.}

Recall \Figref{fig:eval-name-terms} (\Secref{sec:names}), which gives
the dynamics for evaluating name term~$M$ to name term
value~$V$.
Because name terms lack the ability to perform recursion and
pattern-matching, one may expect that they always terminate.
We demonstrate this using a logical relation~(\Theoremref{thmrefstrongnorm-name}).

\paragraph{Program expressions (\Figureref{fig:dynamics-excerpt}).}
Stores hold the mutable state that names dynamically identify.
Big-step evaluation
for expressions relates an initial and final store, and the
``current scope'' and ``current node'', to a program and value.
We define this dynamic semantics, which closely mirrors prior work,
to show that well-typed evaluations always allocate precisely.

To make this theorem meaningful, the dynamics permits programs to
\emph{overwrite} prior allocations with later ones: if a name is used
ambiguously, the evaluation will replace the old store content with
the new store content.
The rules~\textsf{ref} and \textsf{thunk} either extend or
overwrite the store, depending on whether the allocated pointer name
is precise or ambiguous, respectively.
We prove that, in fact, well-typed programs always extend (and never
overwrite) the store in any single derivation.  %
(During change propagation, not
  modeled here, we begin with a store and dependency graph
  from a prior run, and even precise programs overwrite the
  store/graph, as discussed in \Secref{sec:intro}.)

While motivated by incremental computation, we are interested in
precise effects here, not change propagation itself.
Consequently, this semantics is simpler than the dynamics of prior work.
First, the store never caches values from evaluation, that is, it does
not model function caching (memoization).
Next, we do not build the dependency edges required for change
propagation.
Likewise, the ``current node'' is not strictly necessary here, but we
include it for illustration.
Were we modeling change propagation, rules~\textsf{ref},
\textsf{thunk}, \textsf{get} and \textsf{force} would create
dependency edge structure that we omit here.
(These edges relate the current node with the node being observed.)



\section{Metatheory: Type Soundness and Precise Effects}
\label{sec:metatheory}

In this section, we prove that our type system is sound with respect
to evaluation and that the type system enforces precise
effects:
We establish that a well-typed,
terminating program produces a terminal computation of the program's
type, and that the actual dynamic effects are precise (\Defnref{def:rw}).
Specifically, we show that the type system's static effects soundly
approximate this dynamic behavior.
Consequently, sequenced writes never overwrite one another.

We sometimes constrain typing contexts to be \emph{store types},
which type store pointers but not program variables; hence, they only
type \emph{closed} values and programs:
\begin{defn}[Store type]
\Label{def:storetype}
We say that~$\Gamma$ is a \emph{store typing},
written~$\StoreType{\Gamma}$, when each assumption in~$\Gamma$ has the
reference-pointer-type form~$p : A$ or the thunk-pointer-type form~$p
: E$.
\end{defn}


\NotInScope{
(XXX:
Definition of the read sets for Eval-app and Eval-bind had $R_1 \mergeRds (R_2 - W_1)$.
We don't know why we were subtracting $W_1$ out; it means that the definition of a read
set excludes locations written to in the first subderivation, which doesn't match the type system.
So we are removing the subtractions.)
}

\begin{definition}[Precise effects]
\Label{def:rw}
  The precise effect of an evaluation derived by $\D$,
  written $\DHasEffects{\D}{R}{W}$,
  is defined in \Figureref{fig:readswrites} in the appendix.
\end{definition}

This is a (partial) function over derivations.
We call these effects ``precise'' since sibling sub-derivations must have
disjoint write sets.

\vspace*{-0.7ex}
\paragraph{Main theorem:}

We write $<<W'; R'>> \effsub <<W; R>>$
to mean that $W' \subseteq W$ and $R' \subseteq R$.
For proofs, see \Appendixref{apx:proofs}.

\clearpage
\begin{restatable}[Subject Reduction]{theorem}{thmrefsubj}
\Label{thm:refsubj}
~\\
If
$\StoreType{\Gamma_1}$ 
and
$        \Gamma_1 |- M : \namesort @> \namesort$
and
$\Ss$ derives $\Gamma_1 |-^M e : C |> <<W; R>> $
\\ and
$                 |- \St_1 : \Gamma_1$
and
$\Dee$ derives $\PreSt{\St_1}{M}{m} e !! \St_2 ; t$
then
\\
there exists $\Gamma_2 \supseteq \Gamma_1$ such that $\StoreType{\Gamma_2}$ and
$|- \St_2 : \Gamma_2$
and
$\Gamma_2 |- t : C |> <<\emptyset; \emptyset>> $
\\
and
$\DHasEffects{\Dee}{R_\Dee}{W_\Dee}$
and
$<<W_\Dee; R_\Dee>> \effsub <<W; R>>$.
\end{restatable}


\section{Related Work}
\label{sec:relatedwork}
\label{sec:related}
\label{related}

DML \citep{Xi99popl,Xi07} is an influential system of limited dependent types
or \emph{indexed} types.  Inspired by \citet{Freeman91},
who created a system in which datasort refinements were clearly separated from ordinary types,
DML separates the ``weak'' index level of typing from ordinary typing;
the dynamic semantics ignores the index level.

Motivated in part by the perceived burden of type annotations in DML,
liquid types \citep{Rondon08,Vazou13} deploy machinery to infer
more types.  These systems also provide more flexibility:
types are not indexed by fixed tuples.


To our knowledge, \citet{Gifford86} were the first to express effects within (or alongside) types.
Since then, a variety of systems with this power have been developed.
A full accounting of this area is beyond the scope of this paper;
for an overview, see \citet{Henglein05:ATTAPL-Chapter}.
We briefly discuss a type system for regions \citep{Tofte97},
in which allocation is central.
Regions organize subsets of data, so that they can be deallocated together.
The type system tracks each block's region, which in turn requires effects on types:
for example, a function whose effect is to return a block within a given region.
Our type system shares region typing's emphasis on allocation,
but we differ in how we treat the names of allocated objects.
First, names in our system are fine-grained, in contrast to giving all the objects in
a region the same designation.
Second, names have structure---for example, the names
$2{\cdot}n = \NmBin{\NmBin{\leafname}{\leafname}}{n}$
and $1{\cdot}n = \NmBin{\leafname}{n}$
share the right subtree~$n$---which
allows programmers to deterministically assign names.

Type systems for variable binding and fresh name generation,
such as FreshML \citep{Pitts00} and Pure FreshML \citep{Pottier07}, can express that sets
of names are disjoint.
But the names lack internal structure that relates specific names
across disjoint name sets.

%
%
%

Compilers have long used alias analysis to support optimization passes.
\citet{Brandauer15}
extend alias analysis with disjointness domains,
which can express local (as well as global) aliasing constraints.
Such local constraints are more fine-grained than classic region systems;
our work differs in having a rich structure on names.

\paragraph{Techniques for general-purpose incremental computation.}
General-purpose incremental computation techniques provide a
general-purpose \emph{change propagation} algorithm.
In particular, after an initial run of the program, as the input changes dynamically,
change propagation provides a provably sound approach for
recomputing the affected output
\citep{Acar06,AcarLeyWild09,Hammer15}.
Incremental computation can deliver \emph{asymptotic}
speedups for certain algorithms \citep{AcarIhMeSu07,
AcarAhBl08,
SumerAcIhMe11b,
Chen12}.
%
These incremental computing abstractions exist in many languages
\citep{Shankar07,
Hammer09, AcarLeyWild09}.
The type and effect system proposed here complements past work on
self-adjusting computation. In particular, we expect that variations
of the proposed type system can express and verify the use of names in
much of the work cited above.

\paragraph{Imprecise (ambiguous) names in past literature.}
Some past systems dynamically detect ambiguous names, either forcing
the system to fall back to a non-deterministic name
choice~\citep{Acar06,Hammer08}, or to signal an error and
halt~\citep{Hammer15}.
In scenarios with a non-deterministic fall-back mechanism, a name
ambiguity carries the potential to degrade incremental performance,
making it less responsive and asymptotically unpredictable in
general \citep{AcarThesis}.
To ensure that incremental performance gains are predictable, past
work often merely assumes, without enforcement, that names are
precise~\citep{Ley-Wild09,Cicek15,Cicek16}.

%







\section{Conclusion}
\label{sec:futurework}
\label{sec:conclusion}

We define the \emph{precise name} verification problem for programs
that use names to identify (and \emph{cache}) their dynamic data and
computations.
We present the first solution to this problem, in the form of a
refinement type and effect system.

We use the proposed type system to verify algorithms from an
incremental collections library, and other standard incremental
algorithms, such as quickhull.
Drawing closer to an implementation, we derive a bidirectional version
of the type system, and prove that it corresponds to our declarative
type system.  A key challenge in implementing the bidirectional system
is handling constraints over names, name terms and name sets; toward
this goal, we give decidable, syntactic rules to guide these checks.

\section*{Acknowledgments}

We thank Roly Perera and Kyle Headley for their comments and
suggestions on an earlier drafts.
This material is based in part upon work supported by a gift from
Mozilla, a gift from Facebook, and support from the National Science
Foundation under grant number CCF-1619282.
Any opinions, findings, and conclusions or recommendations expressed
in this material are those of the author(s) and do not necessarily
reflect the views of Mozilla, Facebook or the National Science Foundation.

\addtolength{\bibsep}{1pt}
\runonfontsz{9pt}
\bibliography{j,adapton}
\runonfontsz{10pt}

\ifnum\OPTIONAppendix=1
\clearpage
\appendix

\section{Programming Examples: Lists, polymorphism}
\label{sec:polylists}
\label{sec:polymorphism}

Through self-contained running examples that define and compute over
lists, we give a an overview of our type system.
We work towards showing examples of datatypes, and datatype
polymorphism in our proposed system, including how to define and use
abstract types that contain type parameters.
Specifically, we show how to define polymorphic lists that can be
instantiated appropriately to define lists of lists of integers.

We begin with a type for incremental lists of integers.
We use this list type to write \code{list_map}.
We show how this code, and in particular its use of names, induces a
dependence graph with names that derive from the input list.
In particular, we explain two connected activities:
(1) augmenting functional programs to name data and computations, and
(2) generalizing these naming strategies to write composable library
functions.
Toward the latter goal, we show how to generalize lists of integers to
lists of other structures that may contain names, such as other
(sub)-lists.

In support of creating composable incremental computing libraries, we
consider two composition problems, and illustrate how Typed Adapton
introduces additional type-level affordances to overcome them:
\begin{itemize}
\item Names are \emph{not} linear resources: It is natural and common
  to consume a name more than once within a computation.\footnote{As
    anecdotal evidence, consider that functional programs commonly
    exploit sharing in data structures and algorithms, and often do
    not adhere to a linear typing discipline. We still wish to name
    these computation patterns and data structures.} To disambiguate
  multiple writes of the same name, we introduce name functions that
  permit programmers to create distinct dynamic scopes. To streamline
  programs, Typed Adapton employs a scope monad to capture this
  dynamic structure.
\item \emph{Naive type polymorphism} ignores how names relate between
  a structure (e.g.,~a list of lists) and the structures that it
  contains (e.g.,~the sub-lists of the outer list).
  To describe and enforce name-based relationships generically, we
  parameterize generic types with index-transforming functions.  
  In turn, these index functions enforce client-chosen invariants that relate the type indices of sub- and
  super-structures, which may involve arbitrary name sets.
  For instance, we can write \emph{one} definition
  of polymorphic lists, and as clients, 
  instantiate this definition twice to define lists of
  lists which hold integers.
\end{itemize}

\subsection{Lists of integers, with names}

The type below defines incremental lists of integers, $\List{X}{Y}{\Int}$.
This type has two conventional constructors, \keyword{Nil} and
\keyword{Cons}, as well as two additional constructors that use the
two type indices $X$ and~$Y$.  
Each index has \emph{name set} sort~$\NmSet$, which classifies indices
that are sets of names.

The \keyword{Name} constructor permits names from set $X_1$ to appear in
the list sequence; it creates a \keyword{Cons}-cell-like pair holding
a first-class name from $X_1$ (of type~$\Name{X_1}$), and the rest of the
sequence, whose names are from~$X_2$.
The \keyword{Ref} constructor permits injecting reference cells
holding lists into the list type; these pointers' names are drawn from
set $Y = (Y_1 \disj Y_2)$, the disjoint union of the possible pointer
names $Y_1$ for the head reference cell, and the pointer names $Y_2$
contained in this cell's list.
\[
\begin{array}{llcllcl}
\keyword{Nil}  &:& \All{X,Y:\NmSet} &&\unitty&->&\List{X}{Y}{\Int}
\\
\keyword{Cons} &:& \All{X,Y:\NmSet} &\Int -> &\List{X}{Y}{\Int} &->& \List{X}{Y}{\Int}
\\
\keyword{Name} &:& \All{X_1{\bot}X_2,Y:\NmSet} &\Name{X_1} -> &\List{X_2}{Y}{\Int} &->& \List{X_1\disj X_2}{Y}{\Int}
\\
\keyword{Ref}  &:& \All{X,Y_1{\bot}Y_2:\NmSet} 
&\multicolumn{2}{l}{\Ref{Y_1}{} (\List{X}{Y_2}{\Int})} & \rightarrow & \List{X}{Y_1\disj Y_2}{\Int}
\end{array}
\]
For both of these latter forms, the constructor's types enforce that
each name (or pointer name) in the list is disjoint from those in the
remainder of the list.
For instance, when quantifying over these sets, we write $X_1 \disj
X_2$ to impose the constraint that~$X_1$ be disjoint from~$X_2$, and
similarly for $Y_1 \disj Y_2$;
However, for the~\code{Name} constructor, the names in $X_1$ and
pointer names in $Y$ may overlap (or even coincide).
These disjointness constraints consist of syntactic sugar that we expand in \Secref{sec:typesystem}.
The \keyword{Nil} constructor creates an empty sequence with
\emph{any} pointer or name type sets in its resulting type, since,
for practical reasons, we find it helpful to permit type indices to
\emph{over-approximate} name sets.

We distinguish two roles of names: \emph{unallocated names}, which are
unassociated with content of any type, and which need only be
\emph{locally unique} (e.g., unique to a list or other data
structure), and \emph{allocated names}, which dynamically name content
of some fixed type and must be \emph{globally unique}; they each identify
a pointer allocated in the store.

In the \code{Name(n,t)} form, the name~\code{n} is unallocated, and
can be used to allocate a reference or thunk of \emph{any}~type.
In the \code{Ref(r)} form, by contrast, we have a reference
cell~\code{r} that consists of a pointer name whose type is
parameterized by the type of the content it names, in this case an
integer list.

The scope~$s$ distinguishes namespaces, and is explained in further
detail below; intuitively, name funcion~$s$ translates a
locally unique name (unique to the list) into a globally unique name
(unique to global program evaluation).

\subsection{Mapping a list of integers, with names}

\Figref{fig:listmap} (left listing) lists the type, effect and code
for \code{list_map0}, which consumes the list type defined above, but
does not allocate any references or thunks.
We include it here for illustrative purposes, as preparation for
further examples.
The type signature includes abstract name sets~$X$ and $Y$, for the
names and named pointers of the input list~\code{l}.
The resulting list type is indexed by the name set~$X$ and the empty
pointer set~$\emptyset$ since the output contains the same names as
the input~(all drawn from~$X$), but does contain any pointers.
The function's effect, written as~$|> <<\emptyset;Y>>$, indicates that
the code writes no names, but reads the pointers~$Y$ from the input
list~\code{l}.

Turning to the program text for \code{list_map0}, the \code{Nil} and
\code{Cons} cases are entirely conventional: They return~\code{Nil}
and apply \code{f} to the head of each \code{Cons} cell, respectively.
The \code{Name} and \code{Ref} cases handle unallocated names and
(allocated) pointers in the list; both are simple recursive cases.
In the \code{Name} case, \code{list_map0}~maps the input name into
corresponding position of the output list and recurs.
In the \code{Ref} case, \code{list_map0}~uses~\code{get} to observe
the content of the reference cell holding the remainder of the input,
and recurs on this input list.

The right listing gives~\code{list_map1}, whose code and behavior is
identical to \code{list_map0} on the left, except for in the
\code{Name}~case, where the right version allocates a reference cell
to hold the output list, and it allocates (and forces) a thunk to
memoize the recursive call. (The shorthand \code{memo($e_1$,$e_2$)}
expands into \code{force(thunk($e_1$,$e_2$))}).
To name these allocations, \code{list_map1} uses the input list's
name~\keyword{n} for the reference cell and a distinct name for the
thunk, $(\keyword{n}{\cdot}1) :\equiv \NmBin{\keyword{n}}{\leafname}
\ne \keyword{n}$.
Critically, the name~$\keyword{n}{\cdot}1$ is distinct from both
\keyword{n} \emph{and} any name that is distinct from \keyword{n},
including the other names in the list.
That is, for all $m$, we have that $(\keyword{n}\disj m) \Rightarrow
(\keyword{n} \disj n{\cdot}1 \disj m \disj m{\cdot}1)$, meaning
that the four elements in the consequent are pairwise distinct.
Generalizing further, we can use this pattern to systematically add
memoization \emph{and} reference cells to any structurally recursive
algorithm that includes an analog of this~\code{Name} case.
We show examples of such algorithms, below.

Comparing its type with~\code{list_map0}, the returned list type
of~\code{list_map1} now contains names from~$X$ (as before) and
pointers from~$\appscope{X}$.
The effect is also affected: the written name set is now~$\appscope{X
  \disj X{\cdot}1}$, not $\emptyset$.
These sets refer to the names written in the revised~\code{Name} case,
mapped by the current monadic scope, which we denote with the
notation~$\appscope{-}$; recall that the scope is a
dynamically specified name-transforming function that we apply to each
written name, in this case drawn from~$X \disj X{\cdot}1$.
Further, we use shorthand~$X{\cdot}1$ to informally mean ``the names
of set~$X$, mapped by the function~$\Lam{a}{\NmBin{a}{\leafname}}$'';
in \Secref{sec:typesystem}, we make the syntax for name sets, type indices and
effects precise.

\begin{figure}[t]
\begin{tabular}{l@{~}|@{~}l}
\footnotesize
\begin{lstlisting}
list_map0: $\diffbox{\All{X,Y:\namesetsort}}$
  $(\Int -> \Int)-> (\ListDiff{X}{Y}{\Int}) ->$
  $(\ListDiff{X}{\emptyset}{\Int})$
  $\diffbox{|> <<\emptyset; Y>>}$
list_map0 = $\lambda$f. fix rec. $\lambda$l.
 match l with
  Nil       => Nil
  Cons(h,t) => Cons(f h, rec t)
  Name(n,t) => Name(n,   rec t)


  $~$
  Ref(r)    => rec (get r)
\end{lstlisting}
&
\footnotesize
\begin{lstlisting}
list_map1: $\All{X,Y:\namesetsort}$
  $(\Int -> \Int)-> (\List{X}{Y}{\Int}) ->$
  $(\List{X}{\diffbox{\appscope{X}}}{\Int})$
  $|> <<\diffbox{\appscope{X{\disj}X{\cdot}1}}; Y>>$
list_map1 = $\lambda$f. fix rec. $\lambda$l.
 match l with
  Nil       => Nil
  Cons(h,t) => Cons(f h, rec t)
  Name(n,t)
    => Name(n,
        Ref(ref(n, memo(n$\cdot$1,
                    rec t))))
  Ref(r)    => rec (get r)
\end{lstlisting}
\end{tabular}

\caption{\code{list_map0} (left) and \code{list_map1} (right) consist
  of the standard algorithm to map a list of integers~\code{l} using a
  given integer function~\code{f}, augmented with additional cases for
  \code{Ref} and \code{Name}.  
  The left version performs no allocations; the right version uses
  names from the input list to name reference cells and recursive
  calls.
}

\label{fig:listmap}
\end{figure}


\subsection{Names identify nodes in a dynamic dependency graph}

Suppose that we execute \code{list_map1} on an input list with two
integers, two names~$a$ and $b$, and two reference cells~$p$ and
$q$, as shown in the left of~\Figref{fig:listmap}.
Bold boxes denote reference cells in the input and output lists, and
bold circles denote memoized thunks that compute with them.  Each bold
object (thunk or reference) is named precisely.
The two thunks, shown as bold circles named by $a{\cdot}1$ and
$b{\cdot}1$, follow a similar pattern: each has outgoing edges to the
reference that they dereference~(northeast, to~$p$ and $q$,
respectively), the recursive thunk that they force, if any (due west,
to~$b{\cdot}1$ for $a{\cdot}1$, and none for $b{\cdot}1$), and the
reference that they name and allocate~(southeast, to $a$ and $b$,
respectively).

The benefit of using names stems from the nominal indirection that
they afford.  
Operationally, imperative~$O(1)$ changes to the input structure can be
reflected into the output with only $O(1)$ changes to its named
content.
Further, for the memoized thunks, the names also provide a primitive
form of \emph{cache eviction}: Memoized results are \emph{overwritten}
when names are re-associated with new content.
%

\begin{figure}
\begin{tabular}{ll}
\includegraphics[width=0.48\textwidth]{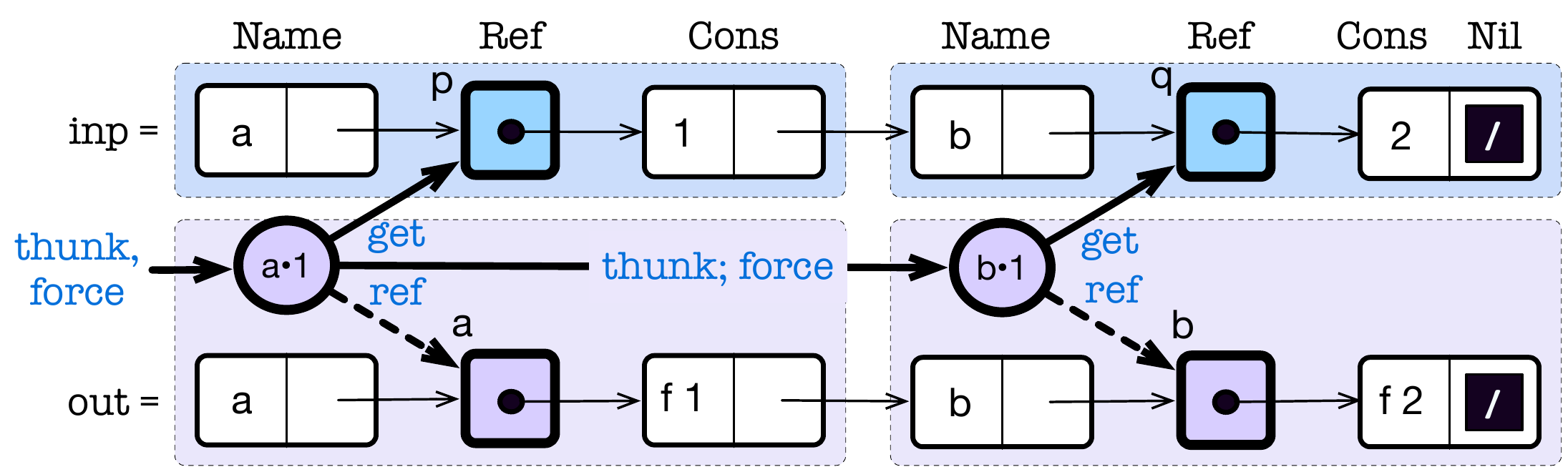}
&
\includegraphics[width=0.48\textwidth]{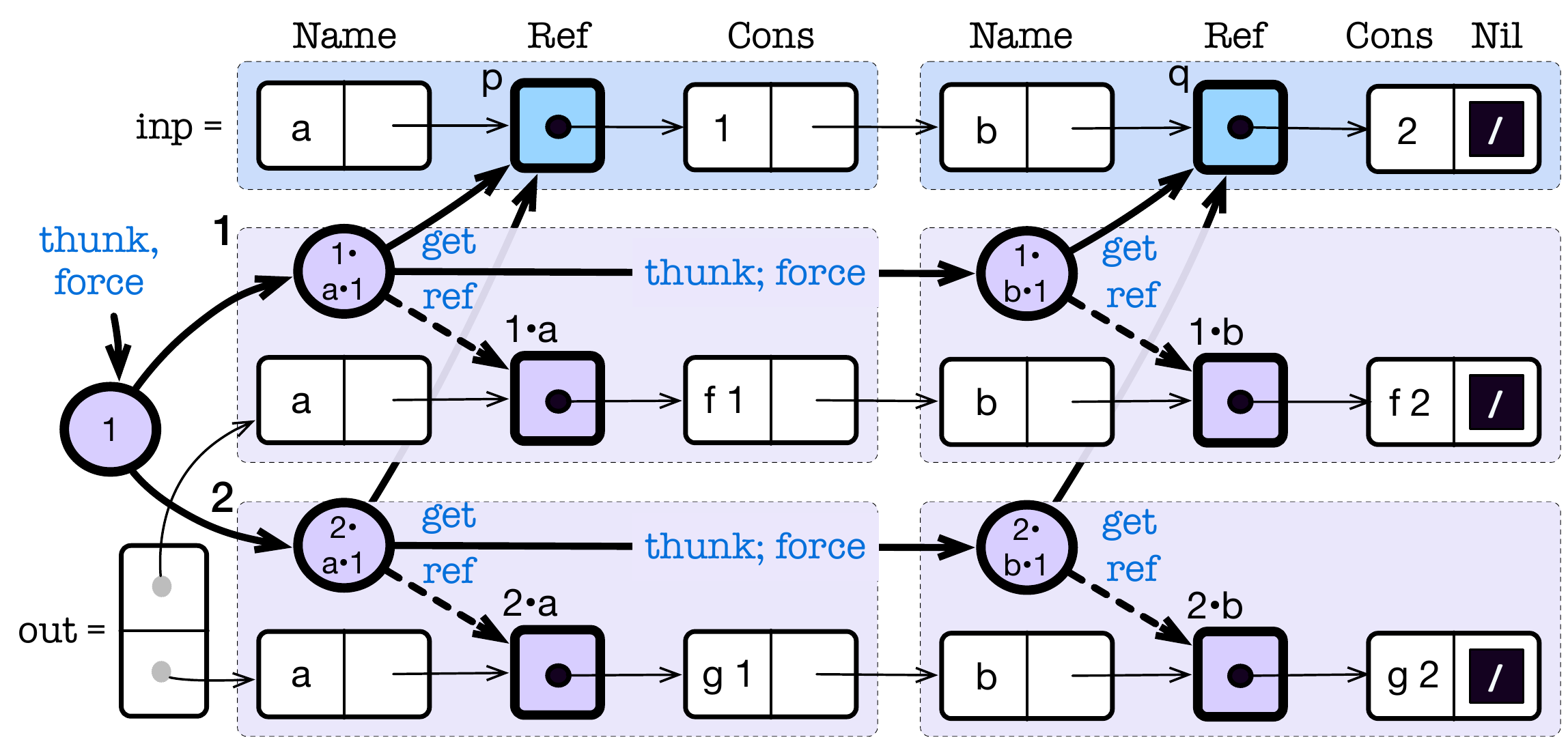}
\end{tabular}

\caption{ \textbf{Left:} The dependence graph produced by evaluating
  function~\code{list_map1} on~\code{inp}, a two-integer, two-name
  input list, shown in the two upper blue regions of the figure.
Its names are drawn from set $X=\{a, b, \ldots\}$, and its pointer
names are drawn from the set~$Y=\{p, q, \ldots\}$.
The two purple sections consist of allocated dependence graph nodes
that depend on the input list: cell~$a$, thunk~$a{\cdot}1$, cell~$b$,
and thunk~$b{\cdot}1$.  
\textbf{Right:} The dependence graph produced by evaluating
\code{map_pair} on the same input list, in two different scopes:~$1$
and~$2$. The scopes give rise to precise names for the output lists
and their associated computations.}
\end{figure}

\subsection{Scopes distinguish distinct sub-computations}

Programs commonly disambiguate two uses of the same name by using two
different write scopes.
Specifically, the write scope monad implicitly threads a name function
through the computation, applying it to each name that the computation
allocates.
Conceptually, the write scope at the outermost layer of the computation is
the identity name function.
As the computation enters subtrees of the dynamic call tree, the write scope
monad nests the write scope by composing name functions.

\begin{lstlisting}
let map_pair l = 
  let xs = scope s1 [ list_map1 f l ]
  let ys = scope s2 [ list_map1 g l ]
    (xs, ys)
let out = memo($1$, map_pair inp)
\end{lstlisting}

For instance, consider two uses of map, with two different
element-mapping functions~\code{f} and \code{g}.
This code uses disjoint write scopes~$s_1 \disj s_2$ to map an input
list~\code{inp} twice within a single incremental program to produce a
\emph{pair} of distinct output lists~\code{out} that each depend on
the common input list~\code{inp}.
To make the example concrete, suppose that $s_1$ and $s_2$ prepend the
name constants~$1$ and~$2$ on to their name argument, respectively.
This program gives rise to the dependence graph shown
in~\Figref{fig:listmap} (right side).
This diagram illustrates how the two sub-computations
of~\code{list_map1} share the input list~\code{inp}: Each of its
reference cells has two incoming~\code{get} edges.
(Recall, names are~\emph{not} linear resources, so this is not problematic).
To distinguish these write scopes, the distinct name functions~$\NmBin{1}{{-}}$ and~$\NmBin{2}{{-}}$ disambiguate the name of
each reference cell and thunk for~\code{list_map1} (e.g., compare the
allocated names of the left-hand and right-hand versions of the figure).

\subsection{Polymorphic lists, with names}

Above, we considered lists of integers.  Now, suppose the archivist
wishes to employ lists of other data structures, such as lists of
lists.  For instance, past literature commonly implements incremental
mergesort using functions over lists of lists.
First, consider what we call \emph{naive type polymorphism}, where we
replace each instance of ``$\Int$'' in type of integer lists
($\List{X}{Y}{\Int}$) with the universally quantified type
variable~$\alpha$:
\newcommand{\diffalpha}{\diffbox{\alpha}}
\[\runonfontsz{9pt}
\begin{array}{l@{\,}l@{\,}l@{\,}l@{\,}l@{\,}l@{\,}c@{\,}l}
\keyword{Nil}  &:& \diffbox{\All{\alpha:\Type}}&\All{X,Y:\NmSet} &&1&->&\List{X}{Y}{\diffalpha}
\\
\keyword{Cons} &:& \diffbox{\All{\alpha:\Type}}&\All{X,Y:\NmSet} &\diffalpha -> &\List{X}{Y}{\diffalpha} &->& \List{X}{Y}{\diffalpha}
\\
\keyword{Name} &:& \diffbox{\All{\alpha:\Type}}&\All{X_1{\bot}X_2,Y:\NmSet} &\Name{X_1} -> &\List{X_2}{Y}{\diffalpha} &->& \List{X_1\disj X_2}{Y}{\diffalpha}
\\
\keyword{Ref}  &:& \diffbox{\All{\alpha:\Type}}&\All{X,Y_1{\bot}Y_2:\NmSet}
&\multicolumn{2}{l}{\Ref{Y_1}{} (\List{X}{Y_2}{\diffalpha})} & \rightarrow & \List{X}{Y_1\disj Y_2}{\diffalpha}
\end{array}
\]
As we demonstrate, naively parameterizing lists by~$\alpha$ does not
afford enough invariants to verify all the programs that we wish to
author.
For instance, suppose the archivist wishes to express the list
transformation~\code{list_join} that appends the sub-lists of a list
of lists; in the \code{Cons} case, it uses auxiliary
function~\code{list_append} to append the sublist at the head with the
remainder of the sublists' elements in the (flattened) tail.

\Figref{fig:listjoin} lists the code for these standard list algorithms,
as well as types for each.
As with \code{list_map0}, we initially consider versions that do not
allocate, for simplicity; later, we can augment the algorithms with
named allocations in the same fashion that we systematically
transformed \code{list_map0} into~\code{list_map1}.
The code for \code{list_append0} checks against the listed type, with
reasoning similar to that of \code{list_map0}; it does not involve
lists of lists, only lists of integers.
As we explain below, the code for \code{list_join0} does not check
against the listed type: The listed type uses polymorphism naively
with respect to name sets.
To overcome this problem, we need to change the refinement type
for~\code{Cons}.
To see why, first consider checking the body against the listed type.

\begin{figure}[t]
\begin{tabular}{l@{\,}|@{\,}l}
\footnotesize
\begin{lstlisting}
            $\checkmark$
list_append0:
   $\All{X_1{\bot}X_2,Y_1{\bot}Y_2:\namesetsort}$
      $\big(\List{X_1}{Y_1}{\Int}\big)$ 
    $-> \big(\List{X_2}{Y_2}{\Int}\big)$
    $-> \big(\List{X_1 \bot X_2}{Y_2}{\Int}\big)$
   $|>  \; <<\emptyset, Y_1>>$
list_append0$\;$=$\;$fix rec. $\lambda$l. $\lambda$r.
 match l with 
  Nil      =>r
  Cons(h,t)=>let t' = rec t r
                Cons(h, t')
  Name(n,t)=>Name(n, rec t)
  Ref(r)   =>rec (get r)
\end{lstlisting}
&
\begin{lstlisting}
          $?$
list_join0:
   $\All{X_1{\bot}X_2,Y_1{\bot}Y_2:\namesetsort}$
      $\big(\List{X_1}{Y_1}{(\List{X_2}{Y_2}{\Int})}\big)$ 
     $-> \big(\List{X_1 \bot X_2}{\emptyset}{\Int}\big)$
    $|>  \; <<\emptyset, Y_1 \bot Y_2>>$

list_join0 = fix rec. $\lambda$l.
 match l with 
  Nil      =>Nil
  Cons(h,t)=>let t' = rec t
              list_append0 h t' 
  Name(n,t)=>Name(n, rec t)
  Ref(r)   =>rec (get r)
\end{lstlisting}
\end{tabular}
\caption{Function~\code{list_append0} copies one list~\code{l},
  replacing its \code{Nil} terminal with a second list~\code{r}.
  Function~\code{list_join0} appends the sub-lists of a list of lists.
  The naive polymorphic list type for~\code{list\_join0} is
  insufficient to type-check the \keyword{Cons} case of its body.
}
\label{fig:listjoin}
\end{figure}

Consider the \code{Cons} case of~\code{list_join0}, whose code is
standard; in the case of a sublist~\code{h}, we append this sublist to
the result of flattening the rest of the input list of lists.
Critically, this step involves reasoning about the relationship between
three structures, each involving a set of names:
\begin{itemize}
\item The head~\code{h} of the list of lists~\code{l}, with names drawn from~$X_2$ (Inner list~\code{h} has type $\List{X_2}{Y_2}{\Int}$),
\item The tail~\code{t} of the list of lists~\code{l}, with outer list names drawn from~$X_1$ and inner list names from~$X_2$, and
\item The flattened tail~\code{t'} of of tail~\code{t}, with names drawn from~$X_1 \disj X_2$.
\end{itemize}
In particular, to prove that the use of \code{list_append0 h t'} is
well-typed, we wish to argue that head~\code{h} (drawn
from~$X_2$) has names that are distinct from those in the flattened
list~\code{t'} (drawn from~$X_1 \disj X_2$).
However, the naive polymorphic list type places no constraints on
names in different sub-lists: It merely says that their names are all
drawn from~$X_2$, not that each distinct sub-list uses
a distinct, non-overlapping subset of~$X_2$.

%
%

Recall the type of \keyword{Cons} given above:
\[
\All{\alpha:\Type}\All{X,Y:\NmSet} \alpha -> (\List{X}{Y}{\alpha}) -> (\List{X}{Y}{\alpha})
\]
In particular, we want the type for \keyword{Cons} to enforce a
relationship of disjointness between the name sets of the new element
and the existing elements in the list, which is does not.
Further, as written, each occurrence of $\alpha$
must be the same, and thus, must over-approximate name sets.  
Instead, the following types for \keyword{Cons} will
permit~\code{list_join} to type-check; we discuss them in turn:

\medskip

\begin{minipage}{0.58\textwidth}
\begin{array}[t]{l}
\textrm{Lists of lists of integers:}
\\
\keyword{Cons} : \\
\All{X{\bot}X_1{\bot}X_2:\NmSet} 
\\
\All{Y{\bot}Y_1{\bot}Y_2:\NmSet}
\\[1ex]
~~~~(\List{X_1}{Y_1}{\Int}) 
\\
-> (\List{X}{Y}{(\List{X_2}{Y_2}{\Int})})
\\
-> \big(
\List{X {\disj} X_1 {\disj} X_2}{Y {\disj} Y_1 {\disj} Y_2}{}
\\ ~~~~~~~{ (\List{X_1 {\disj} X_2}{Y_1 {\disj} Y_2}{\Int}) }
\big)
\end{array}
\end{minipage}
~~
\begin{minipage}{0.5\textwidth}
\begin{array}[t]{l}
\textrm{Lists of $\alpha$ (type~$\alpha$ has kind $\sort -> \Type$):}
\\
\keyword{Cons}:\All{X,Y:\NmSet} \All{i,j : \sort} 
\\
\All{k_1 \equiv f~i~j : \sort}
\\
\All{k_2 \equiv g~X~Y~k_1 : \NmSet ** \NmSet}
\\[1ex]
~~~~~\alpha [[ i ]]
\\
-> (\List{X}{Y}{( \alpha [[ j ]] )})
\\
-> (\Listi{k_2}{( \alpha [[ k_1 ]] )})
\\[0.2ex]
~
\end{array}
\end{minipage}

\medskip

First consider the type on the left, which is specific to lists of
lists of integers.  In particular, this type enforces that the
sublists' names are disjoint from one another, and from the names of
the outer list. 
Consequently, in the \keyword{Cons} case of~\code{list_join}, we can
justify the recursive call, since the outer and inner lists have
distinct names; and, we can justify the call to \code{list_append},
since the names of each inner list are distinct from those of other
inner lists.

Instead of writing a new version of \keyword{Cons} for every list
element type, we want to capture this pattern, and others,
generically.  
The second (rightmost) type of \keyword{Cons} is generic in
type~$\alpha$, but unlike the naive polymorphic type, this version
permits the indices of the type parameter~$\alpha$ to vary across
occurrences ($i$ versus $j$), while enforcing user-defined constraints
over these index occurrences (index function~$f$), and between these
element indices and the indices of the list (index function~$g$).
In particular, the client of this generic list chooses a
type~$\alpha$, with some index sort~$\sort$, and they choose index
functions~$f$ and~$g$:
\[
\begin{array}{lcl}
f &:& \sort @@> \sort @@> \sort
\\
g &:& \NmSet @@> \NmSet @@> \sort @@> (\NmSet ** \NmSet)
\end{array}
\]
As described by its sort, index function~$f$~forms an element index from
two elements' indices~(each of sort~$\sort$).
Likewise, index function~$g$~forms a list index (a pair of name sets)
from a list index (two, curried name sets) and an element index.
By choosing type~$\alpha$, sort~$\sort$ and index functions~$f$
and~$g$, we can recover integer lists, and lists of integer lists as
special cases:

\begin{tabular}{cc}
\begin{minipage}{0.35\textwidth}
\[
\begin{array}{lcl}
\multicolumn{3}{l}{\textit{Integer lists:}}
\\
\alpha &:\equiv& \Int
\\
\sort &:\equiv& 1
\\
f &:\equiv& \Lam{()}{\Lam{()}{()}}
\\
g &:\equiv& \Lam{X}{\Lam{Y}{\Lam{()}{(X,Y)}}}
\end{array}
\]
\end{minipage}
&
\begin{minipage}{0.6\textwidth}
\[
\begin{array}{lcl}
\multicolumn{3}{l}{\textit{Lists of integer lists:}}
\\
\alpha &:\equiv& \List{-}{-}{\Int}
\\
\sort &:\equiv& \NmSet ** \NmSet
\\
f &:\equiv& \Lam{(X_1,Y_1)}{\Lam{(X_2,Y_2)}{(X_1 \disj X_2, Y_1 \disj Y_2)}}
\\
g &:\equiv& \Lam{X_1}{\Lam{Y_1}{\Lam{(X_2,Y_2)}{(X_1 \disj X_2, Y_1 \disj Y_2)}}}
\end{array}
\]
\end{minipage}
\end{tabular}

\medskip

On the left, we express list of integers by choosing~$\alpha :\equiv
\Int$ and $\sort :\equiv 1$, the sort of $\Int$'s (trivial) type
index.
On the right, we express lists of integer lists by choosing
$\alpha :\equiv \List{-}{-}{\Int}$, the type of integer lists, and
define~$f$ to accumulate name sets for the inner lists, and $g$ to
accumulate name sets for \emph{both} the inner and outer lists.  In
this case, the resulting functions~$f$ and~$g$ happen to be
isomorphic.


\clearpage
\section{Omitted Definitions, Figures, and Remarks}
\label{apx:omitted}

\begin{figure}[h]
\begin{mathpar}
\Infer{emp}
    { }
    {
     |- \cdot : \Gamma
    }
~~~
\Infer{ref}
    {
      \St |- \Gamma
      ~~~~
        \Gamma |- v : A
      ~~~~
        \Gamma(p) = A
    }
    {
      |- ( \St, p : v ) : \Gamma
    }
~~~
\Infer{thunk}
    {
      \St |- \Gamma
      ~~~~
        \Gamma |- e : E
      ~~~~
        \Gamma(p) = E
    }
    {
      |- ( \St, p : e ) : \Gamma
    }
\vspace*{-3.5ex}
\end{mathpar}

\caption{Store typing: $\St |- \Gamma$, read ``store $\St$ typed by $\Gamma$''.}
\Label{fig:store-typing}
\end{figure}

\def\DYNSEMFIGMODE{2}

\begin{figure}[h]
  \centering

    \newcommand{\SEP}{\\[1ex]}
    \begin{array}[t]{rl@{~~}lll}
      \DByHasEffects{\D}{\textrm{Eval-term}()}{\emptyset}{\emptyset}
    \SEP
      \DByHasEffects{\D}{\textrm{Eval-app}(\D_1, \D_2)}{R_1 \mergeRds R_2}{W_1 \disjoint W_2}
      &\text{if}&
      \DHasEffects{\D_1}{R_1}{W_1}
      \\ &\text{and}&
      \DHasEffects{\D_2}{R_2}{W_2}
    \SEP
      \DByHasEffects{\D}{\textrm{Eval-bind}(\D_1, \D_2)}{R_1 \mergeRds R_2}{W_1 \disjoint W_2}
      &\text{if}&
      \DHasEffects{\D_1}{R_1}{W_1}
      \\ &\text{and}&
      \DHasEffects{\D_2}{R_2}{W_2}
    \SEP
      \DByHasEffects{\D}{\textrm{Eval-scope}(\D_0)}{R}{W}
      &\text{if}&
      \DHasEffects{\D_0}{R}{W}
    \end{array}
    \medskip
    \begin{array}[t]{rl@{~~}lll}
      \DByHasEffects{\D}{\textsf{Eval-fix}(\D_0)}{R}{W}
      &\text{if}& \DHasEffects{\D_0}{R}{W}
    \\
      \DByHasEffects{\D}{\textsf{Eval-case}(\D_0)}{R}{W}
      &\text{if}& \DHasEffects{\D_0}{R}{W}
    \\
      \DByHasEffects{\D}{\textsf{Eval-split}(\D_0)}{R}{W}
      &\text{if}& \DHasEffects{\D_0}{R}{W}
    \SEP
      \DByHasEffects{\D}{\textsf{Eval-ref}()}{\emptyset}{p}
      &\text{where}&
      e = \Refe{\name{n}}{v} \AND p \equiv M\;n 
    \SEP
      \DByHasEffects{\D}{\textsf{Eval-thunk}()}{\emptyset}{p}
      &\text{where}&
      e = \Thunk{\name{n}}{e_0} \AND p \equiv M\;n 
    \SEP
      \DByHasEffects{\D}{\textsf{Eval-get}()}{p}{\emptyset}
      &\text{where}&
      e = \Get{\refv{p}}
    \SEP
      \DByHasEffects{\D}{\textsf{Eval-force}()}{q, R'}{W'}
      &\text{where}&
      e = \Force{\thunk{q}}
      \\ &\text{and}&
      \DHasEffects{\D'}{R'}{W'}
      \\ &\text{where}&
      \text{$\D'$ is the derivation that computed $t$}
    \end{array}

  \caption{Read- and write-sets of a non-incremental evaluation derivation}
  \FLabel{fig:readswrites}
\end{figure}


%
In \Figureref{fig:readswrites}, we write
\[
\DByHasEffects{\D}{\textit{Rulename (Dlist)}}{R}{W}
\]
to mean that rule $\textit{Rulename}$ concludes $\D$ and has
subderivations $\textit{Dlist}$.  
For example, $\DByHasEffects{\D}{\textsf{Eval-scope}(\D_0)}{R}{W}$
provided that
$\DHasEffects{\D}{R}{W}$, where $\D_0$ derives the only premise of \textsf{Eval-scope}.

\NotInScope{
In the Eval-forceClean case, we refer back to the derivation that (most recently) computed
the thunk being forced (that is, the first subderivation of Eval-computeDep).  A completely formal
definition would take as input a mapping from pointers $q$ to sets $R'$ and $W'$,
return this mapping as output, and modify the mapping in the Eval-computeDep case.
}

\begin{figure}[h]
\small
  \centering

  \judgbox{\Gamma |- A : K}
          {Under $\Gamma$,
            value type $A$ has kind $K$
          }
  \vspace*{-3.0ex}
  \begin{mathpar}
    \Infer{k-typevar}
        {
          (\alpha : K) \in \Gamma
        }
        {
          \Gamma |- \alpha : K
        }
    ~~~
    \Infer{k-tycon}
        {
          (d : K) \in \Gamma
        }
        {
          \Gamma |- d : K
        }
    ~~~~
    \Infer{k-binop}
        {
          \Gamma |- A_1 : \type
          \\
          \Gamma |- A_2 : \type
        }
        {
          \arrayenvbl{
              \Gamma |- (A_1 + A_2) : \type
              \\
              \Gamma |- (A_1 ** A_2) : \type
          }
        }
    \and
    \Infer{\!k-unit}
        {}
        {
          \Gamma |- \unitty : \type
        }
    ~~~
    \Infer{\!k-name}
        {
          \Gamma |- i : \namesetsort
        }
        {
          \Gamma |- \Name{i} : \type
        }
    ~~~
    \Infer{\!k-ref}
        {
          \Gamma |- i : \namesetsort
          \\
          \Gamma |- A : \type
        }
        {
          \Gamma |- (\Ref{i}{A}) : \type
        }
    \and
    \Infer{\!k-thk}
        {
          \arrayenvbl{
          \Gamma |- i : \namesetsort
          \\
          \Gamma |- E \tefftype
        }
        }
        {
          \Gamma |- (\Thk{i}{E}) : \type
        }
    ~~
    \Infer{\!k-app-type}
        {\arrayenvbl{
          \Gamma |- A : (\type => K)\!\!\!\!\!
          \\
          \Gamma |- B : \type
        }}
        {
          \Gamma |- (A\;B) : K
        }
    ~~
    \Infer{\!k-app-index}
        {\arrayenvbl{
          \Gamma |- A : (\sort => K)\!\!\!\!
          \\
          \Gamma |- i : \sort
        }}
        {
          \Gamma |- A[[i]] : K
        }
  \end{mathpar}

  \judgbox{\Gamma |- C \ctype}
          {Under $\Gamma$,
            computation type $C$ is well-formed
          }
  \vspace{-0.7ex}
  \begin{mathpar}
    \Infer{ctype-lift}
         {
             \Gamma |- A : \type
         }
         {
             \Gamma |- (\F A) \ctype
         }
    \and
    \Infer{ctype-arr}
         {
             \Gamma |- A : \type
             \\
             \Gamma |- E \tefftype
         }
         {
             \Gamma |- (A \arr E) \ctype
         }
  \vspace{-0.7ex}
  \end{mathpar}

  \judgbox{\Gamma |- \e \wfeff}{
           Under $\Gamma$,
           effects $\e$ are well-formed}
  \vspace{-1.2ex}
  \begin{mathpar}
    \Infer{wf-eff}
         {
           \Gamma |- W : \namesetsort
           \\
           \Gamma |- R : \namesetsort
         }
         {
           \Gamma |-
           <<
             W; R
           >>
           \wfeff
         }
  \vspace{-1.5ex}
  \end{mathpar}
  
  \judgbox{\Gamma |- P \wfprop}
         {Under $\Gamma$,
           proposition $P$ is well-formed
         }
  \vspace{-3.3ex}
  \begin{mathpar}
    ~~~~~~~~~~
    \Infer{}
         {}
         {\Gamma |- tt \wfprop}
    ~~
    \Infer{}
         {
           \Gamma |- P_1 \wfprop
           ~~~~~
           \Gamma |- P_2 \wfprop
         }
         {
           \Gamma |- (P_1 \andprop P_2) \wfprop
         }
   ~~
   \Infer{}
         {
           \Gamma |- i : \gamma
           \\
           \Gamma |- j : \gamma
         }
         {\arrayenvbl{
             \Gamma |- (i \disj j : \gamma) \wfprop
             \\
             \Gamma |- (i \equiv j : \gamma) \wfprop
             }
         }
  \vspace{-1.0ex}
  \end{mathpar}

  \judgbox{\Gamma |- E \tefftype}
          {Under $\Gamma$,
            type-with-effects $E$ is well-formed
          }
  \begin{mathpar}
    \Infer{\!etype-eff}
        {
          \Gamma |- C \ctype
          \\
          \Gamma |- \e \wfeff
        }
        {
          \Gamma |- (C |> \e) \tefftype
        }
    \and
    \Infer{\!etype-poly}
        {
          \Gamma, \alpha : K |- E \tefftype
        }
        {
          \Gamma |- (\All{\alpha : K} E) \tefftype
        }
    \and
    \Infer{\!etype-idx}
        {
          \Gamma, a : \sort, P |- E  \tefftype
        }
        {
          \Gamma |- (\DAll{a : \sort}{P} E) \tefftype
        }
  \vspace*{-3.0ex}
  \end{mathpar}

  \caption{Kinding and well-formedness for types and effects}
  \label{fig:kinding}
\end{figure}


\clearpage

\subsection{Remarks}
\Label{apx:remarks}

\paragraph{Why distinguish computation types from types-with-effects?}
Can we unify computation types $C$ and types-with-effects $E$?  Not easily.
We have two computation types, $\xF$ and $\arr$.
For $\xF$, the expression being typed could create a thunk, so we must
put that effect somewhere in the syntax.
For $\arr$, applying a function is (per call-by-push-value) just a ``push'': the function carries
no effects of its own (though its codomain may need to have some).
However, suppose we force a thunked function of type $A_1 \arr (A_2 \arr \cdots)$
and apply the function (the contents of the thunk) to one argument.
In the absence of effects, the result would be a computation
of type $A_2 \arr \cdots$, meaning that the computation is waiting for a second argument
to be pushed.  But, since forcing the thunk has the effect of reading the thunk,
we need to track this effect in the result type.  So we cannot return $A_2 \arr \cdots$,
and must instead put effects around $(A_2 \arr \cdots)$.
Thus, we need to associate effects to both $\xF$ and $\arr$, that is, to both computation types.

Now we are faced with a choice:
we could
(1) extend the syntax of each connective with an effect (written next to the connective), or
(2) introduce a ``wrapper'' that encloses a computation type, either $\xF$ or $\arr$.
These seem more or less equally complicated for the present system,
but if we enriched the language with more connectives,
choice (1) would make the new connectives more complicated,
while under choice (2), the complication would already be rolled into the wrapper.
We choose (2), and write the wrapper as $C |> \e$, where $C$ is a computation type
and $\e$ represents effects.

Where should these wrappers live?
We could add $C |> \e$ to the grammar of computation types $C$.  
But it seems useful to have a clear notion of \emph{the} effect associated with a type.
When the effect on the outside of a type is the only effect in the type,
as in $(A_1 \arr \F A_2) |> \e$, ``the'' effect has to be $\e$.
Alas, types like $(C |> \e_1) |> \e_2$ raise awkward questions:
does this type mean the computation does $\e_2$ and then $\e_1$,
or $\e_1$ and then $\e_2$?

We obtain an unambiguous, singular outer effect by distinguishing
types-with-effects $E$ from computation types $C$.
The meta-variables for computation types appear only in the production
$E \bnfas C |> \e$, making types-with-effects $E$ the ``common case'' in the grammar.
Many of the typing rules follow this pattern, achieving some isolation of effect tracking
in the rules.


\section{Omitted Lemmas and Proofs}
\label{apx:proofs}


\begin{lemma}[Index-level weakening]
\Label{lem:index-weak}
~
\begin{enumerate}
\item 
  If   $\Gamma |- M : \sort$
  then $\Gamma, \Gamma' |- M : \sort$.
\item
  If   $\Gamma |- i : \sort$
  then $\Gamma, \Gamma' |- i : \sort$.
\item
  If   $\Gamma |- A : K$
  then $\Gamma, \Gamma' |- A : K$.
\end{enumerate}
\end{lemma}
\begin{proof}
  By induction on the given derivation.
\end{proof}

\begin{lemma}[Weakening]
\Label{type-weak}
~
\begin{enumerate}
\item 
  If   $\Gamma |- e : A$
  then $\Gamma, \Gamma' |- e : A$.
\item 
  If   $\Gamma |-^M e : C$
  then $\Gamma, \Gamma' |-^M e : C$.
\end{enumerate}
\end{lemma}
\begin{proof}
By induction on the given derivation,
using \Lemmaref{lem:weakening-semantic}
(for example, in the case for the value typing rule `name')
and \Lemmaref{lem:index-weak}
(for example, in the case for the computation typing rule `AllIndexElim').
\end{proof}

\begin{lemma}[Substitution]
\Label{lem:subst}
~
\begin{enumerate}
\item
If   $\Gamma |- v : A$
and  $\Gamma, x : A |- e : C$
then $\Gamma |- \big( [v / x] e \big) : C$.
\item
If   $\Gamma |- v : A$
and  $\Gamma, x : A |- v' : B$
then $\Gamma |- \big( [v / x] v' \big) : B$.
\end{enumerate}
\begin{proof}
By mutual induction on the derivation typing $e$ (in part 1)
or $v'$ (in part 2).
\end{proof}
\end{lemma}

\begin{lemma}[Canonical Forms]
\Label{lem:canon}
If $\StoreType{\Gamma}$ and $\Gamma |- v : A$, then
\\
\begin{tabular}{r@{~~}r@{~}c@{~}lll}
1.& If $A$ & $=$ & $1$ &then $v = \unitexp$.
 \\
2.& If $A$ & $=$ & $(B_1 ** B_2)$ &then $v = \Pair{v_1}{v_2}$.
 \\
3.& If $A$ & $=$ & $(B_1 + B_2)$ &then $v = \Inj{i}{v}$ &where $i \in \{1, 2\}$.
 \\
4.& If $A$ & $=$ & $(\Name{X})$ &then $v = \name{n}$ &where $\Gamma |- n \in X$.
 \\
5.& If $A$ & $=$ & $(\Ref{X}{A_0})$ &then $v = \refv{n}$ &where $\Gamma |- n \in X$.
 \\
6.& If $A$ & $=$ & $(\Thk{X}{E})$ &then $v = \thunk{n}$ &where $\Gamma |- n \in X$.
\\
7.& If $A$ & $=$ & $(\namesort @> \namesort)[[M]]$ &then $v = \nametm{M_v}$
           & where $M \conv (\lam{a} M')$
           \\ &&&&&
           and $\cdot |- (\lam{a} M') : (\namesort @> \namesort)$
           \\ &&&&&
           and $M_v \conv M$.
\end{tabular}
\end{lemma}
\begin{proof}
In each part, exactly one value typing rule is applicable,
so the result follows by inversion.
\end{proof}

\begin{lemma}[Application and membership commute]
\Label{app-mem-commute}
If $\Gamma |- n \in i$
and $p \conv \idxapp{M}{n}$
then $\Gamma |- p \in \idxapp{M}{i}$.
\end{lemma}
\begin{proof}
The set $\idxapp{M}{i}$ consists of all elements of $i$,
but mapped by function~$M$.
The name~$p$ is convertible to the name~$\idxapp{M}{n}$.
Since $n \in i$, we have that $p$ is in the
$M$-mapping of $i$, which is $\idxapp{M}{i}$.
\end{proof}

In each case, we write ``\Pointinghand'' to the left of each goal,
as we prove it.

\thmrefsubj*
\begin{proof}
By induction on the typing derivation~$\Ss$.
\begin{itemize}

\DerivationProofCase{ret}
        {
          \Gamma_1 |- v : A
        }
        {
          \Gamma_1 |-^\ambns
          \Ret{v}
          :
          \big(
              (\F A)  |>  <<\emptyset; \emptyset>>
          \big)
        }

\begin{llproof}
\Pf{}{}{(e=t)~\textrm{and}~(\St_1 = \St_2)}{Given}
\Pf{}{}{(R_{\D} = W_{\D} = R = W = \emptyset)}{\ditto}
\Pf{}{}{(\Gamma_2=\Gamma_1)}{Suppose}
\Hand
\Pf{}{|-}{\St_2 : \Gamma_2}{by above equalities}
\Hand
\Pf{\Gamma_2}{|-}{t : C |> <<\emptyset, \emptyset>>}{\ditto}
\Hand
\Pf{}{}{\Dee~\text{reads}~R_\Dee~\text{writes}~W_\Dee}{\ByRWSetRule}
\Hand
\Pf{}{}{<<W_\Dee; R_\Dee>> \effsub <<W; R>>}{All are empty}
\end{llproof}

\DerivationProofCase{get}
        {
          \Gamma_1 |- v : \Ref{X}{A}
        }
        {
          \Gamma_1 |-^\ambns
          \Get{v}
          :
          \big(
             \F A
          \big)
          |>
          <<\emptyset; X>>
        }

\begin{llproof}
\Pf{}{}{(W = \emptyset)~\textrm{and}~(R = X)}{Given}
\Pf{\Gamma_1}{|-}{v : \Ref{X}{A}}{Given}
\Pf{\exists p.}{}{(v = \refv{p})}{\Lemmaref{lem:canon}}
\Pf{\Gamma_1}{|-}{p \in X}{\ditto}
\Pf{}{}{\Gamma_1(p) = A}{By inversion of value typing}
\Pf{\exists v_p.}{}{\St_1(p) = v_p}{Inversion on $|- \St_1 : \Gamma_1$}
\Pf{\Gamma_1}{|-}{v_p:A}{\ditto}
\decolumnizePf
\Pf{}{}{(\Gamma_2=\Gamma_1)~\textrm{and}~(t = \Ret{v_p})}{Suppose}
\Pf{}{}{(R_\Dee = \{ p \})~\textrm{and}~(W_\Dee = \emptyset = W)}{\ditto}
\Hand
\Pf{}{|-}{\St_2 : \Gamma_2}{By above equalities}
\Hand
\Pf{\Gamma_2}{|-}{t : C |> <<\emptyset, \emptyset>>}{\ditto}
\decolumnizePf
\Hand
\Pf{}{}{\Dee~\text{reads}~R_\Dee~\text{writes}~W_\Dee}{\ByRWSetRule}
\Hand
\Pf{}{}{<<W_\Dee; R_\Dee>> \effsub <<W; R>>}{By above equality~$W_\Dee = W = \emptyset$,}
  \Pf{}{}{}{$\ldots$~and inequality for $(R_\Dee = \{ p \}) \subseteq (X = R)$.}
\end{llproof}

\DerivationProofCase{force}
        {
          \Gamma_1 |- v : \Thk{X}{(C |> \e)}
        }
        {
          \Gamma_1 |-^\ambns
          \Force{v}
          :
          \big(
            C |> (<<\emptyset; X>> \effseq \e)
          \big)
        }

\begin{llproof}

\Pf{}{}{\Gamma |- e : \tau}{Given}

\Pf{\Gamma}{|-}{e : \tau}{Given}

\Pf{}{}{(W = \emptyset)~\textrm{and}~(R = X)}{Given}
\Pf{\Gamma_1}{|-}{v : \Thk{X}{(C |> \e)}}{Given}
\Pf{\exists p.}{}{(v = \thunk{p})}{\Lemmaref{lem:canon}}
\Pf{\Gamma_1}{|-}{p \in X}{\ditto}
\Pf{}{}{\Gamma_1(p) = (C |> \e)}  {By inversion of value typing}
\Pf{\exists e_p.}{}{\St_1(p) = e_p}{Inversion on $|- \St_1 : \Gamma_1$}
$\Ss_0 ::~$
\Pf{\Gamma_1}{|-}{e_p : (C |> \e)}{\ditto}
\decolumnizePf
$\D_0 ::~$
\Pf{\St_1}{|-}{{}^M_m~e_p !! \St_2; t}{Inversion of~$\D$}
\Hand
\Pf{}{|-}{\St_2 : \Gamma_2}{By i.h.\ on $\Ss_0$ and $\D_0$}
\Hand
\Pf{\Gamma_2}{|-}{t : C |> <<\emptyset, \emptyset>>}{\ditto}
\Pf{}{}{\Dee_0~\text{reads}~R_\Dee~\text{writes}~W_\Dee}{\ditto}
\Pf{}{}{<<W_{\Dee_0}; R_{\Dee_0}>> \effsub <<W; R>>}{\ditto}
\Hand
\Pf{}{}{\Dee~\text{reads}~R_{\Dee_0}~\text{writes}~W_{\Dee_0}}{\ByRWSetRule}
\Hand
\Pf{}{}{<<W_{\Dee_0}; R_{\Dee_0}>> \effsub <<W; R>>}{by above equality~$W_\Dee = W = \emptyset$,}
  \trailingjust{\hspace*{-9ex}$\ldots$~and inequality $(R_\Dee = \{ p \}) \subseteq (X = R)$.}
\end{llproof}

\DerivationProofCase{scope}
{
  \Gamma_1 |- v : (\namesort @> \namesort)[[M']]
  \\
  \Gamma_1 |-^{M \circ M'} {e_0} : C |> <<W; R>>
}
{
  \Gamma_1 |-^M {\Scope{v}{e_0}}
  :
  C |> <<W; R>>
}

\begin{llproof}
  $\Ss_0 \derives$ \Pf{\Gamma_1}{|-^{M \circ M'}} {e_0 : C |> <<W; R>>}  {Subderivation 2 of $\Ss$}
  $\Dee \derives$ \Pf{}{}{\PreSt{\St_1}{M}{m} \Scope{v}{e_0} !! \St_2 ; t}  {Given}
  $\Dee_0 \derives$ \Pf{}{}{\PreSt{\St_1}{M \circ M'}{m} e_0 !! \St_2 ; t}  {\byinv{scope}}
  \proofsep
  \Pf{\Gamma_1 }{|- }{M : \namesort @> \namesort}{Assumption}
  \Pf{\Gamma_1 }{|- }{v : (\namesort @> \namesort)[[M']]}{Subderivation 1 of $\Ss$}
  \Pf{\Gamma_1 }{|- }{M' : \namesort @> \namesort}{By inversion}
  \Pf{\Gamma_1, x: \namesort }{|- }{M'\,x : \namesort}{\byrule{t-app}}
  \Pf{\Gamma_1, x: \namesort }{|-}{M\,(M'\,x) : \namesort}{\byrule{t-app}}
  \Pf{\Gamma_1 }{|- }{\lam{x} M\,(M'\,x) : \namesort @> \namesort}{\byrule{t-abs}}
  \Pf{\Gamma_1 }{|- }{(M \circ M') : \namesort @> \namesort}{By definition of $M \circ M'$}
  \decolumnizePf
\Hand \Pf{}{|- }{\St_2 : \Gamma_2}{By i.h.\ on $\Ss_0$}
\Hand \Pf{\Gamma_2}{|-}{t : C |> <<\emptyset; \emptyset>>}{\ditto}
      \Pf{}{}{\DHasEffects{\D_0}{R_{\D_0}}{W_{\D_0}}}{\ditto}
      \Pf{}{}{<<W_{\D_0}; R_{\D_0}>> \effsub <<W; R>>}{\ditto}
\Hand \Pf{}{}{\DHasEffects{\D}{R_{\D}}{W_{\D}}}{\ByRWSetRule}
      \Pf{}{}{<<W_{\D}; R_{\D}>> = <<W_{\D_0}; R_{\D_0}>>}{\ditto}
\Hand \Pf{}{}{<<W_{\D}; R_{\D}>> \effsub <<W; R>>}{By above equalities}
\end{llproof}

\DerivationProofCase{thunk}{
 \Gamma_1 |-  v : \Name{X}
 \\
 \Gamma_1 |- e : E
}
{
 \Gamma_1
 |-^M
 { \Thunk{v_1}{v_2} }
 :
 { \F (\Thk{M(X)}{E})
 |>
 <<M(X); \emptyset>>
 }
}

\begin{llproof}
\Pf{}{}{C = \F (\Thk{M(X)}{E}) \AND R=\emptyset \AND W=M(X)}{Given from $\Ss$}
\Pf{}{}{ \Gamma_1 |-  v : \Name{X}}  {Subderivation}
\Pf{}{}{(v = \name{n}) \AND (n \in X)}  {By \Lemref{lem:canon}}
\Pf{}{}{M~n !! p \AND R_\D=\emptyset \AND W_\D=\{p\}}{Given from $\D$}
\Pf{}{}{\St_2 = (\St_1, p : e)}{\ditto}
\decolumnizePf
\Pf{}{}{\Gamma_2 = (\Gamma_1, p : \Thk{p}{E})}{Suppose}
\Hand 
\Pf{}{}{|- \St_2 : \Gamma_2}{\ByStoreTypingRuleApp}
\Pf{}{}{\Gamma_2(p) = E}{By inversion of value typing}
\decolumnizePf
\Pf{}{}{\Gamma_2 |- {\refv{p}} : {\Ref{p}{E}} }{\byrule{thunk}}
\Hand
\Pf{}{}{\Gamma_2 |-^M \Ret{\thunk{p}} : \F{(\Thk{p}{A})} |> <<\emptyset;\emptyset>> }{\byrule{ret}}
\Hand
\Pf{}{}
     {\DHasEffects{\D}{R_\D}{W_\D}
       \AND W_\D = \{ p \}
     }
     {\ByRWSetRule}
 \decolumnizePf
\Pf{}{}{ n \in X}  {Above}
\Pf{}{}{M(n) \in M(X)}  {Name term application is pointwise}
\Pf{}{}{M(n) \in W}  {By above equality}
\Pf{}{}{M(n) = p}  {}
\Pf{}{}{\{ p \}  \subseteq { W }}   {By set theory}
\Hand 
\Pf{}{}{<<W_\Dee; R_\Dee>> \effsub {<<W; R >>}}{}
\end{llproof}

\DerivationProofCase{ref}{
 \Gamma_1 |-  v_1 : \Name{X}
 \\
 \Gamma_1 |- v_2 : { A }
}
{
 \Gamma_1
 |-^M
 { \Refe{v_1}{v_2} }
 :
 { \F (\Ref{M(X)}{A}) 
 |>
 <<M(X); \emptyset>>
 }
}

\begin{llproof}
\Pf{}{}{C = \F (\Ref{M(X)}{A}) \AND R=\emptyset \AND W=M(X)}{Given from $\Ss$}
\decolumnizePf
\Pf{}{}{ \Gamma_1 |-  v_1 : \Name{X}}  {Subderivation}
\Pf{}{}{(v_1 = \name{n}) \AND (n \in X)}  {\Lemmaref{lem:canon}}
\Pf{}{}{M~n !! p \AND R_\D=\emptyset \AND W_\D=\{p\}}{Given from $\D$}
\Pf{}{}{\St_2 = (\St_1, p : v_2)}{\ditto}
\decolumnizePf
\Pf{}{}{\Gamma_2 = (\Gamma_1, p : \Ref{p}{A})}{Suppose}
\Hand 
\Pf{}{}{|- \St_2 : \Gamma_2}{\ByStoreTypingRuleApp}
\Pf{}{}{\Gamma_2(p) = A}{\hspace*{-9ex}By inversion of value typing}
\Pf{}{}{\Gamma_2 |- {\refv{p}} : {\Ref{p}{A}} }{\byrule{ref}}
\Hand
\Pf{}{}{\Gamma_2 |-^M {\Ret{\refv{p}}} : {\Ret{\Ref{p}{A}}} |> <<\emptyset;\emptyset>> }{\byrule{ret}}
\Hand
\Pf{}{}
     {\DHasEffects{\D}{R_\D}{W_\D}
       \AND W_\D = \{ p \}
     }
     {\ByRWSetRule}
\decolumnizePf
\Pf{}{}{ n \in X}  {Above}
\Pf{}{}{M(n) \in M(X)}  {Name term application is pointwise}
\Pf{}{}{M(n) \in W}  {By above equality}
\Pf{}{}{M(n) = p}  {}
\Pf{}{}{\{ p \}  \subseteq { W }}   {By set theory}
\Hand 
\Pf{}{}{<<W_\Dee; R_\Dee>> \effsub {<<W; R >>}}{}
\end{llproof}

\DerivationProofCase{let}
{
  \Gamma_1 |-^M e_1 : (\F A |> \e_1)
  \\
  \Gamma_1, x:A |-^M e_2 : {(C |> \e_2)}
}
{
  \Gamma_1 |-^M {\Let{e_1}{x}{e_2}}
  :
  {C |> (\e_1 \effseq \e_2)}
}

\begin{llproof}
\Pf{}{}{|- \St_1 : \Gamma_1}{Given}
$\Ss_1$ :: \Pf{}{}{\Gamma_1 |-^M {e_1} : {\F A} |> {\e_1}}{Subderivation 1 of $\Ss$}
$\D_1$  :: \Pf{}{}{\PreSt{\St_1}{M}{m} {e_1} !! \St_{12}; {t_1} }{Subderivation 1 of $\D$}
\Pf{}{}{\text{exists}~\Gamma_{12} \supseteq \Gamma_1~\text{such that}~\St_{12} : \Gamma_{12}}{By i.h.\ on $\Ss_1$}
\Pf{}{}{\Gamma_{12} |- { t_1 } : { \F A |> <<\emptyset; \emptyset>> } }{\ditto }
\Pf{}{}{\text{$\D_1$ reads $R_{\D_1}$ writes $W_{\D_1}$}}{\ditto}
\Pf{}{}{<< W_{\D_1}; R_{\D_1} >> \effsub { \e_1 } }{\ditto }
\Pf{}{}{<< W_{\D_1}; R_{\D_1} >> \effsub { <<W_1, R_1>> } }{\ditto }
\decolumnizePf
\Pf{}{}{\Gamma_{12} |- { v } : { A } }{inversion of typing rule \textsf{ret},}
\trailingjust{for terminal computation ${t_1}$}
$\Ss_2$ :: \Pf{}{}{\Gamma_1, x:{A} |-^M {e_2} : {C |> \e_2} }{Subderivation 2 of $\Ss$}
\Pf{}{}{\Gamma_{12}, x:{A} |-^M {e_2} : {C |> \e_2} }{\Lemmaref{type-weak}}
\Pf{}{}{         \Gamma_{12}      |-^M [{v}/x] {e_2} : {C |> \e_2} }{\Lemmaref{lem:subst}}
$\D_2$ :: \Pf{}{}{\PreSt{\St_{12}}{M}{m} {[v/x]e_2} !! \St_{2}; {t_2}}{Subderivation 2 of $\D$}
\Pf{}{}{\text{exists}~\Gamma_{2} \supseteq \Gamma_{12} \supseteq \Gamma_1~\text{such that}}{By i.h.\ on $\Ss_2$}
\Hand
\Pf{}{}{|- \St_2 : \Gamma_{2}}{\ditto}
\Hand
\Pf{}{}{\Gamma_{2} |-^M {t_2} : { C } |> <<\emptyset;\emptyset>>}{\ditto}
\Pf{}{}{\text{$\D_2$ reads $R_{\D_2}$ writes $W_{\D_2}$}}{\ditto}
\Pf{}{}{<< W_{\D_2}; R_{\D_2} >> \effsub { \e_{2} } }{\ditto }
\Pf{}{}{<< W_{\D_2}; R_{\D_2} >> \effsub { <<W_2, R_2>> } }{\ditto }
\decolumnizePf
\Pf{}{}{W_1 \disjoint W_2 \AND R_1 \disjoint W_2}{Definition of $\e_1 \effseq \e_2$}
\Pf{}{}{W_{\D_1} \disjoint W_{\D_2} \AND R_{\D_1} \disjoint W_{\D_2}}{$W_{\D_1} \subseteq {W_1}$; $W_{\D_2} \subseteq {W_2}$; $R_{\D_1} \subseteq {R_1}$}
\Pf{}{}{W_\D = W_{\D_1} \disjoint W_{\D_2}}{\ByRWSetRule}
\Pf{}{}{R_\D = R_{\D_1} \cup (R_{\D_2} - W_{\D_1})}{\ditto}
\Hand
\Pf{}{}{\text{$\D$ reads $R_\D$ writes $W_\D$}}{\ditto}
\Hand
\Pf{}{}{<<W_\D, R_\D>> \effsub {<<W, R>>}}{Since $W_\D \subseteq {W}$ and $R_\D \subseteq {R}$}
\end{llproof}

\DerivationProofCase{app}
        {
          \Gamma |-^\ambns e
              : \big(
                 (A -> E) |> \e_1
                \big)
          \\
          \Gamma |- v : A
        }
        {
          \Gamma |-^\ambns (e\;v) : (E \effcoal \e_1)
        }

Similar to the case for \keyword{let}.

\DerivationProofCase{split}
        {
          \Gamma |-^\ambns v : (A_1 ** A_2)
          \\
          \Gamma, x_1:A_1, x_2:A_2 |-^\ambns e : E
        }
        {
          \Gamma |-^\ambns \Split{v}{x_1}{x_2}{e} : E
        }

Similar to the case for \keyword{let}, using \Lemmaref{lem:canon}.

\DerivationProofCase{case}
        {
          \Gamma |-^\ambns v : (A_1 + A_2)
          \\
          \arrayenvbl{
            \Gamma, x_1:A_1 |-^\ambns e_1 : E
            \\
            \Gamma, x_2:A_2 |-^\ambns e_2 : E
          }
        }
       {
          \Gamma |-^\ambns \Case{v}{x_1}{e_1}{x_2}{e_2} : E
       }

Similar to the case for \keyword{let}, using \Lemmaref{lem:canon}.

\DerivationProofCase{name-app}
        {
          \arrayenvbl{
            \Gamma_1 |- v_M : (\namesort @> \namesort)[[M]]
            \\
            \Gamma_1 |- v : \Name{i}
          }
        }
        {
          \Gamma_1 |- (v_M\;v) : \F (\Name{\idxapp{M}{i}}) |> <<\emptyset; \emptyset>>
        }

\begin{llproof}
\Pf{\Gamma_1}{|-}{v_M : (\namesort @> \namesort)[[M]]}{Given}
\Pf{}{}{v_M = \nametm{M_v}}{\Lemmaref{lem:canon}}
\Pf{}{}{M \conv (\lam{a} M')}  {\ditto}
\Pf{}{}{\cdot |- \lam{a} M' : (\namesort @> \namesort)}{\ditto}
\Pf{}{}{M_v \conv M}  {\ditto}
\decolumnizePf
\Pf{\Gamma_1}{|-}{v : \Name{i}}{Given}
\Pf{}{}{v = \name{n}}{\Lemmaref{lem:canon}}
\Pf{}{}{\Gamma |- n \in i}{\ditto}
\proofsep
\Pf{}{}{M \nteval (\lam{a} M')}{By inversion on $\D$ (name-app)}
\Pf{}{}{[n/a]M' \nteval p}{\ditto}
\convPf{p}{[n/a]M'}{By a property of $\nteval$}
\continueconvPf{\idxapp{(\lam{a} M')}{n}}{By a property of $\conv$}
\continueconvPf{\idxapp{M}{n}}{By a property of $\conv$}
\proofsep
\Pf{}{}{(\Gamma_2 = \Gamma_1),(\St_2 = \St_1)}{Suppose}
\Hand
\Pf{}{}{|- \St_2 : \Gamma_2}{By above equalities and $\St_1 |- \Gamma_1$}
\proofsep
\Pf{\Gamma_1}{|-}{ n \in i}{Above}
\convPf{p}{M(n)}   {Above}
\Pf{\Gamma_1}{|-}{ p \in \idxapp{M}{i}}{By \Lemref{app-mem-commute}}
\decolumnizePf
\Pf{\Gamma_1}{|-}{ \name{p} : \Name{\idxapp{M}{i}})}{\byrule{name}}
\Hand
\Pf{\Gamma_1}{|-}{ \Ret{\name{p}} : \F(\Name{\idxapp{M}{i}}) |> <<\emptyset; \emptyset>>}{\byrule{ret}}
\Hand
\Pf{}{}{\DByHasEffects{\D}{\textrm{Eval-name-app}}{\emptyset}{\emptyset}}{\ByRWSetRule}
\Hand
\Pf{}{}{(R_\D = R = \emptyset), (W_\D = W = \emptyset)}{By above equalities}
\end{llproof}


\NotInScope{
\paragraph{TO DO:} Think about whether/where we need to explicitly show that stuff is well-kinded.
}


\DerivationProofCase{AllIndexIntro}
       {
         \Gamma_1, a:\sort, P |-^\ambns t : E
       }
       {
         \Gamma_1 |-^\ambns t : (\DAll{a : \sort}{P} E)
       }

\begin{llproof}
$\Ss_0 ::$ \Pf{\Gamma_1, a:\sort, P}{|-^\ambns}{t : E}{Subderivation}
$\D_0  ::$ \Pf{\St_1}{|-^\ambns_m}{e !! \St_2 ; t}{Subderivation}
\Pf{\exists \Gamma_2 \subseteq \Gamma_1}{}{}{By i.h.}
\Hand
\Pf{}{|-}{\St_2 : \Gamma_2}{\ditto}
\Pf{}{}{\DHasEffects{\D_0}{R_{\D_0}}{W_{\D_0}}}{\ditto}
\Pf{\Gamma_2}{|-}{t : E}{\ditto}
\Pf{}{}{<<R_{\D_0}; W_{\D_0}>> \effsub <<R; W>>}{\ditto}
\Hand
\Pf{\Gamma_2}{|-^\ambns}{t : (\All{a : \sort} E)}{By typing rule}
\Hand
\Pf{}{}{\DHasEffects{\D}{R_{\D}}{W_{\D}}}{\ByRWSetRule}
\Hand
\Pf{}{}{<<R_{\D}; W_{\D}>> \effsub <<R; W>>}{By set theory}
\end{llproof}


\DerivationProofCase{AllIndexElim}
       {
         \Gamma_1 |-^\ambns e : (\DAll{a : \sort}{P} E)
         \\
         \arrayenvbl{
         \Gamma_1 |- i : \sort
         \\
         \extract{\Gamma_1} ||- [i/a]P
         }
       }
       {
         \Gamma_1 |-^\ambns
         e 
         : [i/a]E
       }

\begin{llproof}
$\Ss_0 ::$ \Pf{\Gamma_1}{|-^\ambns}{e : (\All{a : \sort} E)}{Subderivation}
$\D_0  ::$ \Pf{\St_1}{|-^\ambns_m}{e !! \St_2 ; t}{Subderivation}
\Pf{\exists \Gamma_2 \subseteq \Gamma_1}{}{}{By i.h.}
\Hand
\Pf{}{|-}{\St_2 : \Gamma_2}{\ditto}
\Pf{}{}{\DHasEffects{\D_0}{R_{\D_0}}{W_{\D_0}}}{\ditto}
\Pf{\Gamma_2}{|-}{t : (\All{a : \sort} E)}{\ditto}
\Pf{}{}{<<R_{\D_0}; W_{\D_0}>> \effsub <<R; W>>}{\ditto}
\decolumnizePf
\Pf{\Gamma_1}{|-}{i : \sort}{Subderivation}
\Pf{\Gamma_2}{|-}{i : \sort}{By weakening}
\Hand
\Pf{\Gamma_2}{|-^\ambns}{t : [i/a]E}{By typing rule}
\Hand
\Pf{}{}{\DHasEffects{\D}{R_{\D}}{W_{\D}}}{\ByRWSetRule}
\Hand
\Pf{}{}{<<R_{\D}; W_{\D}>> \effsub <<R; W>>}{By set theory}
\end{llproof}

    \DerivationProofCase{AllIntro}
       {
         \Gamma, \alpha : K |-^\ambns \te : E
       }
       {
         \Gamma |-^\ambns
         \te
         : (\All{\alpha : K} E)
       }

    Similar to the AllIndexIntro case.

    \DerivationProofCase{AllElim}
       {
         \Gamma |-^\ambns e : (\All{\alpha : K} E)
         \\
         \Gamma |- A : K
       }
       {
         \Gamma |-^\ambns
         e 
         : [A/\alpha]E
       }

    Similar to the AllIndexElim case.
\qedhere
\end{itemize}
\end{proof}


\section{Bidirectional Typing}
\label{apx:bidir}

\subsection{Syntax}

As discussed below, bidirectional typing requires some annotations,
so we assume that values $v$ and expressions $e$ have been extended
with annotations $(v : A)$ and $(e : A)$.
We also assume that we have explicit syntactic forms
$\inst{e}{i}$ and $\inst{e}{A}$,
which avoid guessing quantifier instantiations.

\begin{figure}[htbp]
  \centering

  \judgbox{\Gamma |- v => A}{Under $\Gamma$, value $v$ synthesizes type $A$}
  \vspace*{-3.0ex}
  \begin{mathpar}
    \hfill
        \Infer{vsyn-var}
                {
                        (x : A) \in \Gamma
                }
                {
                        \Gamma |- x => A
                }
        ~~~~~~~~
        \Infer{vsyn-anno}
                {
                        \Gamma |- v <= A
                }
                {
                        \Gamma |- (v : A) => A
                }
        ~~~~
  \end{mathpar}

  \judgbox{\Gamma |- v <= A}{Under $\Gamma$, value $v$ checks against type $A$}
  \begin{mathpar}
        \Infer{vchk-unit}
        {}
        {
                \Gamma |- \unitexp <= \Unit
        }
    ~~~~
    \Infer{vchk-pair}
        {
                \Gamma |- v_1 <= A_1
                \\
                \Gamma |- v_2 <= A_2
        }
        { 
                \Gamma |- \Pair{v_1}{v_2} <= (A_1 ** A_2) 
        }
    \\
    \Infer{\!vchk-name}
                {
                        \Gamma |- N \in X
                }
                {
                        \Gamma |- (\name{N}) <= \Name{X}
                }
    ~~
    \Infer{\!vchk-namefn}
                {
                        \Gamma |- M_v => (\namesort @> \namesort)
                        \\
                        M_v \conv M
                }
                {
                        \Gamma |- (\namefn{M_v}) <= (\namesort @> \namesort)[[M]]
                }
        \and
        \Infer{vchk-ref}
                {
                        \Gamma |- N \in X
                        \\
                        \Gamma(N) = A
                }
                {
                        \Gamma |- (\refv{N}) <= \Ref{X}{A}
                }
        \\
        \Infer{vchk-thunk}
                {
                        \Gamma |- N \in X
                        \\
                        \Gamma(N) = E
                }
                {
                        \Gamma
                        |-
                        (\thunk{N})
                        <=
                        (
                          \Thk{X}{E}
                        )
                }
        \and
        \Infer{vchk-conv}
        {
                \Gamma |- v => A_1 
                \\
                A_1 = A_2
        }
        { 
                \Gamma |- v <= A_2 
        }
        \and
        \Infer{vchk-inj1}
        {
          \Gamma |- v <= A_1
        }
        {
          \Gamma |- inj_1\; v <= A_1 + A_2
        }
        \and
        \Infer{vchk-inj2}
        {
          \Gamma |- v <= A_2
        }
        {
          \Gamma |- inj_2\; v <= A_1 + A_2
        }

  \end{mathpar}
  \caption{Bidirectional value typing}
  \label{fig:bidirectional-value-typing}
\end{figure}

\begin{figure}[htbp]
\centering

  \judgbox{\Gamma |-^\ambns e => E}
          {Under $\Gamma$, within namespace~$\ambns$, \\
            computation $e$ synthesizes type-with-effects $E$
          }
  \vspace*{-1ex}
  \begin{mathpar}
    \Infer{syn-anno}
        {
                \Gamma |-^\ambns e <= E
        }
        {
                \Gamma |-^\ambns (e : E) => E
        }
    \and
    \Infer{syn-app}
        {
          \arrayenvbl{
          \Gamma |-^\ambns e
              => \big(
                 (A -> E) |> \e_1
                \big)
          \\
          \Gamma |- v <= A
          ~~~~~
          \Gamma |- E \effcoal \e_1 \equiv E'
          }
        }
        {
          \Gamma |-^\ambns (e\;v) => E'
        }
    \and
    \Infer{syn-force}
        {
          \arrayenvbl{
          \Gamma |- v => \Thk{X}{(C |> \e)}
          \\
          \Gamma |- <<\emptyset; X>> \effseq \e \equiv \e'
          }
        }
        {
                \Gamma |-^\ambns
                \Force{v}
                =>
                C |> \e'
        }
~~~~
    \Infer{syn-get}
        {
                \Gamma |- v => \Ref{X}{A}
        }
        {
                \Gamma |-^\ambns
                \Get{v}
                =>
                (
                  \F A
                )
                |>
                <<\emptyset; X>>
        }
    \and
    \Infer{syn-name-app}
        {
                \Gamma |- v_M => (\namesort @> \namesort)[[M]]
                \\
                \Gamma |- v => \Name{i}
        }
        {
                \Gamma |-^N (v_M\;v) => \F (\Name{M[[i]]}) |> <<\emptyset; \emptyset>>
        }
    \and
    \Infer{syn-AllIndexElim}
        {
                \Gamma |-^\ambns e => (\DAll{a : \sort}{P} E)
                \\
                \Gamma |- i : \sort
                \\
                \extract{\Gamma} ||- [i/a]P
        }
        {
                \Gamma |-^\ambns
                \inst{e}{i} 
                => [i/a]E
        }
    \and
    \Infer{syn-AllElim}
        {
                \Gamma |-^\ambns e => (\All{\alpha : K} E)
                \\
                \Gamma |- A : K
        }
        {
                \Gamma |-^\ambns
                \inst{e}{A} 
                => [A/\alpha]E
              }
  \vspace*{-0.6ex}
  \end{mathpar}

  \judgbox{\Gamma |-^\ambns e <= E}
                  {Under $\Gamma$, within namespace~$\ambns$, \\
           computation $e$ checks against type-with-effects $E$
          }
  \vspace*{-0.5ex}
  \begin{mathpar}
         \Infer{chk-conv}
                {
                        \Gamma |-^\ambns e => E_1 
                        \\
                        E_1 = E_2
                }
                { 
                        \Gamma |-^\ambns e <= E_2 
                }
         ~~~
         \Infer{chk-eff-subsume}
                {
                  \arrayenvbl{
                        \Gamma |-^\ambns e => (C |> \e_1)
                        \\
                        \Gamma |- \e_1 \effsub \e_2
                    }
                }
                {
                        \Gamma |-^\ambns e <= (C |> \e_2)
                }
         \vspace*{-1ex}
         \\
         \Infer{chk-split}
                {
                        \arrayenvbl{
                        \Gamma |-
                        v => (A_1 ** A_2)
                        \\
                        \Gamma, x_1:A_1, x_2:A_2 |-^\ambns e <= E
                        }
                }
                {
                        \Gamma |-^\ambns \Split{v}{x_1}{x_2}{e} <= E
                }
         ~~~
         \Infer{chk-case}
                {
                  \arrayenvbl{
                        \Gamma |-
                        v => (A_1 + A_2)
                        \\
                        \Gamma, x_1:A_1 |-^\ambns e_1 <= E
                        \\
                        \Gamma, x_2:A_2 |-^\ambns e_2 <= E
                        }
                }
                {
                        \Gamma |-^\ambns \Case{v}{x_1}{e_1}{x_2}{e_2} <= E
                }
         \vspace*{-0.3ex}
         \\
         \arrayenvbl{
         \Infer{chk-ret}
                {
                        \Gamma |- v <= A
                }
                {
                        \Gamma |-^\ambns
                        \Ret{v}
                        <=
                        \big(
                        (\F A)  |>  <<\emptyset; \emptyset>>
                        \big)
                }
        \\[0.7ex]
        \Infer{\!chk-lam}
                {
                        \Gamma,x:A |-^\ambns e <= E
                }
                {
                        \Gamma |-^\ambns
                        (\lam{x} e)
                        <=
                        \big(
                        (A{->}E) |> <<\emptyset; \emptyset>>
                        \big)
                }
         }
         \and
         \Infer{chk-let}
                {
                  \arrayenvbl{
                    \Gamma |-^M e_1 => (\F A) |> \e_1
                    \\
                    \Gamma, x:A |-^M e_2 <= (C |> \e_2)
                    \\
                    \Gamma |- \e_1 \effseq \e_2 \equiv \e
                    }
                }
                {
                        \Gamma |-^M \Let{e_1}{x}{e_2}
                        <=
                        C |> \e
                }
        \and
        \Infer{\!chk-thunk}
                {
                        \Gamma |- v <= \Name{X}
                        \\
                        \Gamma |-^\ambns e <= E
                }
                {\small
                        \Gamma |-^\ambns
                        \Thunk{v}{e}
                        <=
                        \big(
                        \F (\Thk{M[[X]]}{E})
                        \big)
                        |>
                        <<M[[X]]; \emptyset>>
                }
    \and
        \Infer{chk-ref}
        {
                \Gamma |- v_1 <= \Name{X}
                \\
                \Gamma |- v_2 <= A
        }
        {
                \Gamma |-^\ambns
                \Refe{v_1}{v_2}
                <=
                \big(
                \F (\Ref{M[[X]]}{A})
                \big)
                |>
                <<M[[X]]; \emptyset>>
        }
    \and
    \Infer{chk-scope}
        {
                \Gamma |- v => (\namesort @> \namesort)[[N']]
                \\
                \Gamma |-^{N \circ N'} e <= (C |> <<W; R>>)
        }
        {
                \Gamma |-^N \Scope{v}{e} <= (C |> <<W; R>>)
        }
    \and
    \Infer{chk-AllIndexIntro}
        {
                \Gamma, a:\sort, P |-^\ambns t <= E
        }
        {
                \Gamma |-^\ambns t <= (\DAll{a : \sort}{P} E)
        }
    \and
    \Infer{chk-AllIntro}
        {
                \Gamma, \alpha : K |-^\ambns t <= E
        }
        {
                \Gamma |-^\ambns
                t
                <= (\All{\alpha : K} E)
        }
  \vspace*{-3.5ex}
  \end{mathpar}
  \caption{Bidirectional computation typing}
  \label{fig:bidirectional-comp-typing}
\end{figure}


\subsection{Bidirectional Typing Rules}

The typing rules in Figures \ref{fig:value-typing}
and \ref{fig:comp-typing}
are declarative:
they define what typings are valid, but not how to derive those typings.
The rules' use of names and effects annotations
means that standard unification-based techniques,
like Damas--Milner inference, are not clearly applicable.
For example, it is not obvious when to apply chk-AllIntro.

Following the tradition of DML,
we obtain an algorithmic version of our typing rules by defining
a bidirectional system \citep{Pierce00}:
we split judgments with a colon into
judgments with an arrow.
Thus, the computation typing judgment
$\cdots e : E$
becomes two judgments.
The first is the \emph{checking} judgment
$\Gamma |-^\ambns e <= E$,
in which the type $E$ is already known---it is an \emph{input} to the algorithm.
The second is the \emph{synthesis} judgment
$\Gamma |-^\ambns e => E$,
in which $E$ is not known---it is an output---and the rules construct $E$
by examining $e$ (and $\Gamma$).

In formulating the bidirectional versions of value and computation typing
(Figures \ref{fig:bidirectional-value-typing}
and \ref{fig:bidirectional-comp-typing}),
we mostly follow the ``recipe'' of \citet{Dunfield04:Tridirectional}:
introduction rules check, and elimination rules synthesize.
More precisely, the \emph{principal judgment}---the judgment, either a premise
or conclusion, that has the connective being introduced or eliminated---is
checking (${<=}$) for introduction rules,
and synthesizing (${=>}$) for elimination rules.
In many cases, once the direction of that premise (or conclusion) is determined,
the direction of the other judgments follows by considering what information
is known (as input, or as the output type of the principal judgment,
if that judgment is synthesizing).
For example, if we commit to checking the conclusion of `chk-lam',
we should check the premise because its type is a subexpression of the
type in the conclusion.
(Checking is more powerful than synthesis: every expression that synthesizes
also checks, but not all expressions that check can synthesize.)

When a synthesis (elimination) premise attempts to type
an expression that is a checking (introduction) form,
the programmer must write a type annotation $(e : E)$.
Thus, following the recipe means that we have a straightforward
\emph{annotation discipline}:
annotations are needed only on redexes.
While we could reduce the number of annotations by adding
synthesis rules---for example, allowing the unit value
$\unitexp$ to synthesize $\unitty$---this makes the system larger
without changing its essential properties;
for a discussion of the implications of such extensions in a different context,
see \citet{Dunfield13}.

Dually, when an expression synthesizes but we are trying to derive a checking judgment,
we use (1) `vchk-conv' for value typing,
or (2) `chk-conv' or `chk-eff-subsume' for computation typing.
The latter rule allows effect subsumption;
together, `chk-conv' and `chk-eff-subsume' encode a subsumption rule,
common in bidirectional systems with subtyping, such as \citet{Dunfield04:Tridirectional}.


\section{Bidirectional Typing Proofs}
\label{apx:bidirectional-proofs}

\begin{restatable}[Soundness of Bidirectional Value Typing]{theorem}{thmrefsoundbival}
	\Label{thmrefsoundbival}
        ~
	\begin{enumerate}
		\item If
		$	\Gamma	|- v => A $, 
		then there exists a value 
		$	v'	$
		such that $	\Gamma	|- v' : A $
		and $	|v| = v'	$.
		
		\item If
		$	\Gamma	|- v <= A $, 
		then there exists a value 
		$	v'	$
		such that $	\Gamma	|- v' : A $
		and $	|v| = v'	$.
	\end{enumerate}
\end{restatable}

\begin{proof}  By induction on the given derivation.

	Part (1):  Proceed by cases on the rule concluding $\Gamma |- v => A$.
        
		\DerivationProofCase{vsyn-var}
			{
				(x : A) \in \Gamma
			}
			{
				\Gamma |- x => A
			}

			\begin{llproof}
				\Pf{}{}{(x : A) \in \Gamma}{Given}
				\Pf{\Gamma}{|-}{x : A}{By rule var}
				\Pf{}{}{|x| = x}{By definition of $|{-}|$}
				\Hand \Pf{\Gamma}{|-}{v' : A ~\textrm{and}~ |v| = v'}{where $v' = x$ and $v = x$}
			\end{llproof}

		\DerivationProofCase{vsyn-anno}
			{
				\Gamma |- v_1 <= A
			}	
			{
				\Gamma |- (v_1 : A) => A
			}

			\hspace*{-4ex}\begin{llproof}
				\Pf{\exists~v_1' ~\textrm{such that}~ \Gamma}{|-}{v_1' : A ~\textrm{and}~ |v_1| = v_1'}{By inductive hypothesis}
				\Pf{}{|(v_1:A)| =}{|v_1| = v_1'}{By definition of $|{-}|$}\trailingjust{and $|v_1| = v_1'$}
				\Hand\!\!\!\!\!\!\!\! \Pf{\Gamma}{|-}{v' : A ~\textrm{and}~ |v| = v'}{where $v' = v_1'$ and $v = (v_1:A)$}
			\end{llproof}

  {\noindent
	Part (2):}  Proceed by cases on the rule concluding $\Gamma |- v <= A$.
		
		\DerivationProofCase{vchk-unit}
			{}
			{
				\Gamma |- \unitexp <= \Unit
			}
			
			\begin{llproof}
				\Pf{\Gamma}{|-}{\unitexp : \Unit}{By rule unit}
				\Pf{}{|\unitexp| =}{\unitexp}{By definition of $|{-}|$}
				\Hand \Pf{\Gamma}{|-}{v' : \unitexp ~\textrm{and}~ |v| = v'}{where $v' = \unitexp$ and $v = \unitexp$}
			\end{llproof}
			
		\DerivationProofCase{vchk-pair}
			{
				\Gamma |- v_1 <= A_1
				\\
				\Gamma |- v_2 <= A_2
			}
			{
				\Gamma |- \Pair{v_1}{v_2} <= (A_1 ** A_2)
			}
		
			\begin{llproof}
				\Pf{\exists~v_1' ~\textrm{such that}~ \Gamma }{|-}{v_1' : A_1 ~\textrm{and}~ |v_1| = v_1'}{By inductive hypothesis}
				\Pf{\exists~v_2' ~\textrm{such that}~ \Gamma}{|-}{v_2' : A_2 ~\textrm{and}~ |v_2| = v_2'}{By inductive hypothesis}
				\Pf{\Gamma}{|-}{\Pair{v_1'}{v_2'} : (A_1 ** A_2)}{By rule pair}
				\Pf{}{|\Pair{v_1}{v_2}| =}{\Pair{|v_1|}{|v_2|}= \Pair{v_1'}{v_2'}}{By definition of $|{-}|$} \trailingjust{$|v_1| = v_1'$ and $|v_2| = v_2'$}
				\Hand \Pf{\Gamma} {|-} {v' : (A_1 ** A_2) ~\textrm{and}~ |v| = v'}{where $v' = \Pair{v_1'}{v_2'}$} \trailingjust{and $v = \Pair{v_1}{v_2}$}
			\end{llproof}
		
		\DerivationProofCase{vchk-name}
   			{
				\Gamma |- n \in X
			}
			{
				\Gamma |- (\name{n}) <= \Name{X}
			}

			\begin{llproof}
				\Pf{\Gamma} {|-} {n \in X}{Given}
				\Pf{\Gamma} {|-} {(\name{n}) : \Name{X}}{By rule name}
				\Pf{}{|(\name{n})| =}{ (\name{n})}{By definition of $|{-}|$}
				\Hand \Pf{\Gamma} {|-} {v' : \Name{X} ~\textrm{and}~ |v| = v'}{where $v' = (\name{n})$}\trailingjust{ and $v = (\name{n})$}
			\end{llproof}
	
		\DerivationProofCase{vchk-namefn}
			{
				\Gamma |- M_v => (\namesort @> \namesort)
				\\
				M_v \conv M
			}
			{	
				\Gamma |- (\namefn{M_v}) <= (\namesort @> \namesort)[[M]]
			}
			
			\begin{llproof}
				\Pf{\exists~M_v' ~\textrm{such that}~ \Gamma} {|-} {M_v' : (\namesort @> \namesort) ~\textrm{and}~ |M_v| = M_v'}{~~By i.h.}
                                \decolumnizePf
				\Pf{}{}{M_v \conv M}{Given}
				\Pf{}{}{|M_v| \conv M}{Type erasure does not affect convertibility}
				\Pf{}{}{M_v' \conv M}{Since $|M_v| = M_v'$}
                                \decolumnizePf
				\Pf{\Gamma} {|-} {(\namefn{M_v'}) : (\namesort @> \namesort)[[M]]}{By rule namefn}
				\Pf{}{|(\namefn{M_v})| =}{ (\namefn{|M_v|}) = (\namefn{M_v'})}{By definition of $|{-}|$}\trailingjust{and $|M_v| = M_v'$}
				\Hand \Pf{\Gamma} {|-} {v' :  (\namesort @> \namesort)[[M]] ~\textrm{and}~ |v| = v'}{where $v' =  (\namefn{M_v'})$}\trailingjust{ and $v = (\namefn{M_v})$}
			\end{llproof}

		\DerivationProofCase{vchk-ref}
			{
				\Gamma |- n \in X
				\\
				\Gamma(n) = A
			}
			{
				\Gamma |- (\refv{n}) <= \Ref{X}{A}
			}
			
			\begin{llproof}
				\Pf{\Gamma} {|-} {n \in X}{Given}
				\Pf{}{}{\Gamma(n) = A}{Given}
				\Pf{\Gamma} {|-} {(\refv{n}) : \Ref{X}{A}}{By rule ref}
				\Pf{}{|(\refv{n})| =}{ \refv{n}}{By definition of $|{-}|$}
				\Hand \Pf{\Gamma} {|-} {v' : \Ref{X}{A} ~\textrm{and}~ |v| = v'}{where $v' =  (\refv{n})$}\trailingjust{ and $v = (\refv{n})$}
			\end{llproof}
			
		\DerivationProofCase{vchk-thunk}
			{
				\Gamma |- n \in X
				\\
				\Gamma(n) = E
			}
			{
				\Gamma
				|-
				(\thunk{n})
				<=
				\big(
				\Thk{X}{E}
				\big)
			}
			
			\begin{llproof}
				\Pf{\Gamma} {|-} {n \in X}{Given}
				\Pf{}{}{\Gamma(n) = E}{Given}
				\Pf{\Gamma} {|-} {(\thunk{n}) : \Thk{X}{E}}{By rule thunk}
				\Pf{}{|(\thunk{n})| =}{ \thunk{n}}{By definition of $|{-}|$}
				\Hand \Pf{\Gamma} {|-} {v' : \Thk{X}{E} ~\textrm{and}~ |v| = v'}{where $v' =  (\thunk{n})$}\trailingjust{and $v = (\thunk{n})$}
			\end{llproof}
		
		\DerivationProofCase{vchk-conv}
			{
				\Gamma |- v_1 => A_1 
				\\
				A_1 = A_2
			}
			{ 
				\Gamma |- v_1 <= A_2 
			}	
		
			\begin{llproof}
				\Pf{\exists~v_1' ~\textrm{such that}~ \Gamma} {|-} {v_1' : A_1 ~\textrm{and}~ |v_1| = v_1'}{By inductive hypothesis}
				\Pf{}{}{A_1 = A_2}{Given}
				\Pf{\Gamma} {|-} {v_1' : A_2}{By rule conv}
				\Hand \Pf{\Gamma} {|-} {v' :  A_2 ~\textrm{and}~ |v| = v'}{where $v' =  v_1'$ and $v = v_1$}
				\end{llproof}
\end{proof}

\begin{restatable}[Completeness of Bidirectional Value Typing]{theorem}{thmrefcomplbival}
	\Label{thmrefcomplbival}
	~\\
	If
	$	\Gamma	|- v : A$
	then there exist values
	$v'$ and $v''$ such that

	\begin{enumerate}
		\item 
		$	\Gamma	|- v' => A $
		and $	|v'| = v	$
		
		\item 
		$	\Gamma	|- v'' <= A $
		and $	|v''| = v	$
	\end{enumerate}
	
\end{restatable}

\begin{proof}
	By induction on the derivation of $	\Gamma |- v : A $.

	\DerivationProofCase{unit}
		{}
		{
			\Gamma |- \unitexp : \Unit
		}
	
		\begin{llproof}
			\Pf{\Gamma} {|-} {\unitexp <= \Unit}{By rule vchk-unit}
			\Pf{}{|\unitexp| =}{ \unitexp}{By definition of $|{-}|$}
			\Hand \Pf{\Gamma} {|-} {v'' <= \unitexp ~\textrm{and}~ |v''| = v}{where $v'' = \unit$ and $v = \unit$}
			\Pf{\Gamma} {|-} {(\unitexp : \Unit) => \Unit}{By rule vsyn-anno}
			\Pf{}{|(\unitexp : \Unit)| =} {\unitexp}{By definition of $|{-}|$}
			\Hand \Pf{\Gamma} {|-} {v' => \unitexp ~\textrm{and}~ |v'| = v}{where $v' = (\unitexp : \Unit)$ and $v = \unit$}
		\end{llproof}
		
	\DerivationProofCase{var}
		{(x : A) \in \Gamma}
		{\Gamma |- x : A}
			
		\begin{llproof}
			\Pf{}{}{(x : A) \in \Gamma}{Given}
			\Pf{\Gamma} {|-} {x => A}{By rule vsyn-var}
			\Pf{}{|x| =}{x}{By definition of $|{-}|$}
			\Hand \Pf{\Gamma} {|-} {v' => A ~\textrm{and}~ |v'| = v}{where $v' = x$ and $v = x$}
			\Pf{\Gamma} {|-} {x <= A}{By rule vchk-conv}
			\Hand \Pf{\Gamma} {|-} {v'' <= A ~\textrm{and}~ |v''| = v}{where $v'' = x$ and $v = x$}\trailingjust{and $|x| = x$}
		\end{llproof}

	\DerivationProofCase{pair}
		{
			\Gamma |- v_1 : A_1
			\\
			\Gamma |- v_2 : A_2
		}
		{\Gamma |- \Pair{v_1}{v_2} : (A_1 ** A_2)}
		
		\hspace*{-4ex}\begin{llproof}
			\Pf{\exists~v_1'' ~\textrm{such that}~ \Gamma} {|-} {v_1'' <= A_1 ~\textrm{and}~ |v_1''| = v_1}{By inductive hypothesis}
			\Pf{\exists~v_2'' ~\textrm{such that}~ \Gamma} {|-} {v_2'' <= A_2 ~\textrm{and}~ |v_2''| = v_2}{By inductive hypothesis}
			\Pf{\Gamma} {|-} {\Pair{v_1''}{v_2''} <= (A_1 ** A_2)}{By rule vchk-pair}
			\Pf{}{|\Pair{v_1''}{v_2''}| =} {\Pair{|v_1''|}{|v_2''|} = \Pair{v_1}{v_2}}{By definition of $|{-}|$; $|v_1''| = v_1$; $|v_2''| = v_2$}
\decolumnizePf
			\Hand\!\!\! \Pf{\Gamma} {|-} {v'' <= (A_1 ** A_2) ~\textrm{and}~ |v''| = v}{where $v'' = \Pair{v_1''}{v_2''}$ and $v = \Pair{v_1}{v_2}$}
			\Pf{\Gamma} {|-} {(\Pair{v_1''}{v_2''}:(A_1 ** A_2)) => (A_1 ** A_2)}{By rule vsyn-anno}
\decolumnizePf
			\Pf{}{|\Pair{v_1''}{v_2''}:(A_1 ** A_2)| = }
			{|\Pair{v_1''}{v_2''}| = \Pair{v_1}{v_2}}{By definition of $|{-}|$; $|\Pair{v_1''}{v_2''}| = \Pair{v_1}{v_2}$}
\decolumnizePf
			\Hand\!\!\! \Pf{\Gamma} {|-} {v' => (A_1 ** A_2) ~\textrm{and}~ |v'| = v}{where $v' = \Pair{v_1''}{v_2''} $ and $v = \Pair{v_1}{v_2}$}
		\end{llproof}

	\DerivationProofCase{name}
		{
			\Gamma |- n \in X
		}
		{
			\Gamma |- (\name{n}) : \Name{X}
		}
	    
	    \begin{llproof}
	    	\Pf{\Gamma} {|-} {n \in X}{Given}
	    	\Pf{\Gamma} {|-} {(\name{n}) <= \Name{X}}{By rule vchk-name}
	    	\Pf{}{|(\name{n})| = }{(\name{n})}{By definition of $|{-}|$}
	    	\Hand \Pf{\Gamma} {|-} {v'' <= \Name{X} ~\textrm{and}~ |v''| = v}{where $v'' = (\name{n})$ and $v = (\name{n})$}
                \decolumnizePf
	    	\Pf{\Gamma} {|-} {(\name{n}:\Name{X}) => \Name{X}}{By rule vsyn-anno}
	    	\Pf{}{|(\name{n}:\Name{X})| = }{|(\name{n})| = (\name{n})}{By definition of $|{-}|$}
                \decolumnizePf
	    	\Hand \Pf{\Gamma} {|-} {v' => \Name{X} ~\textrm{and}~ |v'| = v}{where $v' = (\name{n}:\Name{X}) $ and $v = (\name{n})$}
	    \end{llproof}
            
	\DerivationProofCase{namefn}
		{
			\Gamma |- M_v : (\namesort @> \namesort)
			\\
			M_v \conv M
		}
		{
			\Gamma |- (\namefn{M_v}) : (\namesort @> \namesort)[[M]]
		}

		\begin{llproof}
			\Pf{}{}{\hspace{-28pt} \exists~M_v' ~\textrm{such that}~ }{}
			\Pf{\Gamma} {|-} {M_v' => (\namesort @> \namesort) ~\textrm{and}~ |M_v'| = M_v}{By inductive hypothesis}
			\Pf{}{}{M_v \conv M}{Given}
			\Pf{}{}{|M_v'| \conv M}{Since $|M_v'| = M_v$}
			\Pf{}{}{M_v' \conv M}{Type annotation does not affect convertibility}
                        \decolumnizePf
			\Pf{\Gamma} {|-} {(\namefn{M_v'}) <= (\namesort @> \namesort)[[M]]}{By rule vchk-namefn}
			\Pf{}{}{|(\namefn{M_v'})| = (\namefn{|M_v'|}) = (\namefn{M_v})}{By definition of $|{-}|$}\trailingjust{and $|M_v'| = M_v$}
			\Hand \Pf{\Gamma} {|-} {v'' <= (\namesort @> \namesort)[[M]] ~\textrm{and}~ |v''| = v}{where $v'' = (\namefn{M_v'})$ and $v = (\namefn{M_v})$}
                        \decolumnizePf
			\Pf{\Gamma} {|-} {(\namefn{M_v'} : (\namesort @> \namesort)[[M]]) => (\namesort @> \namesort)[[M]]}{By rule vsyn-anno}
			\Pf{}{}{|(\namefn{M_v'} : (\namesort @> \namesort)[[M]])| 
                          = (\namefn{M_v})}{By definition of $|{-}|$} 
                        \decolumnizePf
			\Hand \Pf{\Gamma} {|-} {v' => (\namesort @> \namesort)[[M]] ~\textrm{and}~ |v'| = v}{where $v' = (\namefn{M_v'} : (\namesort @> \namesort)[[M]]) $}
		\end{llproof}
	
	\DerivationProofCase{ref}
		{
			\Gamma |- n \in X
			\\
			\Gamma(n) = A
		}
		{
			\Gamma |- (\refv{n}) : \Ref{X}{A}
		}
	
		\begin{llproof}
			\Pf{\Gamma} {|-} {n \in X}{Given}
			\Pf{}{}{\Gamma(n) = A}{Given}
			\Pf{\Gamma} {|-} {(\refv{n}) <= \Ref{X}{A}}{By rule vchk-ref}
			\Pf{}{}{|(\refv{n})| =(\refv{n})}{By definition of $|{-}|$}
			\Hand \Pf{\Gamma} {|-} {v'' <= \Ref{X}{A} ~\textrm{and}~ |v''| = v}{where $v'' = (\refv{n})$ and $v = (\refv{n})$}
			\Pf{\Gamma} {|-} {((\refv{n}) : \Ref{X}{A}) => \Ref{X}{A}}{By rule vsyn-anno}
			\Pf{}{}{|(\refv{n}: \Ref{X}{A})| = |(\refv{n})| = (\refv{n})}{By definition of $|{-}|$} 
			\Hand \Pf{\Gamma} {|-} {v' => \Ref{X}{A} ~\textrm{and}~ |v'| = v}{where $v' = (\refv{n}: \Ref{X}{A})$}\trailingjust{ and $v = (\refv{n})$}
		\end{llproof}

	\DerivationProofCase{thunk}
		{
			\Gamma |- n \in X
			\\
			\Gamma(n) = E
		}
		{
			\Gamma
			|-
			(\thunk{n})
			:
			\big(
			\Thk{X}{E}
			\big)
		}
	
		\begin{llproof}
			\Pf{\Gamma} {|-} {n \in X}{Given}
			\Pf{}{}{\Gamma(n) = E}{Given}
			\Pf{\Gamma} {|-} {(\thunk{n}) <= \big(\Thk{X}{E} \big)}{By rule vchk-thunk}
			\Pf{}{}{|(\thunk{n})| = (\thunk{n})}{By definition of $|{-}|$}
			\Hand \Pf{\Gamma} {|-} {v'' <= \big(\Thk{X}{E} \big) ~\textrm{and}~ |v''| = v}{where $v'' = (\thunk{n})$ }\trailingjust{and $v = (\thunk{n})$}
			\Pf{\Gamma} {|-} {(\thunk{n} : \big(\Thk{X}{E} \big)) => \big(\Thk{X}{E} \big)}{By rule vsyn-anno}
			\Pf{}{}{|(\thunk{n} : \big(\Thk{X}{E} \big))| 
                          = \thunk{n}}{By definition of $|{-}|$}
                        \decolumnizePf
			\Hand \Pf{\Gamma} {|-} {v' => \big(\Thk{X}{E} \big) ~\textrm{and}~ |v'| = v}{where $v' = (\thunk{n} : \big(\Thk{X}{E} \big))$}\trailingjust{ and $v = (\thunk{n})$}
		\end{llproof}
\end{proof}

\begin{restatable}[Soundness of Bidirectional Computation Typing]{theorem}{thmrefsoundbicomp}
	\Label{thmrefsoundbicomp}
	~
	\begin{enumerate}
		\item If
		$	\Gamma	|-^\ambns e => E $, 
		then there exists a value 
		$	e'	$
		such that $	\Gamma	|-^\ambns e' : E $
		and $	|e| = e'	$
		
		\item If
		$	\Gamma	|-^\ambns e <= E $, 
		then there exists a value 
		$	e'	$
		such that $	\Gamma	|-^\ambns e' : E $
		and $	|e| = e'	$
	\end{enumerate}
\end{restatable}

\begin{proof}
  By induction on the given derivation.

  {\noindent Part (1):} Proceed by case analysis on the rule concluding
  $\Gamma |-^\ambns e => E$.
	
	\DerivationProofCase{vchk-app}	
		{
			\Gamma |-^\ambns e_1
				=>  \big(
					(A -> E) |> \e_1
					\big)
			\\
			\Gamma |- v <= A
		}
		{
			\Gamma |-^\ambns (e_1\;v) => (E \effcoal \e_1)
		}
	
		\begin{llproof}
			\Pf{
                          \Gamma} {|-^\ambns}{ e_1' : \big(
				(A -> E) |> \e_1
				\big) ~\textrm{and}~ |e_1| = e_1'}{By inductive hypothesis}
			\Pf{
                          \Gamma} {|-} {v' : A ~\textrm{and}~ |v| = v'}{By \Thmref{thmrefsoundbival}}
			\Pf{\Gamma} {|-^\ambns}{ (e_1'\;v') : (E \effcoal \e_1)}{By rule app}
			\Pf{}{}{|(e_1\;v)| = (|e_1|\;|v|) = (e_1'\;v')}{By definition of $|{-}|$}
			\Hand \Pf{\Gamma} {|-^\ambns}{ e' : (E \effcoal \e_1) ~\textrm{and}~ |e| = e'}{where $e' = (e_1'\;v')$ and $e = (e_1\;v)$}
		\end{llproof}	
	
	\DerivationProofCase{syn-force}
		{
			\Gamma |- v => \Thk{X}{(C |> \e)}
		}
		{
			\Gamma |-^\ambns
			\Force{v}
			=>
			\big(
			C |> (<<\emptyset; X>> \effseq \e)
			\big)
		}
		
		\begin{llproof}
			\Pf{
                          \Gamma} {|-} {v' : \Thk{X}{(C |> \e)} ~\textrm{and}~ |v| = v'}{By \Thmref{thmrefsoundbival}}
			\Pf{\Gamma} {|-^\ambns}{ \Force{v'} : \big(C |> (<<\emptyset; X>> \effseq \e) \big)}{By rule force}
			\Pf{}{}{|\Force{v}| = \Force{|v|} = \Force{v'}}{By definition of $|{-}|$}\trailingjust{and $|v| = v'$}
			\Hand \Pf{\Gamma} {|-^\ambns}{ e' : \big(C |> (<<\emptyset; X>> \effseq \e) \big) ~\textrm{and}~ |e| = e'}{where $e' = \Force{v'}$ and $e = \Force{v}$}
		\end{llproof}
	
	\DerivationProofCase{syn-get}
		{
			\Gamma |- v => \Ref{X}{A}
		}
		{
			\Gamma |-^\ambns
			\Get{v}
			=>
			\big(
			\F A
			\big)
			|>
			<<\emptyset; X>>
		}
			
		\begin{llproof}
			\Pf{
                          \Gamma} {|-} {v' : \Ref{X}{A} ~\textrm{and}~ |v| = v'}{By \Thmref{thmrefsoundbival}}
			\Pf{\Gamma} {|-^\ambns}{ \Get{v'} : \big(\F A\big) |> <<\emptyset; X>>}{By rule get}
			\Pf{}{}{|\Get{v}| = \Get{|v|} = \Get{v'}}{By the definition of $|{-}|$}\trailingjust{and $|v| = v'$}
			\Hand \Pf{\Gamma} {|-^\ambns}{ e' : \big(\F A\big) |> <<\emptyset; X>> ~\textrm{and}~ |e| = e'}{where $e' = \Get{v'}$ and $e = \Get{v}$}
		\end{llproof}
		
	\DerivationProofCase{syn-name-app}
		{
			\arrayenvbl{
				\Gamma |- v_M => (\namesort @> \namesort)[[M]]
				\\
				\Gamma |- v => \Name{i}
			}
		}
		{
			\Gamma |-^N (v_M\;v) => \F (\Name{\idxapp{M}{i}}) |> <<\emptyset; \emptyset>>
		}

		\begin{llproof}
			\Pf{
                          \Gamma} {|-} {v_M' : (\namesort @> \namesort)[[M]] ~\textrm{and}~ |v_M| = v_M'}{By \Thmref{thmrefsoundbival}}
			\Pf{
                          \Gamma} {|-} {v' : \Name{i} ~\textrm{and}~ |v| = v'}{By \Thmref{thmrefsoundbival}}
			\Pf{\Gamma} {|-}{^N (v_M'\;v') : \F (\Name{\idxapp{M}{i}}) |> <<\emptyset; \emptyset>>}{By rule name-app}
			\Pf{|(v_M\;v)|}{=} {(|v_M|\;|v|) = (v_M'\;v')}{By definition of $|{-}|$; $|v_M| = v_M'$; $|v| = v'$}
			\Hand\hspace*{-4ex} \Pf{\Gamma} {|-^\ambns}{ e' : \F (\Name{\idxapp{M}{i}}) |> <<\emptyset; \emptyset>> ~\textrm{and}~ |e| = e'}{where $e' = (v_M'\;v')$ and $e = (v_M\;v)$}
		\end{llproof}
			
	\DerivationProofCase{syn-AllIndexElim}
		{
			\Gamma |-^\ambns e => (\All{a : \sort} E)
			\\
			\Gamma |- i : \sort
		}
		{
			\Gamma |-^\ambns
			\inst{e}{i}
			=> [i/a]E
		}
		
		\begin{llproof}
			\Pf{
                          \Gamma} {|-^\ambns}{ e' : (\All{a : \sort} E) ~\textrm{and}~ |e| = e' }{By inductive hypothesis}
			\Pf{\Gamma} {|-} {i : \sort}{Given}
                        \Hand \eqPf{\inst{e}{i}}{|e'|}  {By $|e| = e'$}
			\Hand \Pf{\Gamma} {|-^\ambns}{ e' : [i/a]E}{By rule AllIndexElim}
		\end{llproof}
			
	\DerivationProofCase{syn-AllElim}
		{
			\Gamma |-^\ambns e => (\All{\alpha : K} E)
			\\
			\Gamma |- A : K
		}
		{
			\Gamma |-^\ambns
			\inst{e}{A}
			=> [A/\alpha]E
		}
		
                               Similar to the syn-AllIndexElim case.

	\DerivationProofCase{syn-anno}
		{
			\Gamma |-^\ambns e_1 <= E
		}
		{
			\Gamma |-^\ambns (e_1 : E) => E
		}
	
		\begin{llproof}
			\Pf{
                          \Gamma} {|-^\ambns}{ e_1' : E  ~\textrm{and}~ |e_1| = e_1'}{By inductive hypothesis}
			\Pf{}{}{|e_1:E| = |e_1| = e_1'}{By the definition of $|{-}|$}\trailingjust{and $|e_1| = e_1'$}
			\Hand \Pf{\Gamma} {|-^\ambns}{ e' : E ~\textrm{and}~ |e| = e'}{where $e' = e_1'$ and $e= (e_1 : E)$}
		\end{llproof}

  {\noindent Part (2):} Proceed by case analysis on the rule concluding
  $\Gamma |-^\ambns e <= E$.
	
	\DerivationProofCase{chk-eff-subsume}
		{
			\Gamma |-^\ambns e => (C |> \e_1)
			\\
			\e_1 \effsub \e_2
		}
		{
			\Gamma |-^\ambns e <= (C |> \e_2)
		}
	
		\begin{llproof}
			\Hand \Pf{
                          \Gamma} {|-^\ambns}{ e' : (C |> \e_1) ~\textrm{and}~ |e| = e'}{By inductive hypothesis}
			\Pf{}{}{\e_1 \effsub \e_2}{Given}
			\Hand \Pf{\Gamma} {|-^\ambns}{ e' : (C |> \e_2)}{By rule eff-subsume}
		\end{llproof}
	
	\DerivationProofCase{chk-split}
		{
			\arrayenvbl{
				\Gamma |-
				v => (A_1 ** A_2)
				\\
				\Gamma, x_1:A_1, x_2:A_2 |-^\ambns e_1 <= E
			}
		}
		{
			\Gamma |-^\ambns \Split{v}{x_1}{x_2}{e_1} <= E
		}
	
		\begin{llproof}
			\Pf{
                          \Gamma} {|-} {v' : (A_1 ** A_2) ~\textrm{and}~ |v| = v'}{By \Thmref{thmrefsoundbival}}
			\Pf{
                          \Gamma, x_1:A_1, x_2:A_2} {|-^\ambns}{ e_1' : E ~\textrm{and}~ |e_1| = e_1'}{By inductive hypothesis}
			\Pf{\Gamma} {|-^\ambns}{ \Split{v}{x_1}{x_2}{e_1'} : E}{By rule split}
			\eqPf{|\Split{v}{x_1}{x_2}{e_1}|}{ \Split{|v|}{x_1}{x_2}{|e_1|}}  {}
                        \decolumnizePf
                                                      \continueeqPf{ \Split{v'}{x_1}{x_2}{e_1'}}{By definition of $|{-}|$}\trailingjust{and $|v| = v', |e_1| = e_1'$}
			\Hand \Pf{\Gamma} {|-^\ambns}{ e' : E ~\textrm{and}~ |e| = e'}{where $e' = \Split{v'}{x_1}{x_2}{e_1'}$}
			\Pf{}{}{}{\hspace{8pt} and $e= \Split{v}{x_1}{x_2}{e_1}$}
		\end{llproof}
		
	\DerivationProofCase{chk-case}
		{
			\Gamma |-
			v => (A_1 + A_2)
			\\
			\arrayenvbl{
				\Gamma, x_1:A_1 |-^\ambns e_1 <= E
				\\
				\Gamma, x_2:A_2 |-^\ambns e_2 <= E
			}
		}
		{
			\Gamma |-^\ambns \Case{v}{x_1}{e_1}{x_2}{e_2} <= E
		}
	
		\begin{llproof}
			\Pf{
                          \Gamma} {|-} {v' : (A_1 + A_2) ~\textrm{and}~ |v| = v'}{By \Thmref{thmrefsoundbival}}
			\Pf{
                          \Gamma, x_1:A_1} {|-^\ambns}{ e_1' : E ~\textrm{and}~ |e_1| = e_1'}{By inductive hypothesis}
			\Pf{
                          \Gamma, x_2:A_2} {|-^\ambns}{ e_2' : E ~\textrm{and}~ |e_2| = e_2'}{By inductive hypothesis}
			\Pf{\Gamma} {|-^\ambns}{ \Case{v'}{x_1}{e_1'}{x_2}{e_2'} : E}{By rule case}
			\eqPf{|\Case{v}{x_1}{e_1}{x_2}{e_2}|}{\Case{|v|}{x_1}{|e_1|}{x_2}{|e_2|}}{By definition of $|{-}|$}
			\continueeqPf{\Case{v'}{x_1}{e_1'}{x_2}{e_2'}}{Since $|v| = v', |e_1| = e_1', |e_2| = e_2'$}
                        \decolumnizePf
			\Hand \Pf{\Gamma} {|-^\ambns}{ e' : E ~\textrm{and}~ |e| = e'}{where $e' = \Case{v'}{x_1}{e_1'}{x_2}{e_2'}$}
			\Pf{}{}{}{\hspace{8pt} and $e= \Case{v}{x_1}{e_1}{x_2}{e_2}$}
		\end{llproof}
		
	\DerivationProofCase{chk-ret}
		{
			\Gamma |- v <= A
		}
		{
			\Gamma |-^\ambns
			\Ret{v}
			<=
			\big(
			(\F A)  |>  <<\emptyset; \emptyset>>
			\big)
		}
		
		\begin{llproof}
			\Pf{
                          \Gamma} {|-} {v' : A ~\textrm{and}~ |v| = v'}{By \Thmref{thmrefsoundbival}}
			\Pf{\Gamma} {|-^\ambns}{ \Ret{v'} : \big((\F A)  |>  <<\emptyset; \emptyset>>	\big)}{By rule ret}
			\Pf{}{}{|\Ret{v}| = \Ret{|v|} = \Ret{v'}}{By definition of $|{-}|$}\trailingjust{and $|v| = v'$}
			\Hand \Pf{\Gamma} {|-^\ambns}{ e' : \big((\F A)  |>  <<\emptyset; \emptyset>> \big) ~\textrm{and}~ |e| = e'}{where $e' = \Ret{v'}$ and $e = \Ret{v}$}
		\end{llproof}
		
	\DerivationProofCase{chk-let}
		{
			\Gamma |-^M e_1 => (\F A) |> \e_1
			\\
			\Gamma, x:A |-^M e_2 <= (C |> \e_2)
		}
		{
			\Gamma |-^M \Let{e_1}{x}{e_2}
			<=
			\big(
			C |> (\e_1 \effseq \e_2)
			\big)
		}
		
		\begin{llproof}
			\Pf{
                          \Gamma} {|-}{^M e_1' : (\F A) |> \e_1 ~\textrm{and}~ |e_1| = e_1'}{By inductive hypothesis}
			\Pf{
                          \Gamma, x:A} {|-}{^M e_2' : (C |> \e_2) ~\textrm{and}~ |e_2| = e_2'}{By inductive hypothesis}
			\Pf{\Gamma} {|-}{^M \Let{e_1'}{x}{e_2'} : \big(C |> (\e_1 \effseq \e_2)\big)}{By rule let}
			\eqPf{|\Let{e_1}{x}{e_2}|}{\Let{|e_1|}{x}{|e_2|}}{By definition of $|{-}|$}
			\continueeqPf{\Let{e_1'}{x}{e_2'}}{Since $|e_1| = e_1', |e_2| = e_2'$}
                        \decolumnizePf
			\Hand \Pf{\Gamma} {|-^\ambns}{ e' : \big(C |> (\e_1 \effseq \e_2)\big) ~\textrm{and}~ |e| = e'}{where $e' = \Let{e_1'}{x}{e_2'}$ and $e = \Let{e_1}{x}{e_2}$}
		\end{llproof}
	
	\DerivationProofCase{chk-lam}
		{
			\Gamma,x:A |-^\ambns e_1 <= E
		}
		{
			\Gamma |-^\ambns
			(\lam{x} e_1)
			<=
			\big(
			(A -> E) |> <<\emptyset; \emptyset>>
			\big)
		}

		\begin{llproof}
			\Pf{
                          \Gamma,x:A} {|-^\ambns}{ e_1' : E ~\textrm{and}~ |e_1| = e_1'}{By inductive hypothesis}
			\Pf{\Gamma} {|-^\ambns}{ (\lam{x} e_1') : \big((A -> E) |> <<\emptyset; \emptyset>> \big)}{By rule lam}
			\Pf{}{}{|(\lam{x} e_1)| = (\lam{x} |e_1|) = (\lam{x} e_1')}{By definition of $|{-}|$}\trailingjust{and $|e_1| = e_1'$}
                        \decolumnizePf
			\Hand \Pf{\Gamma} {|-^\ambns}{ e' : \big((A -> E) |> <<\emptyset; \emptyset>> \big) ~\textrm{and}~ |e| = e'}{where $e' = (\lam{x} e_1')$ and $e = (\lam{x} e_1)$}
		\end{llproof}
		
	\DerivationProofCase{chk-thunk}
		{
			\Gamma |- v <= \Name{X}
			\\
			\Gamma |-^\ambns e_1 <= E_1
		}
		{
			\Gamma |-^\ambns
			\Thunk{v}{e_1}
			<=
			\big(
			\F (\Thk{M(X)}{E_1})
			\big)
			|>
			<<M(X); \emptyset>>
		}
		
		\begin{llproof}
			\Pf{\text{Let~}E = }{}{\big(\F (\Thk{M(X)}{E_1}) \big) |> <<M(X); \emptyset>>}{Assumption}
			\Pf{\exists~ v' ~\textrm{such that}~ \Gamma} {|-} {v' : \Name{X} ~\textrm{and}~ |v| = v'}{By \Thmref{thmrefsoundbival}}
			\Pf{\exists~ e_1' ~\textrm{such that}~ \Gamma} {|-^\ambns}{ e_1' : E ~\textrm{and}~ |e_1| = e_1'}{By inductive hypothesis}
			\Pf{\Gamma} {|-^\ambns}{ \Thunk{v'}{e_1'} : E}{By rule thunk}
			\Pf{}{|\Thunk{v}{e_1}| =}{\Thunk{|v|}{|e_1|} = \Thunk{v'}{e_1'}}{By definition of $|{-}|$}\trailingjust{and $|v| = v'$, $|e_1| = e_1'$}
\decolumnizePf
			\Hand \Pf{\Gamma} {|-^\ambns}{ e' : E  ~\textrm{and}~ |e| = e'}{where $e' = \Thunk{v'}{e_1'}$ and $e = \Thunk{v}{e_1}$}
		\end{llproof}

	\DerivationProofCase{chk-ref}
		{
			\Gamma |- v_1 <= \Name{X}
			\\
			\Gamma |- v_2 <= A
		}
		{
			\Gamma |-^\ambns
			\Refe{v_1}{v_2}
			<=
			\big(
			\F (\Ref{M(X)}{A})
			\big)
			|>
			<<M(X); \emptyset>>
		}

		\begin{llproof}
			\Pf{\exists~ v_1' ~\textrm{such that}~ \Gamma} {|-} {v_1' : \Name{X} ~\textrm{and}~ |v_1| = v_1'}{By \Thmref{thmrefsoundbival}}
			\Pf{\exists~ v_2' ~\textrm{such that}~ \Gamma} {|-} {v_2' : A ~\textrm{and}~ |v_2| = v_2'}{By \Thmref{thmrefsoundbival}}
			\Pf{\Gamma} {|-^\ambns}{ \Refe{v_1'}{v_2'} : \big( \F (\Ref{M(X)}{A})	\big) |> <<M(X); \emptyset>>}{By rule ref}
			\Pf{}{|\Refe{v_1}{v_2}| =}{\Refe{|v_1|}{|v_2|} = \Refe{v_1'}{v_2'}}{By definition of $|{-}|$} 
                        \decolumnizePf
			\Hand \Pf{\Gamma} {|-} {^\ambns e' :  \big( \F (\Ref{M(X)}{A})	\big) |> <<M(X); \emptyset>>} {}
                        \Pf{}{}{~\textrm{and}~ |e| = e'}{where $e' = \Refe{v_1'}{v_2'}$ and $e = \Refe{v_1}{v_2}$}
		\end{llproof}
		
                \DerivationProofCase{chk-scope}
                {
                  \Gamma |- v => (\namesort @> \namesort)[[N']]
                  \\
                  \Gamma |-^{N \circ N'} e_1 <= C |> <<W; R>>
                }
                {
                  \Gamma |-^N \Scope{v}{e_1} <= C |> <<W; R>>
                }
        
                \begin{llproof}
                        \Pf{
                          \Gamma} {|-} {v' : \namesort @> \namesort  ~\textrm{and}~ |N'| = v'}{By inductive hypothesis}
                        \Pf{
                          \Gamma} {|-^{N \circ N'}} {e_1' : C |> <<W; R>>  ~\textrm{and}~ |e_1| = e_1'}{By inductive hypothesis}
                        \Pf{\Gamma} {|-^N} {\Scope{v'}{e_1'} : C |> <<W; R>>}{By rule scope}
                        \Pf{}{|\Scope{v}{e_1}| =}{ \Scope{|v|}{|e_1|} = \Scope{v'}{e_1'}}{By definition of $|{-}|$; $|v| = v'$; $|e_1'| = e_1$}
                        \decolumnizePf
                        \Hand \Pf{\Gamma} {|-^\ambns}{ e' : C |> <<W; R>> ~\textrm{and}~ |e| = e'}{where $e' = \Scope{v'}{e_1'}$ and $e = \Scope{v}{e_1}$}
                \end{llproof}
	
	\DerivationProofCase{chk-AllIndexIntro}
		{
			\Gamma, a:\sort |-^\ambns t <= E
		}
		{
			\Gamma |-^\ambns t <= (\All{a : \sort} E)
		}
		
		\begin{llproof}
			\Pf{
                          \Gamma, a:\sort}{|-^\ambns}{ t' : E  ~\textrm{and}~ |t| = t'}{By inductive hypothesis}
			\Pf{\Gamma} {|-^\ambns}{ t' : (\All{a : \sort} E)}{By rule AllIndexIntro}
			\Hand \Pf{\Gamma} {|-^\ambns}{ e' : (\All{a : \sort} E) ~\textrm{and}~ |e| = e'}{where $e' = t'$ and $e = t$}
		\end{llproof}
	
	\DerivationProofCase{chk-AllIntro}
		{
			\Gamma, a:\sort |-^\ambns t <= E
		}
		{
			\Gamma |-^\ambns
			t
			<= (\All{\alpha : K} E)
		}
	
		\begin{llproof}
			\Pf{\exists~ t' ~\textrm{such that}~ \Gamma, a:\sort} {|-^\ambns}{ t' : E  ~\textrm{and}~ |t| = t'}{By inductive hypothesis}
			\Pf{\Gamma} {|-^\ambns}{ t': (\All{\alpha : K} E)}{By rule AllIntro}
			\Hand \Pf{\Gamma} {|-^\ambns}{ e' : (\All{\alpha : K} E) ~\textrm{and}~ |e| = e'}{where $e' = t'$ and $e = t$}
		\end{llproof}	

	\DerivationProofCase{chk-conv}
		{
			\Gamma |-^\ambns e => E_1 
			\\
			E_1 = E_2
		}
		{ 
			\Gamma |-^\ambns e <= E_2 
		}

		\begin{llproof}
			\Hand \Pf{\exists~ e' ~\textrm{such that}~ \Gamma} {|-^\ambns}{ e' : E_1 ~\textrm{and}~ |e| = e'}{By inductive hypothesis}
			\Pf{}{}{E_1 = E_2}{Given}
			\Pf{\Gamma} {|-^\ambns}{ e' : E_2}{Since $E_1 = E_2$}
		\end{llproof}
\end{proof}

\begin{restatable}[Completeness of Bidirectional Computation Typing]{theorem}{thmrefcomplbicomp}
	\Label{thmrefcomplbicomp}
	~\\
	If
	$	\Gamma	|-^\ambns e : E $, 
	then there exists computations
	$	e'	$, $ e''	$ such that
	\begin{enumerate}
		\item 
		$	\Gamma	|-^\ambns e' => E $
		and $	|e'| = e	$
		
		\item 
		$	\Gamma	|-^\ambns e'' <= E $	and $	|e''| = e	$
	\end{enumerate}	
\end{restatable}

\begin{proof}
	By induction on the derivation of $ \Gamma	|-^\ambns e : E $.
	
	\DerivationProofCase{eff-subsume}
  		{
			\Gamma |-^\ambns e : (C |> \e_1)
			\\
			\e_1 \effsub \e_2
		}
		{
			\Gamma |-^\ambns e : (C |> \e_2)
		}
		
		\begin{llproof}
			\Pf{
                          \Gamma}{|-^\ambns}{ e' => (C |> \e_1) ~\textrm{and}~ e = |e'|}{By inductive hypothesis}
			\Pf{}{}{\e_1 \effsub \e_2}{Given}
			\Pf{\Gamma}{|-^\ambns}{ e' <= (C |> \e_2)}{By chk-eff-subsume}
			\Pf{\Gamma}{|-^\ambns}{ (e': (C |> \e_2)) => (C |> \e_2)}{By syn-anno}
			\Pf{}{|(e': (C |> \e_2))| =}{|e'| = e}{By definition of $|{-}|$}
			\Hand \Pf{\Gamma}{|-^\ambns}{ e' => (C |> \e_2) ~\textrm{and}~ |e'| = e}{}
			\Hand \Pf{\Gamma}{|-^\ambns}{ e'' <= (C |> \e_2) ~\textrm{and}~ |e''| = e}{where $e'' = (e': (C |> \e_2))$}
		\end{llproof}
	
	\DerivationProofCase{split}
		{
			\arrayenvbl{
				\Gamma |-
				v : (A_1 ** A_2)
				\\
				\Gamma, x_1:A_1, x_2:A_2 |-^\ambns e_1 : E
			}
		}
		{
			\Gamma |-^\ambns \Split{v}{x_1}{x_2}{e_1} : E
		}
		
		\begin{llproof}
			\Pf{
                          \Gamma, x_1:A_1, x_2:A_2}{|-}{e_1' <= E ~\textrm{and}~ e_1 = |e_1'|}{By inductive hypothesis}
			\Pf{
                          \Gamma}{|-}{v' => (A_1 ** A_2) ~\textrm{and}~ v_1 = |v_1'|}{By \Thmref{thmrefcomplbival}}
			\Pf{\Gamma}{|-^\ambns}{ \Split{v'}{x_1}{x_2}{e_1'} <= E}{By chk-split}
			\Pf{\Gamma}{|-^\ambns}{ (\Split{v'}{x_1}{x_2}{e_1'} : E) => E}{By syn-anno}
                        \decolumnizePf
			\Pf{}{|(\Split{v'}{x_1}{x_2}{e_1'} : E)| =}{ |\Split{v'}{x_1}{x_2}{e_1'}|}{By definition of $|{-}|$}
			\Pf{}{|\Split{v'}{x_1}{x_2}{e_1'}| =}{\Split{|v'|}{x_1}{x_2}{|e_1'|}}{By definition of $|{-}|$}
			\Pf{}{\Split{|v'|}{x_1}{x_2}{|e_1'|} = }{\Split{v}{x_1}{x_2}{e_1}}{Since $|v'| = v$, $|e_1'| = e_1$}
                        \decolumnizePf
			\Hand \Pf{}{}{\Gamma |- e' => E ~\textrm{and}~ |e'| = e}{where $e' = \Split{v'}{x_1}{x_2}{e_1'}$}
			\Pf{}{}{}{\hspace{8pt} and $e = \Split{v}{x_1}{x_2}{e_1}$}
			\Hand \Pf{}{}{\Gamma |- e'' <= E ~\textrm{and}~ |e''| = e}{where $e'' = (\Split{v'}{x_1}{x_2}{e_1'} : E)$}
			\Pf{}{}{}{\hspace{8pt} and $e = \Split{v}{x_1}{x_2}{e_1}$}
		\end{llproof}
	
	\DerivationProofCase{case}
		{
			\Gamma |-
			v : (A_1 + A_2)
			\\
			\arrayenvbl{
				\Gamma, x_1:A_1 |-^\ambns e_1 : E
				\\
				\Gamma, x_2:A_2 |-^\ambns e_2 : E
			}
		}
		{
			\Gamma |-^\ambns \Case{v}{x_1}{e_1}{x_2}{e_2} : E
		}
		
		\begin{llproof}
			\Pf{
                          \Gamma}{ |- } {v' => (A_1 + A_2) ~\textrm{and}~ |v'| = v}{By \Thmref{thmrefcomplbival}}	
			\Pf{
                          \Gamma, x_1:A_1}{|-^\ambns}{ e_1'' <= E ~\textrm{and}~ |e_1''| = e_1}{By inductive hypothesis}
			\Pf{
                          \Gamma, x_2:A_2}{|-^\ambns}{ e_2'' <= E ~\textrm{and}~ |e_2''| = |e_2|}{By inductive hypothesis}
			\Pf{\Gamma}{|-} {^\ambns \Case{v'}{x_1}{e_1''}{x_2}{e_2''} <= E}{By rule chk-case}
			\Pf{\Gamma}{|-} {^\ambns (\Case{v'}{x_1}{e_1''}{x_2}{e_2''} : E) => E}{By rule chk-conv}
                        \decolumnizePf
			\Pf{}{|(\Case{v'}{x_1}{e_1''}{x_2}{e_2''} : E)| = }{|\Case{v'}{x_1}{e_1''}{x_2}{e_2''}|}{By definition of $|{-}|$}
			\Pf{|\Case{v'}{x_1}{e_1''}{x_2}{e_2''}|}{ = }{\Case{|v'|}{x_1}{|e_1''|}{x_2}{|e_2''|}}{By definition of $|{-}|$}
			\Pf{}{\Case{|v'|}{x_1}{|e_1''|}{x_2}{|e_2''|} =}{ \Case{v}{x_1}{e_1}{x_2}{e_2}}{Since $|v'| = v$, $|e_1''| = e_1$, $|e_2''| = e_2$}
\decolumnizePf
			\Hand \Pf{\Gamma}{|-^\ambns}{ e' => E ~\textrm{and}~ |e'| = e}{where $e' = (\Case{v'}{x_1}{e_1''}{x_2}{e_2''} : E)$}
			\Pf{}{}{}{\hspace{8pt} and $e = \Case{v}{x_1}{e_1}{x_2}{e_2}$}
			\Hand \Pf{\Gamma}{|-^\ambns}{ e'' <=  E ~\textrm{and}~ |e''| = e}{where $e'' = \Case{v'}{x_1}{e_1''}{x_2}{e_2''}$}
			\Pf{}{}{}{\hspace{8pt} and $e = \Case{v}{x_1}{e_1}{x_2}{e_2}$}
		\end{llproof}
	
	\DerivationProofCase{ret}
		{
			\Gamma |- v : A
		}
		{
			\Gamma |-^\ambns
			\Ret{v}
			:
			\big(
			(\F A)  |>  <<\emptyset; \emptyset>>
			\big)
		}
	
		\begin{llproof}
			\Pf{\text{Let~}E}{=}{\big( (\F A)  |>  <<\emptyset; \emptyset>> \big)}{Assumption}
			\Pf{\exists~ v'' ~\textrm{such that}~ \Gamma}{|-}{v'' <= A ~\textrm{and}~ |v''| = v}{By \Thmref{thmrefcomplbival}}
			\Pf{\Gamma}{|-^\ambns}{ \Ret{v''} <= E}{By rule chk-ret}
			\Pf{\Gamma}{|-^\ambns}{ (\Ret{v''} : E) => E}{By syn-anno}
			\Pf{}{|(\Ret{v''} : E)| =}{|\Ret{v''}|}{By definition of $|{-}|$}
			\Pf{}{|\Ret{v''}| =}{\Ret{|v''|} = \Ret{v}}{By definition of $|{-}|$}\trailingjust{and $|v''| = v$}
			\Hand \Pf{\Gamma}{|-^\ambns}{ e' => E ~\textrm{and}~ |e'| = e}{where $e' = (\Ret{v''} : E)$ and $e = \Ret{v}$}
			\Hand \Pf{\Gamma}{|-^\ambns}{ e'' <=  E ~\textrm{and}~ |e''| = e}{where $e'' = \Ret{v''}$ and $e = \Ret{v}$}
		\end{llproof}
		
	\DerivationProofCase{let}
		{
			\Gamma |-^M e_1 : (\F A) |> \e_1
			\\
			\Gamma, x:A |-^M e_2 : (C |> \e_2)
		}
		{
			\Gamma |-^M \Let{e_1}{x}{e_2}
			:
			\big(
			C |> (\e_1 \effseq \e_2)
			\big)
		}
	
		\begin{llproof}
			\Pf{\text{Let~}E}{=}{\big(C |> (\e_1 \effseq \e_2)\big)}{Assumption}
			\Pf{
                          \Gamma}{|-}{ ^M e_1' => (\F A) |> \e_1 ~\textrm{and}~ |e_1'| = e_1}{By inductive hypothesis}
			\Pf{
                          \Gamma, x:A}{|-}{^M e_2'' <= (C |> \e_2) ~\textrm{and}~ |e_2''| = e_2}{By inductive hypothesis}
			\Pf{\Gamma}{|-}{^M \Let{e_1'}{x}{e_2''} <= E}{By rule chk-let}
			\Pf{\Gamma}{|-}{^M (\Let{e_1'}{x}{e_2''} : E) => E}{By rule chk-conv}
			\Pf{}{|(\Let{e_1'}{x}{e_2''} : E)| =}{|\Let{e_1'}{x}{e_2''}|}{By definition of $|{-}|$}
			\Pf{}{|\Let{e_1'}{x}{e_2''}| =}{\Let{|e_1'|}{x}{|e_2''|} = \Let{e_1}{x}{e_2}}{By definition of $|{-}|$}\trailingjust{and $|e_1'| = e_1$, $|e_2''| = e_2$}
                        \decolumnizePf
			\Hand \Pf{\Gamma}{|-^\ambns}{ e' => E ~\textrm{and}~ |e'| = e}{where $e' = (\Let{e_1'}{x}{e_2''} : E)$ and $e = \Let{e_1}{x}{e_2}$}
			\Hand \Pf{\Gamma}{|-^\ambns}{ e'' <=  E ~\textrm{and}~ |e''| = e}{where $e'' = \Let{e_1'}{x}{e_2''}$ and $e = \Let{e_1}{x}{e_2}$}
		\end{llproof}
	
	\DerivationProofCase{lam}
		{
			\Gamma,x:A |-^\ambns e_1 : E_1
		}
		{
			\Gamma |-^\ambns
			(\lam{x} e_1)
			:
			\big(
			(A -> E_1) |> <<\emptyset; \emptyset>>
			\big)
		}
		
		\begin{llproof}
			\Pf{\text{Let~}E}{=}{\big((A -> E_1) |> <<\emptyset; \emptyset>>\big)}{Assumption}
			\Pf{\exists~ e_1'' ~\textrm{such that}~ \Gamma,x:A}{|-^\ambns}{ e_1'' <= E_1 ~\textrm{and}~ |e_1''| = e_1}{By inductive hypothesis}
			\Pf{\Gamma}{|-^\ambns}{ (\lam{x} e_1'') <= E}{By rule chk-lam}
			\Pf{\Gamma}{|-^\ambns}{ ((\lam{x} e_1'') : E) => E}{By syn-anno}
			\Pf{}{|(\lam{x} e_1'')| =}{(\lam{x} |e_1''|) = (\lam{x} e_1)}{By definition of $|{-}|$}\trailingjust{and $|e_1''| = e_1$}
\decolumnizePf
			\Hand \Pf{\Gamma}{|-^\ambns}{ e' => E ~\textrm{and}~ |e'| = e}{where $e' = ((\lam{x} e_1'') : E)$ and $e = (\lam{x} e_1)$}
			\Hand \Pf{\Gamma}{|-^\ambns}{ e'' <=  E ~\textrm{and}~ |e''| = e}{where $e'' = (\lam{x} e_1'')$ and $e = (\lam{x} e_1)$}
		\end{llproof}
	
	\DerivationProofCase{app}
		{
			\Gamma |-^\ambns e_1
			: \big(
			(A -> E) |> \e_1
			\big)
			\\
			\Gamma |- v : A
		}
		{
			\Gamma |-^\ambns (e_1\;v) : (E \effcoal \e_1)
		}
	
		\begin{llproof}
			\Pf{\exists~ e_1' ~\textrm{such that}~ \Gamma}{|-^\ambns}{ e_1' => \big((A -> E) |> \e_1 \big) ~\textrm{and}~ |e_1'| = e_1}{By inductive hypothesis}
			\Pf{\exists~ v'' ~\textrm{such that}~ \Gamma}{|-}{v'' <= A ~\textrm{and}~ |v''| = v}{By \Thmref{thmrefcomplbival}}
			\Pf{\Gamma}{|-^\ambns}{ (e_1'\;v'') => (E \effcoal \e_1)}{By rule syn-app}
			\Pf{\Gamma}{|-^\ambns}{ (e_1'\;v'') <= (E \effcoal \e_1)}{By rule chk-conv}
			\Pf{}{|(e_1'\;v'')| = }{(|e_1'|\;|v''|) = (e_1\;v)}{By the definition of $|{-}|$}\trailingjust{and $|e_1'| = e_1$, $|v''| = v$}
\decolumnizePf
			\Hand \Pf{\Gamma}{|-^\ambns}{ e' => E ~\textrm{and}~ |e'| = e}{where $e' = (e_1'\;v'')$ and $e = (e_1\;v)$}
			\Hand \Pf{\Gamma}{|-^\ambns}{ e'' <=  E ~\textrm{and}~ |e''| = e}{where $e'' = (e_1'\;v'')$ and $e = (e_1\;v)$}
		\end{llproof}
		
	\DerivationProofCase{thunk}
		{
			\Gamma |- v : \Name{X}
			\\
			\Gamma |-^\ambns e_1 : E
		}
		{
			\Gamma |-^\ambns
			\Thunk{v}{e_1}
			:
			\big(
			\F (\Thk{M(X)}{E})
			\big)
			|>
			<<M(X); \emptyset>>
		}

		\begin{llproof}
			\Pf{\text{Let~}E}{=}{\big(\F (\Thk{M(X)}{E})	\big) |> <<M(X); \emptyset>>}{Assumption}
			\Pf{\exists~ v'' ~\textrm{such that}~ \Gamma}{|-}{v'' <= \Name{X} ~\textrm{and}~ |v''| = v}{By \Thmref{thmrefcomplbival}}
			\Pf{\exists~ e_1'' ~\textrm{such that}~ \Gamma}{|-^\ambns}{ e_1'' <= E ~\textrm{and}~ |e_1''| = e_1}{By inductive hypothesis}
			\Pf{\Gamma}{|-^\ambns}{ \Thunk{v''}{e_1''} <= E}{By rule chk-thunk}
			\Pf{\Gamma}{|-^\ambns}{ (\Thunk{v''}{e_1''} : E) => E}{By rule syn-anno}
			\Pf{}{|(\Thunk{v''}{e_1''} : E)| = }{|\Thunk{v''}{e_1''}|}{By definition of $|{-}|$}
			\Pf{|\Thunk{v''}{e_1''}|} {=} {\Thunk{|v''|}{|e_1''|} = \Thunk{v}{e_1}}{By definition of $|{-}|$}\trailingjust{and $|v''| = v$, $|e_1''| = e_1$}
                        \decolumnizePf
			\Hand \Pf{\Gamma}{|-^\ambns}{ e' => E ~\textrm{and}~ |e'| = e}{where $e' = (\Thunk{v''}{e_1''} : E)$ and $e = \Thunk{v}{e_1}$}
			\Hand \Pf{\Gamma}{|-^\ambns}{ e'' <=  E ~\textrm{and}~ |e''| = e}{where $e'' = \Thunk{v''}{e_1''}$ and $e = \Thunk{v}{e_1}$}
		\end{llproof}
	
	\DerivationProofCase{force}
		{
			\Gamma |- v : \Thk{X}{(C |> \e)}
		}
		{
			\Gamma |-^\ambns
			\Force{v}
			:
			\big(
			C |> (<<\emptyset; X>> \effseq \e)
			\big)
		}
		
	    \begin{llproof}
	    	\Pf{\text{Let~}E}{=}{\big(C |> (<<\emptyset; X>> \effseq \e)
	    		\big)}{Assumption}
	    	\Pf{
                  \Gamma}{|-}{v' => \Thk{X}{(C |> \e)} ~\textrm{and}~ |v'| = v}{By \Thmref{thmrefcomplbival}}
	    	\Pf{\Gamma}{|-^\ambns}{ \Force{v'} => E}{By rule syn-force}
	    	\Pf{\Gamma}{|-^\ambns}{ \Force{v'} <= E}{By chk-conv}
	    	\Pf{}{|\Force{v'}| =}{\Force{|v'|} = \Force{v}}{By definition of $|{-}|$}\trailingjust{and $|v'| = v$}
                \decolumnizePf
	    	\Hand \Pf{\Gamma}{|-^\ambns}{ e' => E ~\textrm{and}~ |e'| = e}{where $e' = \Force{v'}$ and $e = \Force{v}$}
	    	\Hand \Pf{\Gamma}{|-^\ambns}{ e'' <=  E ~\textrm{and}~ |e''| = e}{where $e'' = \Force{v'}$ and $e = \Force{v}$}
		\end{llproof}
	
	\DerivationProofCase{ref}
		{
			\Gamma |- v_1 : \Name{X}
			\\
			\Gamma |- v_2 : A
		}
		{
			\Gamma |-^\ambns
			\Refe{v_1}{v_2}
			:
			\big(
			\F (\Ref{M(X)}{A})
			\big)
			|>
			<<M(X); \emptyset>>
		}
		
		\begin{llproof}
			\Pf{\text{Let~}E}{=}{\big( \F (\Ref{M(X)}{A}) \big) |> <<M(X); \emptyset>>}{Assumption}
			\Pf{
                          \Gamma} {|-} {v_1'' <= \Name{X} ~\textrm{and}~ |v_1''| = v_1}{By \Thmref{thmrefcomplbival}}
			\Pf{
                          \Gamma} {|-} {v_2'' <= A ~\textrm{and}~ |v_2''| = v_2}{By \Thmref{thmrefcomplbival}}
			\Pf{\Gamma} {|-} {^\ambns \Refe{v_1''}{v_2''} <= E}{By rule chk-ref}
			\Pf{\Gamma} {|-} {^\ambns (\Refe{v_1''}{v_2''} : E) => E}{By rule syn-anno}
                        \decolumnizePf
			\Pf{}{|(\Refe{v_1''}{v_2''}:E)| = }{|\Refe{v_1''}{v_2''}|}{By definition of $|{-}|$}
			\Pf{}{|\Refe{v_1''}{v_2''}| = }{\Refe{|v_1''|}{|v_2''|} = \Refe{v_1}{v_2}}{By definition of $|{-}|$}\trailingjust{and $|v_1''| = v_1$, $|v_2''| = v_2$}
                        \decolumnizePf
			\Hand \Pf{\Gamma}{|-^\ambns}{ e' => E ~\textrm{and}~ |e'| = e}{where $e' = (\Refe{v_1''}{v_2''} : E)$ and $e = \Refe{v_1}{v_2}$}
			\Hand \Pf{\Gamma}{|-^\ambns}{ e'' <=  E ~\textrm{and}~ |e''| = e}{where $e'' = \Refe{v_1''}{v_2''}$ and $e = \Refe{v_1}{v_2}$}
		\end{llproof}
	
	\DerivationProofCase{get}
		{
			\Gamma |- v : \Ref{X}{A}
		}
		{
			\Gamma |-^\ambns
			\Get{v}
			:
			\big(
			\F A
			\big)
			|>
			<<\emptyset; X>>
		}
	
		\begin{llproof}
			\Pf{
                          \Gamma} {|-} {v' => \Ref{X}{A} ~\textrm{and}~ |v'| = v}{By \Thmref{thmrefcomplbival}}
			\Pf{\Gamma} {|-} {^\ambns \Get{v'} => \big( \F A \big) |> <<\emptyset; X>>}{By rule syn-get}
			\Pf{\Gamma} {|-} {^\ambns \Get{v'} <= \big( \F A \big) |> <<\emptyset; X>>}{By rule chk-conv}
			\Pf{}{|\Get{v'}| = }{\Get{|v'|} = \Get{v}}{By definition of $|{-}|$}\trailingjust{and $|v'| = v$}
                        \decolumnizePf
			\Hand \Pf{\Gamma}{|-^\ambns}{ e' =>  \big( \F A \big) |> <<\emptyset; X>> ~\textrm{and}~ |e'| = e}{where $e' = \Get{v'}$ and $e = \Get{v}$}
			\Hand \Pf{\Gamma}{|-^\ambns}{ e'' <=  \big( \F A \big) |> <<\emptyset; X>> ~\textrm{and}~ |e''| = e}{where $e'' = \Get{v'}$ and $e = \Get{v}$}
		\end{llproof}	
	
	\DerivationProofCase{name-app}
		{
			\arrayenvbl{
				\Gamma |- v_M : (\namesort @> \namesort)[[M]]
				\\
				\Gamma |- v : \Name{i}
			}
		}
		{
			\Gamma |-^N (v_M\;v) : \F (\Name{\idxapp{M}{i}}) |> <<\emptyset; \emptyset>>
		}
	
		\begin{llproof}
			\Pf{
                          \Gamma} {|-} {v_M' => (\namesort @> \namesort)[[M]] ~\textrm{and}~ |v_M'| = v_M}{By \Thmref{thmrefcomplbival}}
			\Pf{
                          \Gamma} {|-} {v' => \Name{i} ~\textrm{and}~ |v'| = v}{By \Thmref{thmrefcomplbival}}
			\Pf{\Gamma} {|-} {^N (v_M'\;v') => \F (\Name{\idxapp{M}{i}}) |> <<\emptyset; \emptyset>>}{By rule syn-name-app}
			\Pf{\Gamma} {|-} {^N (v_M'\;v') <= \F (\Name{\idxapp{M}{i}}) |> <<\emptyset; \emptyset>>}{By rule chk-conv}	
			\Pf{}{|(v_M'\;v')| =}{(|v_M'|\;|v'|) = (v_M\;v)}{By definition of $|{-}|$}\trailingjust{and $|v_M'| = v_M$, $|v'| = v$}
\decolumnizePf
			\Hand \Pf{\Gamma}{|-^N} {e' => \F (\Name{\idxapp{M}{i}}) |> <<\emptyset; \emptyset>> ~\textrm{and}~ |e'| = e}{where $e' = (v_M'\;v')$ and $e = (v_M\;v)$}
			\Hand \Pf{\Gamma}{|-^N} {e'' <= \F (\Name{\idxapp{M}{i}}) |> <<\emptyset; \emptyset>> ~\textrm{and}~ |e''| = e}{where $e'' = (v_M'\;v')$ and $e = (v_M\;v)$}
		\end{llproof}
		
        \DerivationProofCase{scope}
                {
                        \Gamma |- v : (\namesort @> \namesort)[[N']]
                        \\
                        \Gamma |-^{N \circ N'} e_1 : C |> <<W; R>>
                }
                {
                        \Gamma |-^N \Scope{v}{e_1} : C |> <<W; R>>
                }
                
                \begin{llproof}
                		\Pf{\text{Let~}E}{=}{C |> <<W; R>>}{Assumption}
                        \Pf{\exists v'' ~\textrm{such that}~, \Gamma}{|-}{v'' => (\namesort @> \namesort)[[N']] ~\textrm{and}~ |v''| = v}{By \Thmref{thmrefcomplbival}}
                        \Pf{\exists e_1' ~\textrm{such that}~, \Gamma}{|-^{N \circ N'}}{e_1' <= E ~\textrm{and}~ |e_1'| = e_1}{By inductive hypothesis}
                        \Pf{\Gamma}{|-^N}{\Scope{v''}{e_1'} <= E}{By rule chk-scope}
                        \Pf{\Gamma}{|-^N}{(\Scope{v''}{e_1'} : E) => E}{By rule syn-anno}
                        \decolumnizePf
                        \Pf{}{|(\Scope{v''}{e_1'} : E)|}{= |(\Scope{v''}{e_1'}|}{By definition of $|{-}|$}
                        \Pf{}{|(\Scope{v''}{e_1'}| = \Scope{|v''|}{|e_1'|}}{=\Scope{v}{e_1}}{By definition of $|{-}|$}
                        \decolumnizePf
                        \Hand \Pf{\Gamma}{|-^\ambns}{ e' => E ~\textrm{and}~ |e'| = e}{where $e' = (\Scope{v''}{e_1'} : E)$ and $e = \Scope{v}{e_1}$}
                        \Hand \Pf{\Gamma}{|-^\ambns}{ e'' <= E ~\textrm{and}~ |e''| = e}{where $e'' =\Scope{v''}{e_1'}$ and $e = \Scope{v}{e_1}$}
                \end{llproof}
	
	\DerivationProofCase{AllIndexIntro}
		{
			\Gamma, a:\sort |-^\ambns t : E
		}
		{
			\Gamma |-^\ambns t : (\All{a : \sort} E)
		}
	
		\begin{llproof}
			\Pf{\exists~ t'' ~\textrm{such that}~ \Gamma, a:\sort} {|-} {^\ambns t'' <= E ~\textrm{and}~ |t''| = t}{By inductive hypothesis}
			\Pf{\Gamma}{|-^\ambns}{ t'' <= (\All{a : \sort} E)}{By rule chk-AllIndexIntro}
			\Pf{\Gamma}{|-^\ambns}{ (t'' : (\All{a : \sort} E)) => (\All{a : \sort} E)}{By rule syn-anno}
			\Pf{}{|(t'' : (\All{a : \sort} E))| =}{|t''| = t}{By definition of $|{-}|$}\trailingjust{and $|t''| = t$}
                        \decolumnizePf
			\Hand \Pf{\Gamma}{|-^\ambns}{ e' => (\All{a : \sort} E) ~\textrm{and}~ |e'| = e}{where $e' = (t'' : (\All{a : \sort} E))$ and $e = t$}
			\Hand \Pf{\Gamma}{|-^\ambns}{ e'' <= (\All{a : \sort} E) ~\textrm{and}~ |e''| = e}{where $e'' = t''$ and $e = t$}
		\end{llproof}	
	
	\DerivationProofCase{AllIndexElim}
		{
			\Gamma |-^\ambns e : (\All{a : \sort} E)
			\\
			\Gamma |- i : \sort
		}
		{
			\Gamma |-^\ambns
			e
			: [i/a]E
		}
	
		\begin{llproof}
			\Pf{\exists~ e' ~\textrm{such that}~ \Gamma} {|-} {^\ambns e' => (\All{a : \sort} E) ~\textrm{and}~ |e'| = e}{By inductive hypothesis}
			\Pf{\Gamma}{|-}{i : \sort}{Given}
			\Pf{\Gamma}{|-^\ambns}{ e' => [i/a]E}{By rule syn-AllIndexElim}
			\Pf{\Gamma}{|-^\ambns}{ e' <= [i/a]E}{By rule chk-conv}	
			\Hand \Pf{\Gamma}{|-^\ambns}{\inst{e'}{i} => [i/a]E ~\textrm{and}~ |\inst{e'}{i}| = e}{}
			\Hand \Pf{\Gamma}{|-^\ambns}{\inst{e'}{i} <= [i/a]E ~\textrm{and}~ |\inst{e'}{i}| = e}{}
		\end{llproof}
	
	\DerivationProofCase{AllIntro}
		{
			\Gamma, a:\sort |-^\ambns t : E
		}
		{
			\Gamma |-^\ambns
			t
			: (\All{\alpha : K} E)
		}

                Similar to the AllIndexIntro case.
	
	
	\DerivationProofCase{AllElim}
		{
			\Gamma |-^\ambns e : (\All{\alpha : K} E)
			\\
			\Gamma |- A : K
		}
		{
			\Gamma |-^\ambns
			e 
			: [A/\alpha]E
		}

                Similar to the AllIndexElim case.                 \qedhere
\end{proof}


\section{Name Term Language}
\label{sec:nametermlang}

We define a restricted \emph{name term} language for computing larger names from smaller names.
This language consists of the following:
\begin{itemize}
\item Syntax for \emph{names}, \emph{name terms} and \emph{sorts}.
\item Name term sorting: A judgment that assigns sorts to name terms.
\item Big-step evaluation for name terms: A judgment that assigns \emph{name term values} to \emph{name terms}.
\item Semantic definition of equivalent and disjoint name terms.
\item Logical proof rules for equivalent and disjoint name terms: Two judgements that should be sound with respect to the semantic definitions of equivalence and disjointness.
\end{itemize}

%
%
%
%
%
%

\subsection{Name term equivalence and apartness}



 For instance, the following two functions have the same sort, and are
 apart:
\[
\begin{array}{ll}
\Lam{a}{\Pair{a}{a}} &: \Nm @> \Nm * \Nm
\\
\Lam{b}{\Pair{b}{\NmBin{\leafname}{b}}} &: \Nm @> \Nm * \Nm
\end{array}
\]
 These functions have a common sort~$\Nm @> \Nm * \Nm$, which says that
 they map a single name (of sort~$\Nm$) to a pair of names (of
 sort~$\Nm * \Nm$).
 These functions are apart, and this apartness ultimately
 follows from reasoning about binary composition of names: any name $n$ is distinct from
 the binary composition of the name constant~$\leafname$ with name~$n$.

 These two name functions also have the same sort, but are \emph{not}
 apart:
\[
\begin{array}{cc}
\Lam{a}\Pair{a}{{\NmBin{a}{\leafname}}} : \Nm @> \Nm * \Nm
\\
\Lam{b}{\Pair{a}{\NmBin{\leafname}{b}}} : \Nm @> \Nm * \Nm
\end{array}
\]
 In particular, when $b = a = \leafname$, they will produce the same
 pair, namely~$\Pair{\leafname}{\NmBin{\leafname}{\leafname}}$.

 Notice that these two functions are obviously not equivalent either:
 Given any name $n$ such that $n \ne \leafname$, these two functions
 will produce two distinct names.  This follows from another reasoning
 principle for name composition: binary composition with $\leafname$
 in the first function is in a distinct order from that in the second
 function, precisely when $n \ne \leafname$.

Finally, these two functions are equivalent: Given equivalent
arguments, they will always produce equivalent results:
\[
\begin{array}{rl}
  \Lam{a}{\Lam{b}{a~b}}
  &: (\namesort @> \namesort) @> \namesort @> \namesort
  \\
  \Lam{a_1}{\Lam{b_1}{\big(\Lam{a_2}{\Lam{b_2}{a_2~b_2}}\big)~a_1~b_1 }}
  &: (\namesort @> \namesort) @> \namesort @> \namesort
\end{array}
\]
To see this equivalence more clearly, imagine reducing the inner
$\beta$-reduxes by substituting $a_1$ and $b_1$ for $a_2$ and $b_2$,
respectively. After this reduction, the two functions are
$\alpha$-equivalent.


Below, we give a semantic definition of both equivalence and
disjointness, in terms of the sorting and operational definitions
given above.

These definitions are clear, but not immediately practical: for
function sorts, they universally quantify over all possible arguments
for the function's argument.
Since these arguments can include names, which consist of arbitrary
finite binary trees, as well as other functions, there is not an
obvious finite set of arguments to test while still being sound with
respect to these definitions.
For a practical implementation of Typed Adapton, we seek definitions
of equivalence and disjointness that admit decision procedures, and
are sound (and, ideally, complete) with respect to these semantic definitions.
For this purpose, we give decidable logical rules that induct over the
syntax of the two terms.

\subsection{Semantic equivalence and disjointness}

Below, we define semantic equivalence and disjointness of (sorted)
name terms.
We define these semantic properties inductively, based on the common
sort of the name terms.
In this sense, these definitions can be viewed as instances of logical
relations.

We define contexts~$\Gamma$ that relate two variables, either
asserting that they are equivalent, or disjoint:
\[\begin{array}[t]{lcll}
\textrm{Substitutions}
& \sigma & ::= & \cdot ~|~ \sigma, (a \mapsto N)
\\
\textrm{Relational sorting contexts}
& \Gamma &  ::= & \cdot
\\
\textrm{~~~(Hypothetical variable equivalence)}
&        &  ~|~ & \Gamma, \left( a \equiv b : \sort \right)
\\
\textrm{~~~(Hypothetical variable apartness)}
&        &  ~|~ & \Gamma, \left(  a \disj b  : \sort \right)
\end{array}
\]
\[
\begin{array}[t]{p{3mm}rcl}
&(\cdot).1 & = & \cdot
\\
&(\Gamma, a \equiv b : \sort).1 & = & (\Gamma).1, a : \sort
\\
&(\Gamma, a \disj b : \sort).1 & = & (\Gamma).1, a : \sort
\end{array}
~
\begin{array}[t]{p{3mm}rcl}
&(\cdot).2 & = & \cdot
\\
&(\Gamma, a \equiv b : \sort).2 & = & (\Gamma).2, b : \sort
\\
&(\Gamma, a \disj b : \sort).2 & = & (\Gamma).2, b : \sort
\\[2px]
\end{array}
\]
\begin{defn}[Closing substitutions]~\\
We define closing substitution pairs related by equivalence and
  disjointness assumptions in a context~$\Gamma$.
These definitions use and are used by the definitions
below for equivalence and apartness of open terms.
\begin{itemize}
\item
  $||- \sigma_1 \equiv \sigma_2 : \Gamma$ means that
  $(x \equiv y : \sort) \in \Gamma$
  implies
  \big(%
  $\sigma_1(x) = N$
  and $\sigma_2(y) = M$
  and $\cdot ||- N \equiv M : \sort$%
  \big)

\item
  $||- \sigma_1 \disj \sigma_2 : \Gamma$ means that
  $(x \disj y : \sort) \in \Gamma$
  implies
  \big($\sigma_1(x) = N$
  and $\sigma_2(y) = M$
  and $\cdot ||- N \disj M : \sort$%
  \big)

\item
  $||- \sigma_1 \sim \sigma_2 : \Gamma$ means that
  $||- \sigma_1 \equiv \sigma_2 : \Gamma$
  and
  $||- \sigma_1 \disj \sigma_2 : \Gamma$
\end{itemize}
\end{defn}

\begin{defn}[Semantic equivalence]
\label{def:semantic-equivalence-M}
  We define $\Gamma ||- M_1 \equiv M_2 : \sort$ as follows:
  \\
  $(\Gamma).1 |- M_1 : \sort$
  and $(\Gamma).2 |- M_2 : \sort$
  and,
  \\
  for all $\sigma_1$, $\sigma_2$
  such that $||- \sigma_1 \equiv \sigma_2 : \Gamma$ and $[\sigma_1]M_1 !! V_1$ and $[\sigma_2]M_2 !! V_2$,
  we have the following about $V_1$ and $V_2$:

  \medskip

 \begin{tabular}{|c|l|}
\hline
 Sort ($\sort$) & Values~$V_1$ and $V_2$ of sort~$\sort$ are equivalent, written $||- V_1 \equiv V_2 : \sort$
\\
\hline
  $\unitsort$            & Always
\\
  $\Nm$ & When $V_1 = n_1$ and $V_2 = n_2$ and $n_1 = n_2$ (identical binary trees)
\\
  $\sort_1 * \sort_2$ & When $V_1 = (V_{11}, V_{12})$ and $V_1 = (V_{21}, V_{22})$
  \\ & ~~and $||- V_{11} \equiv V_{21} : \sort_1$ and $||- V_{21} \equiv V_{22} : \sort_2$
\\
  $\sort_1 @> \sort_2$ &
 When $V_1 = \Lam{a_1}{M_1}$ and $V_2 = \Lam{a_2}{M_2}$,
 \\ & ~~and
 for all name terms $||- N_1 \equiv N_2 : \sort_1$,
\\
& ~~~~$[N_1/a_1] M_1 !! W_1$ and  $[N_2 / a_2] M_2 !! W_2$ implies $||- W_1 \equiv W_2 : \sort_2$
\\
\hline
\end{tabular}
\end{defn}

\medskip

\begin{defn}[Semantic apartness]
\label{def:semantic-apartness-M}
  We define $\Gamma ||- M_1 \disj M_2 : \sort$ as follows:
  \\
  $(\Gamma).1 |- M_1 : \sort$
  and $(\Gamma).2 |- M_2 : \sort$
  and,
  \\
  for all $\sigma_1$, $\sigma_2$ such that
  $||- \sigma_1 \sim \sigma_2 : \Gamma$ and $[\sigma_1]M_1 !! V_1$ and $[\sigma_2]M_2 !! V_2$,
  we have
  the following about $V_1$ and $V_2$:

  \medskip

 \begin{tabular}{|c|l|}
\hline
 Sort ($\sort$) & Values~$V_1$ and $V_2$ of sort~$\sort$ are apart, written $||- V_1 \disj V_2 : \sort$
\\
\hline
  $\unitsort$            & Always
\\
  $\Nm$ & When $V_1 = n_1$ and $V_2 = n_2$ and $n_1 \ne n_2$ (distinct binary trees)
\\
  $\sort_1 * \sort_2$
  & When $V_1 = (V_{11}, V_{12})$
   and $V_1 = (V_{21}, V_{22})$
  \\
  & ~~ and $||- V_{11} \disj V_{21} : \sort_1$
  and $||- V_{21} \disj V_{22} : \sort_2$
\\
  $\sort_1 @> \sort_2$ &
 When $V_1 = \Lam{a_1}{M_1}$ and $V_2 = \Lam{a_2}{M_2}$, \\
 & ~~and
 for all name terms $||- N_1 \equiv N_2 : \sort_1$,
\\ ~~
 & ~~~~$[N_1/a_1] M_1 !! W_1$ and  $[N_2 / a_2] M_2 !! W_2$ implies $||- W_1 \disj W_2 : \sort_2$
\\
\hline
\end{tabular}
\end{defn}

\begin{figure}
\centering
  \judgbox{\Gamma |- M \equiv N : \sort}
          {The name terms $M$ and $N$ are \emph{equivalent} at sort $\gamma$}

\begin{mathpar}
\Infer{Eq-Var}
{(M \equiv N : \sort) \in \Gamma}
{\Gamma |- M \equiv N : \sort}
~~~
\Infer{E-Refl}
      { 
      (\Gamma).1 |- M : \sort
      \\\\
      (\Gamma).2 |- M : \sort
      }
      {\Gamma |- M \equiv M : \sort}
~~~
\Infer{E-Sym}
      {\Gamma |- N \equiv M : \sort}
      {\Gamma |- M \equiv N : \sort}
\and
\Infer{Eq-Trans}
{
  \Gamma  |- M_1 \equiv M_2 : \sort
  \\\\
  \Gamma  |- M_2 \equiv M_3 : \sort
}
{ \Gamma |- M_1 \equiv M_3 : \sort }
\\
\Infer{Eq-Pair}
{\Gamma |- M_1 \equiv N_1 : \sort_1
  \\
 \Gamma |- M_2 \equiv N_2 : \sort_2
}
{\Gamma |- (M_1, M_2) \equiv (N_1, N_2) : \sort_1 * \sort_2}
\and
\Infer{Eq-Bin}
{\Gamma |- M_1 \equiv N_1 : \namesort
  \\
 \Gamma |- M_2 \equiv N_2 : \namesort
}
{\Gamma |- \NmBin{M_1}{M_2} \equiv \NmBin{N_1}{N_2} : \namesort}
\\
\Infer{Eq-Lam}
{\Gamma, \big(a \equiv b : \sort_1\big) |- M \equiv N                   : \sort_2 }
{\Gamma                                 |- \Lam{a}{M} \equiv \Lam{b}{N} : \sort_1 @> \sort_2 }
\and
\Infer{Eq-App}
{
\arrayenvcl{
 \Gamma |- M_1 \equiv N_1 : \sort_1 @> \sort_2
  \\
 \Gamma |- M_2 \equiv N_2 : \sort_1
}
}
{\Gamma |- M_1 (M_2) \equiv N_1 (N_2) : \sort_2}
\and
\Infer{Eq-$\beta$}
{
  \Gamma  |- M_2 \equiv M_2' : \sort_1
  \\
  \Gamma, a \equiv a : \sort_1  |- M_1 \equiv M_1' : \sort_2
}
{ \Gamma |- (\Lam{a}M_1)M_2 \equiv [M_2'/a]M_1' : \sort_2 }
\end{mathpar}

\caption{Deductive rules for showing that two name terms are equivalent}
\label{fig:equiv-rules}

\end{figure}

\begin{figure}
\centering
  \judgbox{\Gamma |- M \disjoint N : \sort}
          {The name terms $M$ and $N$ are \emph{apart} at sort $\gamma$}

\begin{mathpar}
\Infer{Var}
{(a \disjoint b : \sort) \in \Gamma}
{\Gamma |- a \disjoint b : \sort}
\and
\Infer{D-Sym}
{\Gamma |- N \disjoint M : \sort}
{\Gamma |- M \disjoint N : \sort}
\and
\Infer{D-trans} 
{
  \Gamma  |- M_1 \equiv M_2 : \sort
  \\
  \Gamma  |- M_2 \disj M_3 : \sort
}
{ \Gamma |- M_1 \disj M_3 : \sort }
\and
\Infer{D-Proj$_1$}
{\Gamma |- M_1 \disjoint N_1 : \sort_1}
{\Gamma |- (M_1, M_2) \disjoint (N_1, N_2) : \sort_1 * \sort_2}
\and
\Infer{D-Proj$_2$}
{\Gamma |- M_2 \disjoint N_2 : \sort_2}
{\Gamma |- (M_1, M_2) \disjoint (N_1, N_2) : \sort_1 * \sort_2}
\and
\Infer{D-Bin$_1$}
{\Gamma |- M_1 \disj N_1 : \namesort
}
{\Gamma |- \NmBin{M_1}{M_2} \disj \NmBin{N_1}{N_2} : \namesort}
\and
\Infer{D-Bin$_2$}
{
  \Gamma |- M_2 \disj N_2 : \namesort
}
{\Gamma |- \NmBin{M_1}{M_2} \disj \NmBin{N_1}{N_2} : \namesort}
\and
\Infer{D-EqTag$_1$}
{
  \Gamma |- M_1 \equiv M_2 : \namesort
}
{\Gamma |- \NmBin{M_2}{N} \disj M_1 : \namesort}
\and
\Infer{D-EqTag$_2$}
{
  \Gamma |- N_1 \equiv N_2 : \namesort
}
{\Gamma |- \NmBin{M}{N_1} \disj N_2 : \namesort}
\and
\Infer{D-Lam}
{\Gamma, \big(a \equiv b : \sort_1\big) |- M \disjoint N                   : \sort_2 }
{\Gamma                                 |- \Lam{a}{M} \disjoint \Lam{b}{N} : \sort_1 @> \sort_2 }
\and
\Infer{D-App}
{\Gamma |- M_1 \disjoint N_1 : \sort_1 @> \sort_2
  \\
 \Gamma |- M_2 \equiv    N_2 : \sort_1
}
{\Gamma |- M_1 (M_2) \disjoint N_1 (N_2) : \sort_2}
\and
\Infer{D-$\beta$}
{
  \Gamma.1 |- M_2 : \gamma_2
  \\
  \Gamma.1, a : \gamma_2 |- M_1 : \gamma
  \\
  \Gamma |- [M_2/a] M_1 \disj N : \gamma
}
{
  \Gamma |- (\Lam{a}{M_1})\;M_2 \disj N : \gamma
}
\end{mathpar}
\caption{Deductive rules for showing that two name terms are apart}
\label{fig:disj-rules}
\end{figure}


\subsection{Metatheory of name term language}
\label{sec:nameterm-rules-soundness}
\label{sec:nameterm-rules-completeness}

Some lemmas in this section are missing complete proofs and should be
considered conjectures
(\Lemmaref{lem:syn-eq-proj}---\Lemmaref{lem:sub-eval-sem-eq}).

\begin{lemma}[Projections of syntactic equivalence]~\\
  \label{lem:syn-eq-proj}
  If $\Gamma |- M_1 \equiv M_2 : \gamma$, then $\Gamma.1 |- M_1 : \gamma$ and
  $\Gamma.2 |- M_2 : \gamma$.
\end{lemma}

\begin{lemma}[Compatibility of substitution with name term structure]
\label{lem:compat-sub-ntm}
~
\begin{enumerate}
\item We have $[\sigma](M_1 , M_2) = ([\sigma]M_1,[\sigma]M_2)$.
\item We have $[\sigma]\NmBin{M_1}{M_2} = \NmBin{[\sigma]M_1}{[\sigma]M_2}$.
\item We have $[\sigma](MN) = ([\sigma]M)([\sigma]N)$.
\end{enumerate}
\end{lemma}

\begin{lemma}[Determinism of evaluation up to substitution]~\\
  \label{lem:sub-eval-det}
  If $\Gamma.1 |- M : \gamma$, $\Gamma.2 |- M : \gamma$, $||- \sigma_1 \equiv
  \sigma_2 : \Gamma$, $[\sigma_1]M !! V_1$, and $[\sigma_2]M !! V_2$, then there
  exists a $V$ such that $V_1 = [\sigma_1]V$ and $V_2 = [\sigma_2]V$.
\end{lemma}

\begin{lemma}[Reflexivity of semantic equivalence]
  ~
\label{lem:refl-sem}
\begin{enumerate}
\item If $|- M : \gamma$, then $||- M \equiv M : \gamma$.
\item If $\Gamma.1 |- V : \gamma$, $\Gamma.2 |- V : \gamma$, $||- \sigma_1
  \equiv \sigma_2 : \Gamma$, $V_1 = [\sigma_1]V$, and $V_2 = [\sigma_2]V$, then
  $||- V_1 \equiv V_2 : \gamma$.
\end{enumerate}
\end{lemma}

\begin{lemma}[Type safety]~\\
  \label{lem:type-safety}
  If $\Gamma |- M : \gamma$, $[\sigma]M !! [\sigma]V$, then
  $\Gamma |- V : \gamma$.
\end{lemma}

\begin{lemma}[Symmetry of semantic equivalence]
  ~
\label{lem:sym-sem}
\begin{enumerate}
  \item If $||- \sigma_1 \equiv \sigma_2 : \Gamma$, then $||- \sigma_2 \equiv \sigma_1 : \Gamma^\mathsf{T}$.
  \item If $||- V_1 \equiv V_2 : \gamma$, then $||- V_2 \equiv V_1 : \gamma$.
  \item If $\cdot ||- M_1 \equiv M_2 : \gamma$, then $\cdot ||- M_2 \equiv M_1
    : \gamma$.
  \item If $\Gamma ||- M_1 \equiv M_2 : \gamma$, then $\Gamma^\mathsf{T} ||-
    M_2 \equiv M_1 : \gamma$.
    \end{enumerate}
\end{lemma}

\begin{lemma}[Evaluation respects semantic equivalence]~\\
\label{lem:eval-resp-sem-eq}
  If $\Gamma ||- M \equiv N : \gamma$, $||- \sigma_1 \equiv \sigma_2 : \Gamma$,
  and $[\sigma_1] M !! V_1$, then there exists $V_2$ such that $[\sigma_2]N !!
  V_2$ and $||- V_1 \equiv V_2 : \gamma$.
\end{lemma}

\begin{lemma}[Closing substitutions respect syntactic equivalence]~\\
\label{lem:fund-sub}
  If $||- \sigma_1 \equiv \sigma_2 : \Gamma$ and $\Gamma |- M_1 \equiv M_2 :
  \gamma$, then $\cdot ||- [\sigma_1]M_1 \equiv [\sigma_2]M_2 : \gamma$.
\end{lemma}


\begin{lemma}[Reflexivity of name term evaluation]~\\
\label{lem:sub-eval-sem-eq}
  If $\Gamma.1 |- M : \gamma$, $\Gamma.2 |- M : \gamma$, $||- \sigma_1 \equiv
  \sigma_2 : \Gamma$, $[\sigma_1]M !! V_1$, and $[\sigma_2]M !! V_2$, then $||-
  V_1 \equiv V_2 : \gamma$.
\end{lemma}

\begin{lemma}[Transitivity of value equivalence]~\\
\label{lem:trans-val}
  If $|- V_1 : \gamma$, $|- V_2 : \gamma$, $||- V_1 \equiv V_2 : \gamma$, and
  $||- V_2 \equiv V_3 : \gamma$, then $||- V_1 \equiv V_3 : \gamma$.
\end{lemma}
\begin{proof}
  Uses strong normalization.
\end{proof}

\begin{conjecture}[Soundness of deductive equivalence]
  If
  $\Gamma |- M_1 \equiv M_2 : \sort$
  then
  $\Gamma ||- M_1 \equiv M_2 : \sort$.
\end{conjecture}
\begin{proof} By induction on the given derivation.
\DerivationProofCase{Eq-Var}
{(a \equiv b : \sort) \in \Gamma}
{\Gamma |- a \equiv b : \sort}

By definition of closing substitutions.

\DerivationProofCase{Eq-Refl}
      { 
      (\Gamma).1 |- M : \sort
      \\\\
      (\Gamma).2 |- M : \sort
      }
      {\Gamma |- M \equiv M : \sort}

By \Lemmaref{lem:sub-eval-det}, \Lemmaref{lem:type-safety}, and \Lemmaref{lem:refl-sem}.
      
\DerivationProofCase{Eq-Sym}
      {\Gamma |- N \equiv M : \sort}
      {\Gamma |- M \equiv N : \sort}

By \Lemmaref{lem:sym-sem}.

\DerivationProofCase{Eq-Trans}
{
  \Gamma  |- M_1 \equiv M_2 : \sort
  \\\\
  \Gamma  |- M_2 \equiv M_3 : \sort
}
{ \Gamma |- M_1 \equiv M_3 : \sort }

By idempotency of flipping relational contexts, \Lemmaref{lem:eval-resp-sem-eq},
inductive hypotheses on the two given subderivations, \Lemmaref{lem:sym-sem}, and \Lemmaref{lem:trans-val}.

\DerivationProofCase{Eq-Pair}
{\Gamma |- M_1 \equiv N_1 : \sort_1
  \\
 \Gamma |- M_2 \equiv N_2 : \sort_2
}
{\Gamma |- (M_1, M_2) \equiv (N_1, N_2) : \sort_1 * \sort_2}

By \Lemmaref{lem:compat-sub-ntm} and inductive hypothesis.

\DerivationProofCase{Eq-Bin}
{\Gamma |- M_1 \equiv N_1 : \namesort
  \\
 \Gamma |- M_2 \equiv N_2 : \namesort
}
{\Gamma |- \NmBin{M_1}{M_2} \equiv \NmBin{N_1}{N_2} : \namesort}

By \Lemmaref{lem:compat-sub-ntm} and inductive hypothesis.

\DerivationProofCase{Eq-Lam}
{\Gamma, \big(a \equiv b : \sort_1\big) |- M \equiv N                   : \sort_2 }
{\Gamma                                 |- \Lam{a}{M} \equiv \Lam{b}{N} : \sort_1 @> \sort_2 }

By Barendregt's Substitution Lemma and inductive hypothesis.

\DerivationProofCase{Eq-App}
{
\arrayenvcl{
 \Gamma |- M_1 \equiv N_1 : \sort_1 @> \sort_2
  \\
 \Gamma |- M_2 \equiv N_2 : \sort_1
}
}
{\Gamma |- M_1 (M_2) \equiv N_1 (N_2) : \sort_2}

By \Lemmaref{lem:compat-sub-ntm} and inversion (teval-app) of resulting derivations, the inductive hypothesis on
the two given syntactic equivalence subderivations (of Eq-App), and definition
of semantic equivalence of arrow-sorted values, we get the result.

\DerivationProofCase{Eq-$\beta$}
{
  \Gamma  |- M_2 \equiv M_2' : \sort_1
  \\
  \Gamma, a \equiv a : \sort_1  |- M_1 \equiv M_1' : \sort_2
}
{ \Gamma |- (\Lam{a}M_1)M_2 \equiv [M_2'/a]M_1' : \sort_2 }

Fix $||- \sigma_1 \equiv \sigma_2 : \Gamma$. Suppose $[\sigma_1]((\lam{a}
M_1)M_2) !! V_1$ and $[\sigma_2]([M_2'/a]M_1') !! V_2$. We need to show $||- V_1
\equiv V_2 : \gamma_2$. By \Lemmaref{lem:compat-sub-ntm} and inversion of
teval-app, $[\sigma_1]M_2 !! V$ and $[V/a]([\sigma_1]M_1) !! V_1$ for some $V$.
Hence, because $\Gamma, a \equiv a : \gamma_1$, we have $[\sigma_1, a \mapsto
V]M_1 !! V_1$. Rewrite $[\sigma_2]([M_2'/a]M_1') !! V_2$ as $[\sigma_2, a \mapsto
[\sigma_2]M_2']M_1' !! V_2$. By \Lemmaref{lem:fund-sub}, $\cdot |- V \equiv
[\sigma_2]M_2' : \gamma_1$. Therefore,
$$
||- \sigma_1, a \mapsto V \equiv \sigma_2, a \mapsto [\sigma_2]M_2' : \Gamma, a
\equiv a : \gamma_1
$$
By the inductive hypothesis on $\Gamma, a \equiv a : \gamma_1 |- M_1 \equiv M_1'
: \gamma_2$, we get $||- V_1 \equiv V_2 : \gamma_2$.
\end{proof}

\begin{conjecture}[Soundness of deductive disjointness]
  If
  $\Gamma |- M_1 \disj M_2 : \sort$
  then
  $\Gamma ||- M_1 \disj M_2 : \sort$.
\end{conjecture}


\begin{conjecture}[Completeness of deductive equivalence]
  If
  $\Gamma ||- M_1 \equiv M_2 : \sort$
  then
  $\Gamma |- M_1 \equiv M_2 : \sort$.
\end{conjecture}

\begin{conjecture}[Completeness of deductive disjointness]
  If
  $\Gamma ||- M_1 \disj M_2 : \sort$
  then
  $\Gamma |- M_1 \disj M_2 : \sort$.
\end{conjecture}



\section{Index term language}
\label{sec:indextermlang}

%
%

%

\begin{figure}[htbp]
  \centering
  \small
  \begin{align*}
    \extractassns{\cdot}
      &= \cdot
    \\
    \extractassns{\Gamma, P}
      &= \extractassns{\Gamma}, P
    \\
    \extractassns{\Gamma, \trueprop}
      &= \extractassns{\Gamma}
    \\
    \extractassns{\Gamma, (P_1 \andprop \dots \andprop P_n)}
      &= \extractassns{\Gamma}, P_1, \dots, P_n
\\ &\text{for $n \geq 1$, where each $P_k$}
   \\ &\text{~~~has the  form $i \disj j : \sort$ or $i \equiv j : \sort$}
    \\
    \extractassns{\Gamma, \mathcal{Z}}
      &= \extractassns{\Gamma}~\text{where $\mathcal{Z}$ is not a proposition}
  \end{align*}
  \begin{align*}
    \extractctx{\cdot}
      &= \cdot
    \\
    \extractctx{\Gamma, a : \sort}
      &= \extractctx{\Gamma}, (a \equiv a : \sort)
    \\
    \extractctx{\Gamma, \alpha : \type}
      &= \extractctx{\Gamma}
    \\
    \extractctx{\Gamma, d : K}
      &= \extractctx{\Gamma}
    \\
    \extractctx{\Gamma, p : \cdots}
      &= \extractctx{\Gamma}
    \\
    \extractctx{\Gamma, x : A}
      &= \extractctx{\Gamma}
    \\
    \extractctx{\Gamma, P}
     & = \extractctx{\Gamma}
  \end{align*}
  \[
  \extract{\Gamma} ~=~
  (\extractassns{\Gamma}; \extractctx{\Gamma})
  \]
  
  \caption{Extraction function on typing contexts}
  \label{fig:extract}
\end{figure}


\begin{figure}
  \judgbox{i \Value}
          {Index $i$ is a value (it evaluates to itself)}
\begin{mathpar}
\Infer{val-singleton}
{ }
{ \{ N \} \Value }
~~~
\Infer{val-disj}
{ i \Value 
  \\
  j \Value 
}
{ i \disj j \Value }
~~~
\Infer{val-pair}
{ i \Value 
  \and
  j \Value 
}
{ \Pair{i}{j} \Value }
~~~
\Infer{val-abs}
{ }
{ \lam{a}{i} \Value }
\end{mathpar}

\vspace{1ex}

\judgbox{i !! j}
        {Index~$i$ evaluates to index~$j$}
\vspace{-2ex}
\begin{mathpar}
\Infer{\!value}
    {i \Value}
    { i !! i }
~~~~
\Infer{\!pair}
    {
      i_1 !! j_1
      ~~~~
      i_2 !! j_2
    }
    {\Pair{i_1}{i_2} !! \Pair{j_1}{j_2}}
~~~~
\Infer{\!proj}
    { i !! (j_1, j_2) }
    { \Proj{b}{i} !! j_b}
~~~~
\Infer{\!union}
    {
      \arrayenvbl{
        i !! i' 
        \\
        j !! j'
      }
    }
    {(i \union j) !! (i' \union j')}
\and
\Infer{\!disj}
    {
      \arrayenvbl{
        i !! i' 
        \\
        j !! j'
      }
      ~~~~~
      ||- i' \disj j' : \namesetsort
    }
    {(i \disj j) !! (i' \disj j')}
\and
\Infer{\!app}
{ i !! \lam{a}{i'}
\\
j !! j'
\\
[j'/a] i' !! k
}
{i(j) !! k}
\and
\Infer{\!map-set}
{ j !! j' 
  \\
\mapset{M}{j'}{j''} 
}
{ M [[ j ]] !! j'' }
\end{mathpar}

\judgbox{ \mapset{M}{X}{Y} }
        {Name term function $M$, applied to each member of $X$, yields name set $Y$}
\begin{mathpar}
\Infer{Single}
{ M ( N ) \nteval V }
{ \mapset{M}{ \{ N \} }{ \{ V \} } }
\and
\Infer{Apart}
{ \mapset{ M }{ X_1 }{ Y_1 } 
  \\
  \mapset{ M }{ X_2 }{ Y_2 } 
}
{ \mapset{ M }{ X_1 \disj X_2 }{ Y_1 \disj Y_2 } }
\vspace*{-1.5ex}
\end{mathpar}

\caption{Evaluation rules for indices}
\label{fig:index-eval}
\end{figure}


\begin{figure}
\judgbox{ \Gamma |- M \in X}
        {Name term~$M$ is a member of name set~$X$, 
          assuming $X \Value$
        }
\begin{mathpar}
\Infer{Apart$_1$}
{ \Gamma |- M \in X          }
{ \Gamma |- M \in (X \disj Y) }
\and
\Infer{Apart$_2$}
{ \Gamma |- M \in Y          }
{ \Gamma |- M \in (X \disj Y)  }
\and
\Infer{Union$_1$}
{ \Gamma |- M \in X          }
{ \Gamma |- M \in (X \cup Y) }
\and
\Infer{Union$_2$}
{ \Gamma |- M \in Y          }
{ \Gamma |- M \in (X \cup Y)  }
\and
\Infer{Single}
{ \Gamma |- M \equiv N : \namesort }
{ \Gamma |- M \in \{ N \}  }
\and
\Infer{EqualNameSet}
{
  \extractassns \Gamma |- X \equiv Y : \NmSet
  \\
  \Gamma |- N \in Y
}
{
  \Gamma |- N \in X
}
\end{mathpar}

\judgbox{\Gamma |- M \notin X}
        {The name of name term $M$ is \emph{not} a member of name set $X$,
          assuming $X \Value$}

\begin{mathpar}
\Infer{Apart}
{ \Gamma |- M \notin X \\
  \Gamma |- M \notin Y }
{ \Gamma |- M \notin \left( X \disj Y \right)  }
\and
\Infer{Single}
{ \Gamma |- M \disj N : \namesort }
{ \Gamma |- M \notin \{ N \} }
\and
\Infer{Empty}
{  }
{ \Gamma |- M \notin \emptyset }
\vspace*{-2.3ex}
\end{mathpar}

\caption{Name term membership}
\label{fig:name-term-membership}
\end{figure}


%
%


\subsection{Semantic equivalence and apartness of index terms}

\begin{defn}[Closing substitutions for index terms]~\\ 
We define closing substitution pairs related by equivalence and
  disjointness assumptions in a context~$\Gamma$.
These definitions use and are used by the definitions
below for equivalence and apartness of open terms.

\begin{itemize}
\item
  $||- \sigma_1 \equiv \sigma_2 : \Gamma$ means that 
  $(x \equiv y : \sort) \in \Gamma$ 
  implies 
  \big( 
  $\sigma_1(x) = i$ 
  and $\sigma_2(y) = j$ 
  and $\cdot ||- i \equiv j : \sort$
  \big)  
  
\item
  $||- \sigma_1 \disj \sigma_2 : \Gamma$ means that
  $(x \disj y : \sort) \in \Gamma$
  implies
  \big($\sigma_1(x) = i$ 
  and $\sigma_2(y) = j$ 
  and $\cdot ||- i \disj j : \sort$
  \big)

\item
  $||- \sigma_1 \sim \sigma_2 : \Gamma$ means that
  \big(
  $||- \sigma_1 \equiv \sigma_2 : \Gamma$
  and
  $||- \sigma_1 \disj \sigma_2 : \Gamma$
  \big)
\end{itemize}
\end{defn}

\begin{defn}[Semantic equivalence of index terms]
\label{def:semantic-equivalence-i}
  We define $\Gamma ||- i_1 \equiv i_2 : \sort$ as follows:
  \\
  $(\Gamma).1 |- i_1 : \sort$
  and $(\Gamma).2 |- i_2 : \sort$
  and, 
  \\
  for all $\sigma_1$, $\sigma_2$
  such that
  $||- \sigma_1 \equiv \sigma_2 : \Gamma$ and $[\sigma_1]i_1 !! j_1$ and $[\sigma_2]i_2 !! j_2$,
  we have
  the following about $j_1$ and $j_2$:
  
  \medskip

 \begin{tabular}{|c|l|}
\hline
 Sort~$\sort$ & Indices~$j_1$ and $j_2$ of sort~$\sort$ are equivalent, written $||- j_1 \equiv j_2 : \sort$
\\
\hline
  $\unitsort$            & Always
\\
  $\NmSet$ & When \big( $||- M \in j_1$  if and only if~~$||- M \in j_2$  \big)
\\
  $\sort_1 * \sort_2$ &
  When $j_1 = (j_{11}, j_{12})$
  and $j_1 = (j_{21}, j_{22})$
  \\
  & ~~and $||- j_{11} \equiv j_{21} : \sort_1$
  \\
  & ~~and $||- j_{12} \equiv j_{22} : \sort_2$
\\
  $\sort_1 @@> \sort_2$ &
 When $j_1 = \Lam{a_1}{X_1}$ and $j_2 = \Lam{a_2}{X_2}$,
 \\ & ~~ and
 for all name terms $||- Y_1 \equiv Y_2 : \sort_1$,
\\
&  $~~~~~~~~(\Lam{a_1}{X_1})(Y_1) !! Z_1$ and  $(\Lam{a_2}{X_2})(Y_2) !! Z_2$
\\
& $~~~~~$implies ~$||- Z_1 \equiv Z_2 : \sort_2$
\\
\hline
\end{tabular}
\end{defn}

\medskip

\begin{defn}[Semantic apartness of index terms]
\label{def:semantic-apartness-i}
  We define $\Gamma ||- i_1 \disj i_2 : \sort$ as follows:
  \\
  $(\Gamma).1 |- i_1 : \sort$
  and $(\Gamma).2 |- i_2 : \sort$
  and, 
  \\
  for all $\sigma_1$, $\sigma_2$
  such that
  $||- \sigma_1 \sim \sigma_2 : \Gamma$ and $[\sigma_1] i_1 !! j_1$ and $[\sigma_2] i_2 !! j_2$,
  we have the following about $j_1$ and $j_2$:
  
  \medskip

 \begin{tabular}{|c|l|}
\hline
 Sort ($\sort$) & Index values $j_1$ and $j_2$ of sort~$\sort$ are apart, written $||- j_1 \disj j_2 : \sort$
\\
\hline
  $\unitsort$            & Always
\\
  $\NmSet$ &
  When
  \tabularenvl{
  \big( $||- M \in j_1$~implies~~$||- M \notin j_2$  \big)
  \\ and
  \big( $||- M \in j_2$~implies~~$||- M \notin j_1$  \big)
  }
\\
  $\sort_1 * \sort_2$ &
  When
  \tabularenvl{
  $j_1 = (j_{11}, j_{12})$
  and $j_1 = (j_{21}, j_{22})$
  \\
  and $||- j_{11} \disj j_{21} : \sort_1$
  and $||- j_{12} \disj j_{22} : \sort_2$
  }
\\
  $\sort_1 @@> \sort_2$ &
 When $j_1 = \Lam{a_1}{X_1}$ and $V_2 = \Lam{a_2}{X_2}$,
 \\ & ~~and
 for all name terms $||- Y_1 \equiv Y_2 : \sort_1$,
\\
&  ~~~~~~$(\Lam{a_1}{X_1})(Y_1) !! Z_1$ and  $(\Lam{a_2}{X_2})(Y_2) !! Z_2$
\\
& ~~~~implies $||- Z_1 \disj Z_2 : \sort_2$
\\
\hline
\end{tabular}
\end{defn}

\medskip

The next two definitions bridge the gap with the type system,
in which contexts $\Gamma_T$ also include propositions $P$.
It is defined assuming that $\extract{\Gamma_T}$
(defined in \Figureref{fig:extract})
has given us some propositions $P_1, \dots, P_n$ and 
a relational context $\Gamma$.

\begin{defn}[Extended semantic equivalence of index terms]
\label{def:super-semantic-equivalence-i}
~\\
  We define $P_1, \dots, P_n ; \Gamma ||- i \equiv j  : \sort$
  to hold if and only if
  \begin{center}
      $\mathcal{J}(P_1)$
      and $\cdots$
      and $\mathcal{J}(P_n)$
      implies
      $\Gamma ||- i \equiv j : \sort$    
  \end{center}
  where $\mathcal{J}(i \mathrel{\Theta} j : \sort) = (\Gamma ||- i \mathrel{\Theta} j : \sort)$.
\end{defn}

\begin{defn}[Extended semantic apartness of index terms]
\label{def:super-semantic-apartness-i}
~\\
  We define $P_1, \dots, P_n ; \Gamma ||- i \disj j  : \sort$
  to hold if and only if
  \begin{center}
      $\mathcal{J}(P_1)$
      and $\cdots$
      and $\mathcal{J}(P_n)$
      implies
      $\Gamma ||- i \disj j : \sort$    
  \end{center}
  where $\mathcal{J}(i \mathrel{\Theta} j : \sort) = (\Gamma ||- i \mathrel{\Theta} j : \sort)$.
\end{defn}

When a typing context is weakened, semantic equivalence and apartness
under the extracted context continue to hold:

\begin{lemma}[Weakening of semantic equivalence and apartness]
\label{lem:weakening-semantic}
  ~\\
  If
  $\extract{\Gamma_T} ||- i_1 \equiv i_2 : \sort$
  (respectively $i_1 \disj i_2 : \sort$)
  then
  $\extract{\Gamma_T, \Gamma_T'} ||- i_1 \equiv i_2 : \sort$  
  (respectively $i_1 \disj i_2 : \sort$.
\end{lemma}
\begin{proof}  
  By induction on $\Gamma_T'$.

  We prove the $\equiv$ part; the $\disj$ part is similar.

  \begin{itemize}
  \item 
  If $\Gamma_T' = \cdot$,
  we already have the result.

  \item
  If $\Gamma_T' = (\Gamma', P)$
  then:

    By i.h., $\extract{\Gamma_T, \Gamma'} ||- i_1 \equiv i_2 : \sort$.
    
    That is, $\extractassns{\Gamma_T, \Gamma'};
    \extractctx{\Gamma_T, \Gamma'} ||- i_1 \equiv i_2 : \sort$.

    By its definition, $\extractctx{\Gamma_T, \Gamma', P} = 
    \extractctx{\Gamma_T, \Gamma', P}$.
    \\
    Therefore, we have
    $\extractassns{\Gamma_T, \Gamma'};
    \extractctx{\Gamma_T, \Gamma', P} ||- i_1 \equiv i_2 : \sort$.
    \\
    Adding an assumption before the semicolon only supplements
    the antecedent in \Defnref{def:super-semantic-equivalence-i},
    so
    \[
    \extractassns{\Gamma_T, \Gamma', P};
    \extractctx{\Gamma_T, \Gamma', P} ||- i_1 \equiv i_2 : \sort
    \]
    which was to be shown.
    
  \item
    If $\Gamma_T' = (\Gamma', a : \sort)$
    then by i.h.,
    \[
    \extract{\Gamma_T, \Gamma'} ||- i_1 \equiv i_2 : \sort
    \]
    By definition of $\extractctxsym$,
    \[
    \extractctx{\Gamma_T, \Gamma', a : \sort}
    =
    \extractctx{\Gamma_T, \Gamma'}, a \equiv a : \sort
    \]
    By the i.h.\ and \Defnref{def:semantic-equivalence-i},
    $(\extractctx{\Gamma_T, \Gamma'}).1 |- i_1 : \sort$
    We need to show that
    $(\extractctx{\Gamma_T, \Gamma', a : \sort}).1 |- i_1 : \sort$,
    which follows by weakening on sorting.
    The ``$.2$'' part is similar.

    Since $a$ does not occur in $i_1$ and $i_2$,
    applying longer substitutions that include $a$
    to $i_1$ and $i_2$ does not change them; thus, we get the same
    $j_1$ and $j_2$ as for $\Gamma_T, \Gamma'$.

  \item In the remaining cases of $\mathcal{Z}$ for $\Gamma_T' = (\Gamma', \mathcal{Z})$, neither $\extractassnssym$ nor $\extractctxsym$ change,
    and the i.h.\ immediately gives the result. \qedhere
  \end{itemize}
\end{proof}

\clearpage
\subsection{Deductive equivalence and apartness for index terms}

\begin{figure}[h]
\centering
  \judgbox{\Gamma |-i \equiv j : \sort}
          {The name terms $i$ and $j$ are \emph{equivalent} at sort $\gamma$}

\begin{mathpar}
\Infer{Eq-Var}
{(i \equiv j : \sort) \in \Gamma}
{\Gamma |- i \equiv j : \sort}
\and
\Infer{E-Refl}
      { (\Gamma).1 |-i : \sort
      \\
      (\Gamma).2 |-i : \sort
      }
      {\Gamma |-i \equiv i : \sort}
\and
\Infer{E-Sym}
      {\Gamma |- j \equiv i : \sort}
      {\Gamma |-i \equiv j : \sort}
\\
\Infer{Eq-Pair}
{\Gamma |-i_1 \equiv j_1 : \sort_1
  \\
 \Gamma |-i_2 \equiv j_2 : \sort_2
}
{\Gamma |- (i_1,i_2) \equiv (j_1, j_2) : \sort_1 * \sort_2}
\and
\Infer{Eq-Lam}
{\Gamma, \big(a \equiv b : \sort_1\big) |-i \equiv j                   : \sort_2 }
{\Gamma                                 |- \Lam{a}{i} \equiv \Lam{b}{j} : \sort_1 @> \sort_2 }
\and
\Infer{Eq-App}
{
\arrayenvcl{
 \Gamma |-i_1 \equiv j_1 : \sort_1 @> \sort_2
  \\
 \Gamma |-i_2 \equiv j_2 : \sort_1
}
}
{\Gamma |-i_1 (i_2) \equiv j_1 (j_2) : \sort_2}
\and
\Infer{Eq-$\beta$}
{
\arrayenvcl{
  (\Gamma).1, a : \sort_2  |-i_1 : \sort
  \\
  (\Gamma).1  |-i_2 : \sort_2
  \\
  \Gamma      |- [i_2/a]i_1 \equiv j : \sort
}
}
{ \Gamma |- (\Lam{a}i_1)i_2 \equiv j : \sort }
\and
\Infer{Eq-Empty}
{ }
{ \Gamma |- \emptyset \equiv \emptyset : \NmSet }
\and
\Infer{Eq-Single}
{ \Gamma |- M \equiv N : \Nm }
{ \Gamma |- \{ M \} \equiv \{ N \} : \NmSet }
\and
\Infer{Eq-Apart}
{ \Gamma |- X_1 \equiv X_2 : \NmSet 
  \\
  \Gamma |- Y_1 \equiv Y_2 : \NmSet
}
{ \Gamma |- (X_1 \disj Y_1) \equiv (X_2 \disj Y_2) : \NmSet }
\and
\Infer{Eq-Perm}
{ \Gamma |- (X_2 \disj X_1) \equiv Y : \NmSet }
{ \Gamma |- (X_1 \disj X_2) \equiv Y : \NmSet }
\and
\Infer{Eq-Map}
{ \Gamma |- M \equiv N : \Nm @> \Nm
  \\
  \Gamma |- X \equiv Y : \NmSet
}
{ \Gamma |- M [[ X ]] \equiv N [[ Y ]] : \NmSet }
\vspace*{-1.5ex}
\end{mathpar}

\caption{Deductive rules for showing that two index terms are equivalent}
\label{fig:idx-equiv-rules}

\end{figure}

\begin{figure}
\centering
  \judgbox{\Gamma |-i \disjoint j : \sort}
          {The index terms $i$ and $j$ are \emph{apart} at sort $\gamma$}

\begin{mathpar}
\Infer{Var}
{(a \disjoint b : \sort) \in \Gamma}
{\Gamma |- a \disjoint b : \sort}
\and
\Infer{D-Sym}
{\Gamma |- j \disjoint i : \sort}
{\Gamma |-i \disjoint j : \sort}
\and
\Infer{D-Proj$_1$}
{\Gamma |-i_1 \disjoint j_1 : \sort_1}
{\Gamma |- (i_1,i_2) \disjoint (j_1, j_2) : \sort_1 * \sort_2}
\and
\Infer{D-Proj$_2$}
{\Gamma |-i_2 \disjoint j_2 : \sort_2}
{\Gamma |- (i_1,i_2) \disjoint (j_1, j_2) : \sort_1 * \sort_2}
\and
\Infer{D-Lam}
  {\Gamma, (a \equiv b : \sort_1) |-i \disjoint j                   : \sort_2 }
  {\Gamma                                 |- \Lam{a}{i} \disjoint \Lam{b}{j} : \sort_1 @> \sort_2 }
\and
\Infer{D-App}
  {
    \Gamma |-i_1 \disjoint j_1 : \sort_1 @> \sort_2
    \\
    \Gamma |-i_2 \equiv    j_2 : \sort_1
  }
  {\Gamma |-i_1 (i_2) \disjoint j_1 (j_2) : \sort_2}
\and
\Infer{D-$\beta$}
  {
    \Gamma  |- [i_2/a]i_1 \disj j : \sort
    \\
    \arrayenvbl{
      (\Gamma).1  |-i_2 : \sort_2
      \\
      (\Gamma).1, a : \sort_2  |-i_1 : \sort
    }
  }
  { \Gamma |- (\Lam{a}i_1)i_2 \disj j : \sort }
\\
\Infer{D-Empty}
  { (\Gamma).2 |- X : \NmSet }
  { \Gamma |- \emptyset \disj X : \NmSet }
\and
\Infer{D-Single}
  { \Gamma |- M \disj N : \Nm }
  { \Gamma |- \{ M \} \disj \{ N \} : \NmSet }
\and
\Infer{\!D-Apart}
  {
    \Gamma |- X_1 \disj Y : \NmSet 
    ~~~~~
    \Gamma |- X_2 \disj Y : \NmSet
  }
  { \Gamma |- (X_1 \disj X_2) \disj Y : \NmSet }
\and
\Infer{\!D-Map}
  { 
    \Gamma |- M \disj N : \Nm @> \Nm
    ~~~~~
    \Gamma |- X \equiv Y : \NmSet
  }
  { \Gamma |- M [[ X ]] \disj N [[ Y ]] : \NmSet }
\vspace*{-1.5ex}
\end{mathpar}

\caption{Deductive rules for showing that two index terms are apart}
\label{fig:idx-disj-rules}
\end{figure}







\clearpage
\section{Strong Normalization Proof for the Name Term Language}
\label{apx:strongnormproof-nameterm}

\begin{definition}[$R_{\sort}(e)$]~
	\label{def:strongnormop-name}
	\begin{enumerate}
		\item $R_{\sort}(e)$ if and only if $e$ halts, and $\sort \ne (\sort_1 @> \sort_2)$
		\item $R_{(\sort_1 @> \sort_2)}(e)$ if and only if
                  (1) $e$ halts and
                  (2) $R_{\sort_1}(e')$ implies $R_{\sort_2}(e~e')$.
	\end{enumerate}
\end{definition}

\begin{lemma}[Termination]
	\label{lem:strongnorm-lemma-1-name}
	If $R_{\sort}(e)$, then $e$ halts.
\end{lemma}

\begin{proof}
	By definition of $R_{\sort}(e)$.
\end{proof}

\begin{lemma}[Preservation]
	\label{lem:strongnorm-lemma-2-name}
	If $\Gamma |- e : \sort$ and $\sigma |- \Gamma$, then $ |- [\sigma] e : \sort$.
\end{lemma}

\begin{proof}
	By induction on the size of $\sigma$. When $\text{size}(\sigma) = 1$, the Lemma reduces to the Substitution Lemma.
\end{proof}

\begin{lemma}[Normalization]
	\label{lem:strongnorm-lemma-3-name}
	If $\Gamma |- e : \sort$ and $\sigma |- \Gamma$, then $R_{\sort}([\sigma]e)$.
\end{lemma}

\begin{proof} By induction on the given derivation.

	\DerivationProofCase{M-unit}
		{}
		{\Gamma |- \unitexp : \unitsort}
	
		\begin{llproof}
			\Pf{[\sigma]\unitexp}{=}{\unitexp}{By definition of $[\sigma](-)$}
			\Pf{\unitexp}{\nteval}{\unitexp}{By Rule teval-value}
			\Pf{[\sigma]\unitexp}{\nteval}{\unitexp}{Since $[\sigma]\unitexp = \unitexp$, $\unitexp \nteval \unitexp$}
			\Pf{\Gamma}{|-}{\unitexp : \unitsort}{Given}
			\Pf{}{|-}{[\sigma] \unitexp : \unitsort}{By \Lemmaref{lem:strongnorm-lemma-2-name}}
			\Hand \Pf{}{}{R_{\unitsort}([\sigma] \unitexp)}{Since $[\sigma]\unitexp \nteval \unitexp$}\trailingjust{and $|- [\sigma] \unitexp : \unitsort$}			
		\end{llproof}
	
	\DerivationProofCase{M-const}
		{ }
		{\Gamma |- n : \namesort}
	
		Similar to the M-unit case.                 
	
	\DerivationProofCase{M-var}
		{(a : \sort) \in \Gamma}
		{\Gamma |- a : \sort}
	
		\begin{llproof}
			\Pf{\exists v_i.([\sigma]a}{\nteval}{v_i)}{By definition of $[\sigma](-)$}
			\Pf{}{|-}{[\sigma] a : \sort}{From \Lemmaref{lem:strongnorm-lemma-2-name}}
			\Hand \Pf{}{}{R_{\sort}([\sigma] a)}{Since $[\sigma]a \nteval v_i$}\trailingjust{and $|- [\sigma] a : \sort$}
		\end{llproof}
	
	\DerivationProofCase{M-abs}
		{\Gamma, a : \sort_1 |- M : \sort_2}
		{
			\arrayenvbl{
				\Gamma |- (\lam{a} M) : (\sort_1 @> \sort_2)
				\\
				~
			}
		}
	
		\begin{llproof}
			\Pf{[\sigma, a -> \sort_1]}{|-}{\Gamma'}{Assumption, where $\Gamma' = \Gamma, a : \sort_1$}
			\Pf{|-}{([\sigma, a -> \sort_1](\lam{a} M))} {: (\sort_1 @> \sort_2)}{By \Lemmaref{lem:strongnorm-lemma-2-name}}
			\Pf{[\sigma, a -> \sort_1] (\lam{a} M)}{=}{(\lam{a} M)}{By definition of $[\sigma](-)$}
			\Hand \Pf{R_{\sort}([\sigma, a -> \sort_1] (\lam{a} M))}{}{}{Since $[\sigma, a -> \sort_1](\lam{a} M) \nteval  (\lam{a} M)$}\trailingjust{and $|- [\sigma, a -> \sort_1]  (\lam{a} M) : (\sort_1 @> \sort_2)$}
		\end{llproof}

	\DerivationProofCase{\!M-app}
		{
			\Gamma |- M_1 : (\sort' @> \sort)
			\\
			\Gamma |- M_2 : \sort'
		}
		{\Gamma |- (M_1\;M_2) : \sort}
		
		\begin{llproof}
			\Pf{}{}{R_{(\sort' @> \sort)}([\sigma] M_1)}{By inductive hypothesis}
			\Pf{}{}{R_{\sort'}([\sigma] M_2)}{By inductive hypothesis}
			\Pf{}{}{R_{\sort}(([\sigma]M_1\;[\sigma]M_2))}{By definition of $R_{-}(-)$}	
			\Hand \Pf{}{}{R_{\sort}([\sigma](M_1\;M_2))}{By definition of $[\sigma](-)$}	
		\end{llproof}
	
	\DerivationProofCase{\!M-pair}
		{
			\Gamma |- M_1 : \sort_1
			\\
			\Gamma |- M_2 : \sort_2
		}
		{\Gamma |- \Pair{M_1}{M_2} : (\sort_1 ** \sort_2)}
		
		\begin{llproof}
			\Pf{}{}{R_{\sort_1}([\sigma] M_1)}{By inductive hypothesis}
			\Pf{}{}{R_{\sort_2}([\sigma] M_2)}{By inductive hypothesis}
			\Pf{\exists v_1.([\sigma] M_1}{\nteval}{v_1)}{Since $R_{\sort_1}([\sigma] M_1)$}
			\Pf{\exists v_2.([\sigma] M_2}{\nteval}{v_2)}{Since $R_{\sort_2}([\sigma] M_2)$}
			\Pf{([\sigma] M_1, [\sigma] M_2)}{\nteval}{(v_1, v_2)}{By Rule teval-tuple}
			\Pf{[\sigma](M_1, M_2)}{\nteval}{(v_1, v_2)}{By definition of $[\sigma](-)$}	
			\Pf{}{|-}{[\sigma] (M_1, M_2) : (\sort_1 ** \sort_2)}{From \Lemmaref{lem:strongnorm-lemma-2-name}}
			\Hand \Pf{}{}{R_{(\sort_1 ** \sort_2)}([\sigma](M_1\;M_2))}{Since $[\sigma](M_1\;M_2) \nteval (v_1, v_2)$}\trailingjust{and $|- [\sigma] (M_1\;M_2) : (\sort_1 ** \sort_2)$}
		\end{llproof}
	
	\DerivationProofCase{\!M-bin}
		{
			\Gamma |- M_1 : \namesort
			\\\\
			\Gamma |- M_2 : \namesort
		}
		{\Gamma |- \NmBin{M_1}{M_2} : \namesort}
		
		\begin{llproof}
			\Pf{}{}{R_{\namesort}([\sigma] M_1)}{By inductive hypothesis}
			\Pf{}{}{R_{\namesort}([\sigma] M_2)}{By inductive hypothesis}
			\Pf{\exists v_1.([\sigma] M_1}{\nteval}{v_1)}{Since $R_{\sort_1}([\sigma] M_1)$}
			\Pf{\exists v_2.([\sigma] M_2}{\nteval}{v_2)}{Since $R_{\sort_2}([\sigma] M_2)$}
			\Pf{\NmBin{[\sigma] M_1}{[\sigma] M_2}}{\nteval}{(v_1, v_2)}{By Rule teval-bin}
			\Pf{[\sigma]\NmBin{M_1}{M_2}}{\nteval}{(v_1, v_2)}{By definition of $[\sigma](-)$}	
			\Pf{}{|-}{[\sigma] \NmBin{M_1}{M_2} : \namesort}{From \Lemmaref{lem:strongnorm-lemma-2-name}}
			\Hand \Pf{}{}{R_{\namesort}([\sigma] \NmBin{M_1}{M_2})}{Since $[\sigma] \NmBin{M_1}{M_2} \nteval (v_1, v_2)$}\trailingjust{and $|- [\sigma] \NmBin{M_1}{M_2} : \namesort$}
		\end{llproof}
\end{proof}

\begin{restatable}[Strong Normalization]{theorem}{thmrefstrongnormname}
	\label{thmrefstrongnorm-name}
	If $\Gamma |- e : \sort$, then $e$ is normalizing.
\end{restatable}

\begin{proof}~\\
	\begin{llproof}
		\Pf{}{}{R_{\sort}(e)}{By \Lemmaref{lem:strongnorm-lemma-3-name}}
		\Hand \Pf{}{}{e ~\textrm{halts}~}{By \Lemmaref{lem:strongnorm-lemma-1-name}}
	\end{llproof}
\end{proof}


\section{Strong Normalization Proof for the Index Term Language}
\label{apx:strongnormproof-indexterm}

\begin{definition}[$R_{\sort}(e)$]~
	\label{def:strongnormop-index}
	\begin{enumerate}
                  \item $R_{\NmSet}(e)$ iff $e$ halts.
                  \item $R_{(\sort_1 @> \sort_2)}(e)$ if and only if (1) $e$ halts and (2) if $R_{\sort_1}(e')$ then $R_{\sort_2}(e~e')$.
                  \item $R_{\sort_1 ** \sort_2}(e)$ if and only if $e$ halts and $R_{\sort_1}(e)$ and $R_{\sort_2}(e)$.
	\end{enumerate}
\end{definition} 

\begin{lemma}[Termination]
	\label{lem:strongnorm-lemma-1-index}
	If $R_{\sort}(e)$, then $e$ halts.
\end{lemma}

\begin{proof}
	By definition of $R_{\sort}(e)$.
\end{proof}

\begin{lemma}[Preservation]
	\label{lem:strongnorm-lemma-2-index}
	If $\Gamma |- e : \sort$ and $\sigma |- \Gamma$, then $ |- [\sigma] e : \sort$.
\end{lemma}

\begin{proof}
	By induction on the size of $\sigma$. When size([$\sigma$]) = 1, the Lemma reduces to the Substitution Lemma.
\end{proof}

\begin{lemma}[Normalization]
	\label{lem:strongnorm-lemma-3-index}
	If $\Gamma |- e : \sort$ and $\sigma |- \Gamma$, then $R_{\sort}([\sigma]e)$.
\end{lemma}

\begin{proof} By induction on the given derivation.
	 \DerivationProofCase{sort-var}
		{
			(a : \sort) \in \Gamma
		}
		{
			\Gamma |- a : \sort
		}
	
		\begin{llproof}
			\Pf{\exists v_i.([\sigma]a}{!!}{v_i)}{By definition of $[\sigma](-)$}
			\Pf{}{|-}{[\sigma] a : \sort}{By \Lemmaref{lem:strongnorm-lemma-2-index}}
			\Hand \Pf{}{}{R_{\sort}([\sigma] a)}{Since $[\sigma]a \nteval v_i$ and $|- [\sigma] a : \sort$}
		\end{llproof}
	
	\DerivationProofCase{sort-unit}
		{}
		{
			\Gamma |- \unitindex : \unitsort
		}
	
		\begin{llproof}
			\Pf{[\sigma]\unitindex}{=}{\unitindex}{By definition of $[\sigma](-)$}
			\Pf{\unitindex}{!!}{\unitindex}{By Rule value; Since $\unitindex$ val}
			\Pf{[\sigma]\unitindex}{!!}{\unitindex}{Since $[\sigma]\unitindex = \unitindex$, $\unitindex !! \unitindex$}
			\Pf{\Gamma}{|-}{\unitindex : \unitsort}{Given}
			\Pf{}{|-}{[\sigma] \unitindex : \unitsort}{By \Lemmaref{lem:strongnorm-lemma-2-index}}
			\Hand \Pf{}{}{R_{\unitsort}([\sigma] \unitindex)}{Since $[\sigma]\unitindex !! \unitindex$ and $|- [\sigma] \unitindex : \unitsort$}			
		\end{llproof}
	
	\DerivationProofCase{sort-pair}
		{
			\Gamma |- i_1 : \sort_1
			\\
			\Gamma |- i_2 : \sort_2
		}
		{
			\Gamma |- \Pair{i_1}{i_2} : (\sort_1 ** \sort_2)
		}
		
		\begin{llproof}
			\Pf{}{}{R_{\sort_1}([\sigma] i_1)}{By inductive hypothesis}
			\Pf{}{}{R_{\sort_2}([\sigma] i_2)}{By inductive hypothesis}
			\Pf{\exists v_1.([\sigma] i_1}{!!}{v_1)}{Since $R_{\sort_1}([\sigma] i_1)$}
			\Pf{\exists v_2.([\sigma] i_2}{!!}{v_2)}{Since $R_{\sort_2}([\sigma] i_2)$}
			\Pf{([\sigma] i_1, [\sigma] i_2)}{!!}{(v_1, v_2)}{By Rule pair}
			\Pf{[\sigma](i_1, i_2)}{!!}{(v_1, v_2)}{By definition of $[\sigma](-)$}	
			\Pf{}{|-}{[\sigma] (i_1, i_2) : (\sort_1 * \sort_2)}{By \Lemmaref{lem:strongnorm-lemma-2-index}}
			\Hand \Pf{}{}{R_{(\sort_1 * \sort_2)}([\sigma](i_1\;i_2))}{Since $[\sigma](i_1\;i_2) !! (v_1, v_2)$}\trailingjust{and $~|- [\sigma] (i_1\;i_2) : (\sort_1 ** \sort_2)$}
		\end{llproof}
	
	\DerivationProofCase{sort-proj}
		{
			\Gamma |- i : \sort_1 * \sort_2
		}
		{
			\Gamma |- \Proj{b}{i} : \sort_b
		}

		\begin{llproof}
			\Pf{}{}{R_{(\sort_1 * \sort_2)}([\sigma] i)}{By inductive hypothesis}
			\Pf{}{}{R_{\sort_b}([\sigma] ~\Proj{b}{i})}{By definition of $R_{-}(-)$}
		\end{llproof}
	
	\DerivationProofCase{sort-empty}
		{
			\quad
		}
		{
			\Gamma |- \emptyset : \namesetsort
		}

        Similar to the sort-unit case.
        
	\DerivationProofCase{sort-singleton}
		{
			\Gamma |- N : \namesort
		}
		{
			\Gamma |- \{ N \} : \namesetsort
		}
	
		\begin{llproof}
			\Pf{[\sigma]\{ N \}}{=}{\{ N \}}{By definition of $[\sigma](-)$}
			\Pf{}{}{\{ N \} ~\textrm{val}~}{By Rule val-singleton}
			\Pf{\Gamma}{|-}{\{ N \} : \namesetsort}{Given}
			\Pf{}{|-}{[\sigma] \{ N \} : \namesetsort}{By \Lemmaref{lem:strongnorm-lemma-2-index}}
			\Hand \Pf{}{}{R_{\namesetsort}([\sigma] \{ N \})}{Since $[\sigma] \{ N \}$ val,$~|- [\sigma] \{ N \} : \namesetsort$}			
		\end{llproof}

	\DerivationProofCase{sort-union}
		{
			\arrayenvbl{
				\Gamma |- X : \namesetsort
				\\
				\Gamma |- Y : \namesetsort
			}
		}
		{
			\Gamma |- (X \union Y) : \namesetsort
		}
	
		\begin{llproof}
			\Pf{}{}{R_{\namesetsort}([\sigma] X)}{By inductive hypothesis}
			\Pf{}{}{R_{\namesetsort}([\sigma] Y)}{By inductive hypothesis}
			\Pf{\exists v_1.([\sigma] X}{!!}{v_1)}{Since $R_{\namesetsort}([\sigma] X)$}
			\Pf{\exists v_2.([\sigma] Y}{!!}{v_2)}{Since $R_{\namesetsort}([\sigma] Y)$}
			\Pf{([\sigma] X \union [\sigma] Y)}{!!}{(v_1 \union v_2)}{By Rule union}
			\Pf{[\sigma](X \union Y)}{!!}{(v_1  \union v_2)}{By definition of $[\sigma](-)$}	
			\Pf{}{|-}{[\sigma] (X \union Y) : (\namesetsort)}{By \Lemmaref{lem:strongnorm-lemma-2-index}}
			\Pf{\xHand}{}{R_{\namesetsort}([\sigma](X \union Y))}{Since $[\sigma](X \union Y) !! (v_1 \union v_2)$}\trailingjust{and $~|- [\sigma] (X \union Y) : \namesetsort$}
		\end{llproof}
	
	\DerivationProofCase{sort-sep-union}
		{
			\arrayenvbl{
				\Gamma |- X : \namesetsort
				\\
				\Gamma |- Y : \namesetsort
			}
			\\
			\extract{\Gamma} ||- X \disj Y : \namesetsort
		}
		{
			\Gamma |- (X \disj Y) : \namesetsort
		}
	
		\begin{llproof}
			\Pf{}{}{R_{\namesetsort}([\sigma] X)}{By inductive hypothesis}
			\Pf{}{}{R_{\namesetsort}([\sigma] Y)}{By inductive hypothesis}
			\Pf{\exists v_1.([\sigma] X}{!!}{v_1)}{Since $R_{\namesetsort}([\sigma] X)$}
			\Pf{\exists v_2.([\sigma] Y}{!!}{v_2)}{Since $R_{\namesetsort}([\sigma] Y)$}
			\Pf{([\sigma] X \disj [\sigma] Y)}{!!}{(v_1 \disj v_2)}{By Rule union}
			\Pf{[\sigma](X \disj Y)}{!!}{(v_1 \disj v_2)}{By definition of $[\sigma](-)$}	
			\Pf{}{|-}{[\sigma] (X \disj Y) : (\namesetsort)}{By \Lemmaref{lem:strongnorm-lemma-2-index}}
			\Hand \Pf{}{}{R_{\namesetsort}([\sigma](X \disj Y))}{Since $[\sigma](X \disj Y) !! (v_1 \disj v_2)$}\trailingjust{and $~|- [\sigma] (X \disj Y) : \namesetsort$}
		\end{llproof}

	\DerivationProofCase{sort-abs}
		{
			\Gamma, a : \sort_1 |- i : \sort_2
		}
		{
			\Gamma |- (\Lam{a}{i}) : (\sort_1 @@> \sort_2)
		}
	 
	 	\begin{llproof}
	 		\Pf{[\sigma, a -> \sort_1]}{|-}{\Gamma'}{Assumption, where $\Gamma' = \Gamma, a : \sort_1$}
	 		\Pf{|-}{([\sigma, a -> \sort_1](\lam{a} i))} {: (\sort_1 @@> \sort_2)}{By \Lemmaref{lem:strongnorm-lemma-2-index}}
	 		\Pf{[\sigma, a -> \sort_1] (\lam{a} i)}{=}{(\lam{a} i)}{By definition of $[\sigma](-)$}
\decolumnizePf
	 		\Hand \Pf{R_{\sort}([\sigma, a -> \sort_1] (\lam{a} i))}{}{}{Since $[\sigma, a -> \sort_1](\lam{a} i) !! (\lam{a} i)$} \trailingjust{and $ |- [\sigma, a -> \sort_1]  (\lam{a} i) : (\sort_1 @@> \sort_2)$}
	    \end{llproof}
	    
	\DerivationProofCase{sort-apply}
		{
			\Gamma |- i : \sort_1 @@> \sort_2
			\\
			\Gamma |- j : \sort_1
		}
		{
			\Gamma |- i(j) : \sort_2
		}        
		
		\begin{llproof}
				\Pf{}{}{R_{(\sort_1 @@> \sort_2)}([\sigma] i)}{By inductive hypothesis}
				\Pf{}{}{R_{\sort_1}([\sigma] j)}{By inductive hypothesis}
				\Pf{}{}{R_{\sort_2}(([\sigma]i([\sigma]j)))}{By definition of $R_{-}(-)$}	
				\Hand \Pf{}{}{R_{\sort_2}([\sigma](i(j)))}{By definition of $[\sigma](-)$}	
		\end{llproof}
	\DerivationProofCase{sort-map-set}
		{
			\Gamma |- M : \namesort @> \namesort
			\\
			\Gamma |- j : \namesetsort
		}
		{
			\Gamma |- M[[j]] : \namesetsort
		}
	
		\begin{llproof}
			\Pf{}{}{R_{\namesetsort}(j)}{By inductive hypothesis}

			\Pf{\exists v.(M[[j]]}{!!}{v)}{By \Thmref{thmrefstrongnorm-name}}
			\Pf{\exists v'.(M[[j]]}{\leadsto}{v)}{By Rule Single}
			\Pf{M[[j]]}{!!}{v'}{By Rule map-set}
			\Pf{|-}{([\sigma](M[[j]])} {: \namesetsort}{By \Lemmaref{lem:strongnorm-lemma-2-index}}
                        \decolumnizePf
			\Hand \Pf{}{}{R_{\namesetsort}([\sigma](M[[j]])}{Since $M[[j]] =v'$ and $|- ([\sigma](M[[j]]) : \namesetsort$}

		\end{llproof}	
\end{proof}

\begin{restatable}[Strong Normalization]{theorem}{thmrefstrongnormindex}
	\label{thmrefstrongnorm-index}
	If $\Gamma |- e : \sort$, then $e$ is normalizing.
\end{restatable}

\begin{proof}~\\
	\begin{llproof}
		\Pf{}{}{R_{\sort}(e)}{By \Lemmaref{lem:strongnorm-lemma-3-index}}
		\Hand \Pf{}{}{e ~\textrm{halts}~}{By \Lemmaref{lem:strongnorm-lemma-1-index}}
	\end{llproof}
\end{proof}


\fi
\end{document}